\pdfoutput=1
\documentclass[acmsmall]{acmart}

\def\setnoreview{}
\def\setnocomments{} 
\newif\ifnocomments
\newif\ifoverpage
\newif\ifnoreview
\newif\ifnofull

\ifdefined\setnocomments
	\nocommentstrue
\else
	\nocommentsfalse
\fi
\ifdefined\setoverpage
	\overpagetrue
\else
	\overpagefalse
\fi
\ifdefined\setnoreview
	\noreviewtrue
\else
	\noreviewfalse
\fi
\ifdefined\setnofull
	\nofulltrue
\else
	\nofullfalse
\fi

\setcopyright{rightsretained}
\acmDOI{10.1145/3632875}
\acmYear{2024}
\copyrightyear{2024}
\acmSubmissionID{popl24main-p166-p}
\acmJournal{PACMPL}
\acmVolume{8}
\acmNumber{POPL}
\acmArticle{33}
\acmMonth{1}
\received{2023-07-11}
\received[accepted]{2023-11-07}

\usepackage{hyperref}
\usepackage{booktabs}
\usepackage{subcaption}
\usepackage{bm}
\usepackage{cleveref}
\usepackage{paralist}
\usepackage{enumitem}
\usepackage{colortbl}
\usepackage{multirow}
\usepackage{multicol}
\usepackage{thmtools}
\usepackage{sidecap}
\usepackage{xparse}
\usepackage{array}
\usepackage{wrapfig}
\usepackage{listings}
\usepackage{tikz}
\usetikzlibrary{arrows}
\usetikzlibrary{automata}
\usetikzlibrary{positioning}
\usetikzlibrary{tikzmark}
\usetikzlibrary{fit}
\usetikzlibrary{shapes.geometric}

\usepackage{arydshln}
\usepackage{framed}
\usepackage{empheq}  

\crefname{definition}{Def.}{Def.}
\crefname{figure}{Fig.}{Fig.}
\crefname{example}{Ex.}{Ex.}
\crefname{section}{Sec.}{Sec.}
\crefname{appendix}{App.}{App.}
\crefname{theorem}{Thm.}{Thm.}
\crefname{table}{Tab.}{Tab.}
\crefname{equation}{eq.}{eqs.}

\lstdefinelanguage{ivy}{
	keywords=[1]{module, sort, type, types, relation, individual,
		definition, transition, invariant, function, axiom,
		assume, init, require},
	keywordstyle=[1]{\bfseries},
	keywords=[2]{true,false},
	keywordstyle=[2]{\bfseries\color{purple}},
	keywords=[3]{mutable,immutable},
	keywordstyle=[3]{\bfseries},
	comment=[l]{\#}
}

\let\origthelstnumber\thelstnumber
\makeatletter
\newcommand*\SuppressNumber{\lst@AddToHook{OnNewLine}{\let\thelstnumber\relax \advance\c@lstnumber-\@ne\relax }}

\newcommand*\ReactivateNumber{\lst@AddToHook{OnNewLine}{\let\thelstnumber\origthelstnumber \advance\c@lstnumber\@ne\relax}}
\makeatother

\tikzset{tnode/.style={circle,draw,minimum size=23pt,inner sep=0pt}}
\tikzset{vnode/.style={regular polygon,regular polygon sides=4,draw,minimum size=20pt}}

\newif\iflong
\longfalse

\newcommand{\sharon}[1]{{\textcolor{purple}{SS: {\em #1}}}}
\newcommand{\neta}[1]{{\textcolor{teal}{NE: \em #1}}}
\newcommand{\oded}[1]{{\textcolor{blue}{OP: \em #1}}}

\ifnocomments
\renewcommand{\sharon}[1]{\ifdim\lastskip>0pt\ignorespaces\fi}
\renewcommand{\neta}[1]{\ifdim\lastskip>0pt\ignorespaces\fi}
\renewcommand{\oded}[1]{\ifdim\lastskip>0pt\ignorespaces\fi}
\fi
\newcommand{\commentout}[1]{\ifdim\lastskip>0pt\ignorespaces\fi}

\newcommand{\Naturals}{{\mathbb{N}}}
\newcommand{\Nat}{{\Naturals}}
\newcommand{\Integers}{{\mathbb{Z}}}
\newcommand{\Templates}{{\mathbb{\struct}}}
\newcommand{\Heuristic}{{\mathbb{H}}}

\newcommand{\Templatesepr}{{\Templates}}
\newcommand{\TemplatesExtended}{{\Templates^{++}}}

\newcommand{\Sorts}{{\mathcal{S}}}
\newcommand{\Domain}{{\mathcal{D}}}
\newcommand{\Interp}{{\mathcal{I}}}
\newcommand{\Bounds}{{\mathcal{B}}}
\newcommand{\ElemsOf}{{\mathcal{E}_\mathcal{\Bounds}}}
\newcommand{\Concrete}{{\mathcal{E}}}
\newcommand{\Variables}{{\mathcal{X}}}
\newcommand{\Logic}{{\mathcal{L}}}
\newcommand{\Order}{{\mathcal{O}}}
\newcommand{\Function}{{\mathcal{F}}}

\newcommand{\FinderOf}[1]{{#1^\sharp}}
\newcommand{\finder}{\FinderOf{\struct}}
\newcommand{\InterpF}{\FinderOf{\Interp{}}}
\newcommand{\BoundsF}{\FinderOf{\Bounds{}}}

\newcommand{\BoundsH}{\FinderOf{\Bounds_\Heuristic}}
\newcommand{\InterpH}{\FinderOf{\Interp_\Heuristic}}
\newcommand{\SkeletonFreeVariables}{SFV}

\newcommand{\varphikey}{{\varphi_{\finder}}}
\newcommand{\varphiaux}{{\varphi_{\textit{aux}}}}

\newcommand{\sort}{{\bm{s}}}

\newcommand{\TemplateTransform}{{\text{trans}}}
\newcommand{\FinderTransform}{{\text{transF}}}

\newcommand{\LIA}{{\text{LIA}}}
\newcommand{\Terms}{{\mathcal{T}}}
\newcommand{\QF}{{\text{QF}}}
\newcommand{\FV}{{\text{FV}}}

\newcommand{\Prop}{{\text{Prop}}}

\newcommand{\powerset}[1]{{\mathcal{P}(#1)}}
\newcommand{\parens}[1]{{\left( #1 \right)}}
\newcommand{\braces}[1]{{\left\{ #1 \right\}}}
\newcommand{\brackets}[1]{{\left[ #1 \right]}}
\newcommand{\abs}[1]{{\left| #1 \right|}}
\newcommand{\pair}[1]{{\left\langle #1 \right\rangle}}

\newcommand{\llt}{\mathrel{\prec}}
\newcommand{\nllt}{\mathrel{\not\prec}}
\newcommand{\lleq}{\mathrel{\preccurlyeq}}
\newcommand{\lgt}{\mathrel{\succ}}
\newcommand{\Allt}{A_{\prec}}
\newcommand{\foleq}{\mathrel{{\approx}}}
\newcommand{\liaeq}{\mathrel{{\approx}}}

\newcommand{\subLIA}{{_\LIA}}
\newcommand{\supLIA}{{\text{\tiny LIA}}}

\newcommand{\Next}{{\FunDef{next}}}
\newcommand{\Sorted}{{\Relation{sorted}}}
\newcommand{\List}{{\Relation{list}}}
\newcommand{\ListSegment}{{\Relation{lseg}}}

\newcommand{\GuardFlag}[1]{{b_{ #1 }}}
\newcommand{\RawIntFlag}{d}
\NewDocumentCommand{\IntFlag}{G{}G{}}{{\RawIntFlag^{ #2 }_{ #1 }}
}
\DeclareMathOperator{\Eq}{eq}

\newcommand{\Sort}[1]{{\texttt{\small\bfseries{#1}}}}
\newcommand{\Const}[1]{{\texttt{\small #1}}}
\newcommand{\Relation}[1]{\texttt{\small #1}}
\newcommand{\FunDef}[1]{\texttt{\small #1}}
\newcommand{\Action}[1]{\texttt{\small #1}}
\newcommand{\code}[1]{\texttt{\small #1}}
\newcommand{\Previous}{prev}
\newcommand{\PreviousFun}{{\FunDef{\Previous{}}}}
\newcommand{\TinyPrevious}{{\FunDef{\figLfont{\Previous{}}}}}

\newcommand{\liff}{{\leftrightarrow}}

\newcommand{\PreStateColor}{magenta}
\newcommand{\PostStateColor}{cyan}
\newcommand{\OrderColor}{gray}

\newcommand{\FragmentShortName}{OSC}
\newcommand{\FragmentLongName}{Ordered Self Cycle}

\newcommand{\ToolShortName}{FEST}
\newcommand{\ToolLongName}{Find and Evaluate Symbolic structures via Templates}

\newcommand{\Zzz}{Z3}
\newcommand{\cvc}{\textsc{cvc5}}
\newcommand{\Solvers}{SMT solvers}

\AtEndPreamble{\newtheorem{remark}[theorem]{Remark}}

\newcommand{\para}[1]{\vspace{2pt}\textit{#1.}}

\newcolumntype{H}{>{\setbox0=\hbox\bgroup}c<{\egroup}@{}}

\newcommand{\BellNumber}{b}

\newcommand{\concrete}{explicit}

\newcommand{\aConcrete}{\aBeforeConcrete{} explicit}
\newcommand{\AConcrete}{\ABeforeConcrete{} explicit}
\newcommand{\concretization}{explication}
\newcommand{\Concretization}{Explication}
\newcommand{\concretize}{explicate}

\newcommand{\aBeforeConcrete}{an}
\newcommand{\ABeforeConcrete}{An}

\newcommand{\structure}{symbolic structure}
\newcommand{\structAdj}{symbolic-structure}
\newcommand{\StructAdj}{Symbolic-structure}
\newcommand{\aStructAdj}{\aBeforeStructure{} \structAdj{}}
\newcommand{\AStructAdj}{\ABeforeStructure{} \structAdj{}}
\newcommand{\structures}{\structure{}s}
\newcommand{\Structure}{Symbolic structure}
\newcommand{\Structures}{\Structure{}s}
\newcommand{\STructure}{Symbolic Structure}
\newcommand{\STructures}{\STructure{}s}
\newcommand{\template}{template}
\newcommand{\templates}{\template{}s}
\newcommand{\Template}{Template}
\newcommand{\TemplatesWord}{\Template{}s}
\newcommand{\aBeforeStructure}{a}
\newcommand{\ABeforeStructure}{A}
\newcommand{\aStruct}{\aBeforeStructure{} \structure{}}
\newcommand{\AStruct}{\ABeforeStructure{} \structure{}}
\newcommand{\aBeforeTemplate}{a}
\newcommand{\ABeforeTemplate}{A}
\newcommand{\aTemplate}{\aBeforeTemplate{} \template{}}
\newcommand{\ATemplate}{\ABeforeTemplate{} \template{}}
\newcommand{\struct}{S}
\newcommand{\structAss}{v_S}
\newcommand{\barStructAss}{\bar v_S}
\newcommand{\structAssignment}{symbolic assignment}
\newcommand{\structAssignments}{\structAssignment{}s}
\newcommand{\StructAssignment}{Symbolic assignment}
\newcommand{\StructAssignments}{\StructAssignment{}s}

\newcommand{\aBeforeStructAssignment}{a}
\newcommand{\aStructAssignment}{\aBeforeStructAssignment{} \structAssignment{}}
\newcommand{\structModel}{symbolic model}
\newcommand{\structModels}{\structModel{}s}
\newcommand{\aStructModel}{\aBeforeStructure{} \structModel{}}
\newcommand{\templatization}{symbolization}

\newcommand{\BeyondFormula}{\varphi^{+}}

\newcommand{\figLfont}[1]{{\fontsize{7pt}{7pt}\selectfont #1}}

\begin{document}
\title[An Infinite Needle in a Finite Haystack]{
An Infinite Needle in a Finite Haystack: Finding Infinite Counter-Models in Deductive Verification
}         \author{Neta Elad}
\orcid{0000-0002-5503-5791}
\affiliation{\institution{Tel Aviv University}
	\city{Tel Aviv}
	\country{Israel}
}
\email{netaelad@mail.tau.ac.il}

\author{Oded Padon}
\orcid{0009-0006-4209-1635}
\affiliation{\institution{VMware Research}
	\city{Palo Alto}
	\country{USA}
}
\email{oded.padon@gmail.com}

\author{Sharon Shoham}
\orcid{0000-0002-7226-3526}
\affiliation{\institution{Tel Aviv University}
	\city{Tel Aviv}
	\country{Israel}
}
\email{sharon.shoham@cs.tau.ac.il} 

\begin{abstract}
\neta{this is a small comment to make sure I compiled without comments}

First-order logic, and quantifiers in particular, are widely used in
deductive verification of programs and systems. Quantifiers are
essential for describing systems with unbounded domains, but prove
difficult for automated solvers. Significant effort has been dedicated
to finding quantifier instantiations that establish unsatisfiability
of quantified formulas, thus ensuring validity of a system's
verification conditions. However, in many cases the formulas are
satisfiable---this is often the case in intermediate steps of the
verification process, e.g., when an invariant is not yet
inductive. 
For such cases, existing tools are limited to finding
\emph{finite} models as counterexamples.  
Yet, some quantified formulas are satisfiable but only
have \emph{infinite} models,
which current solvers are unable to find.
Such infinite counter-models are especially typical when 
first-order logic
is used to approximate the natural numbers, the integers, 
or other inductive definitions such as linked lists, 
which is common in
deductive verification.
The inability of solvers to find infinite models
makes them diverge 
in these cases,
providing little feedback 
to the user
as they try to make progress 
in their verification attempts.

In this paper,
we tackle the problem of finding such infinite models,
specifically, finite representations thereof
that can be presented to the user of a deductive verification tool.
These models give insight into the verification failure,
and allow the user to identify
and fix bugs in the modeling of the system and its properties.
Our approach consists of three parts.
First, we introduce \emph{\structures{}} 
as a way to represent
certain infinite models,
and show
they admit an efficient model checking procedure.
Second, we describe an effective model finding procedure that
symbolically explores a given (possibly infinite) family of \structures{}
in search of an infinite model for a given formula.
Finally, we identify a new decidable fragment of first-order logic
that extends and subsumes the many-sorted variant of EPR,
where satisfiable formulas always have a model
representable by \aStruct{}
within a known family, making our model finding procedure 
a decision procedure for that fragment.

\commentout{
We implement our approach
in a tool called \ToolShortName{}
(\ToolLongName{}),
and evaluate it
on examples from a variety of domains:
distributed consensus protocols, \oded{maybe remove ``consensus''}
linked lists,
and axiomatic arithmetic.
Our results show that \ToolShortName{} quickly finds
infinite counter-models that distill the source of failure
and demonstrate it in a simple way,
even in cases beyond the decidable fragment,
while state-of-the-art \Solvers{} such as \Zzz{} and \cvc{}
and resolution-based theorem provers such as Vampire
diverge or return ``unknown''.
}

We evaluate our approach
on examples from 
the domains of distributed consensus protocols
and of heap-manipulating programs (specifically, linked lists).
Our implementation quickly finds
infinite counter-models that demonstrate the source of verification failures
in a simple way,
while state-of-the-art \Solvers{} and theorem provers such as \Zzz{}, \cvc{}, and Vampire
diverge or return ``unknown''.

 \end{abstract}
  \maketitle

\section{Introduction}
A plethora of tools and techniques
have emerged in recent years
that use first-order logic (FOL) with quantifiers for verification~\cite{paxos-made-epr,mcmillan-decidable-ivy,linked-lists-epr,modularity-for-decidability,verification-modulo-axioms,bounded-horizon,ic3po,natural-proofs,fossil,distai,duoai,updr,swiss,ic3po-nfm,i4,p-fol-ic3,induction-duality,vericon-networks}.
These works all use
quantified formulas in FOL
to encode the behavior of systems
and specify desired properties and inductive invariants.
Using FOL to model
(or abstract)
verification problems
has the advantage of completeness,
meaning that every formula has either
a refutation or a satisfying model.
Furthermore, FOL allows quantification,
which is essential for
reasoning about unbounded domains, such as
the set of
processes participating in a distributed protocol,
the set of
messages exchanged in a network, or
the set of
objects in the heap. Following previous work,
we target FOL without theories,
which
focuses the verification challenge on dealing with quantifiers,
while side-stepping
the difficulty of combining theories
with quantifiers and uninterpreted symbols.

Quantifiers provide expressivity,
but pose a significant challenge
for automated verification.
\Solvers{}~\cite{z3-smt-solver,cvc5-smt-solver,yices-smt-solver}
and
theorem provers~\cite{vampire,spass}
often struggle and diverge
while searching
for a refutation or a satisfying model,
in which case they time out or return ``unknown''.

In many cases this divergence is not simply bad luck,
but a result of a more fundamental problem:
formulas that are satisfiable but only have infinite models.
Though infinite models typically
do not represent concrete states of a system,
their existence may arise from
the use of FOL to approximate a stronger logic,
for example least fixed-point.
In such cases,
the solution is to add additional axioms
that eliminate spurious behaviors~\cite{natural-proofs,fossil,vampire-integer-induction,induction-for-smt}.
Infinite counter-models
could provide useful guidance
on how to refine the FOL modeling with additional axioms,
and ultimately allow the user
to successfully complete the verification task.
However, existing solvers cannot find infinite counter-models.
As a result, when the FOL modeling
is still missing some axiom
(e.g., an induction principle),
solvers diverge,
leaving the user
with an inconclusive result
and no actionable information.

While a significant amount of research
has been dedicated to
developing algorithms and heuristics for
finding quantifier instantiations for
refutation~\cite{complete-instantiation-smt,array-property-fragment,local-theory-ematching,incremental-instance-generation,axioms-with-triggers,ematching-for-smt,simplify,
refinement-reflection},
ensuring that finite models exist~\cite{paxos-made-epr,linked-lists-epr,modularity-for-decidability},
and discovering finite models~\cite{finite-model-finding,quantifier-finite-model-finding,constructing-bachmair-models,finite-models-vampire},
less attention has been given to finding infinite models
in the context of verification.
Although there are works on proving the existence or constructing infinite models
(e.g., based on a saturated set of clauses
in resolution-based theorem proving~\cite{bachmair-original,bachmair-infinite-models},
or using the model evolution calculus~\cite{model-evolution-calculus,model-evolution-implementation}; see \Cref{sec:related} for further discussion),
finding infinite models has not (yet) made its way into common verification frameworks based on SMT solvers.
In this paper,
we aim for a technique for finding infinite models
that can work well in the context of SMT-based deductive verification.
To that end, we seek a representation of first-order structures
that captures models that appear in real-world problems,
admits an effective search procedure,
and can be communicated to users.

This paper develops a finite representation
of (certain) infinite models,
and an effective search procedure
for such models.
Our key observation is that many of the infinite models arising in practice have
repeating patterns
that can be summarized in a finite way.
We are motivated by examples from various domains
such as
heap-manipulating programs
(in particular usage of linked lists)~\cite{natural-proofs,fossil}
and distributed protocols~\cite{ivy-original,
paxos-made-epr}
where a linearly ordered set is used to model
reachability of objects in the heap
or consecutive rounds of execution
(respectively).
Importantly, these linear orders are intended to be finite,
but may be infinite under FOL semantics.
Accordingly, we introduce \emph{\structures{}}
as finite representations of possibly infinite models.
\Structures{} use finitely many
\emph{summary nodes} to capture
possibly infinite sets of elements,
and use the language of Linear Integer Arithmetic (LIA)
to describe repeating patterns. This allows us to check if
\aStruct{} satisfies an FOL formula
by translating the formula into a (quantified) LIA formula,
which solvers such as \Zzz{}~\cite{z3-smt-solver}
or \cvc{}~\cite{cvc5-smt-solver}
can solve quickly.
Our LIA-based \structures{}
are particularly useful when verification uses FOL
to (soundly) abstract common linear orders
such as the natural numbers or the integers,
or to model
(semi-) linear data structures such as linked lists.

Given that checking
if a specific \structure{} satisfies a formula
is reducible to LIA,
we can obtain a procedure
that searches for \aStructModel{} (i.e., a satisfying \structure{})
for a given formula by enumerating \structures{}.
However, such a procedure is inefficient.
Instead of enumerating all possible \structures{},
we introduce an algorithm that can efficiently search
a (possibly infinite) space of \structures{},
given by \aBeforeTemplate{} \emph{\template{}}.
\ATemplate{} specifies the number of
\structAdj{}
nodes,
as well as
a subset of the LIA language
to be used within the \structures{}.
The search algorithm avoids
explicitly constructing all potential \structures{}.
Rather, it symbolically encodes
the entire family of \structures{}
described by the \template{}
into a single (quantified) LIA formula
that, when satisfiable,
induces \aStructModel{}
for the FOL formula at hand.
This search procedure is sound
for
establishing the satisfiability of
any formula in FOL:
if \aStruct{} is found, the formula is satisfiable.
Further, we suggest a simple heuristic \template{},
parameterized by size,
that captures non-trivial infinite models
for a variety of examples.
As a notable example,
using such \aTemplate{},
we are able to find
an infinite counter-model showing that
the original
inductive invariant of the
Paxos distributed consensus protocol~\cite{lamport-paxos}
is not inductive in its
FOL formalization given in~\cite{ic3po}.

Although not every satisfiable FOL formula
has \aStructModel{}
(as expected,
given that satisfiability in FOL
is not recursively enumerable),
we identify a new decidable fragment of FOL
where that is the case,
i.e., each satisfiable formula
has \aStructModel{}.
We further prove a ``small \structModel{} property,''
\begin{inparaenum}\item
	bounding the size of potentially satisfying \structures{}
	for formulas in the fragment, and
	\item
	characterizing a finite subset of the LIA language
	that is sufficient for them.
\end{inparaenum}
This makes our search algorithm
a decision procedure for the fragment.
Interestingly, our decidability proof departs from the
widely-used approach of establishing
a
finite model property,
as the underlying models we consider are infinite.

The new decidable fragment, called \FragmentLongName{}
(\FragmentShortName{}), is an extension of the
many-sorted variant of the decidable
Effectively PRopositional fragment
(EPR, also known as the
Bernays-Sch\"onfinkel-Ramsey class)~\cite{bsr-epr,epr-decidable}\footnote{Following previous papers, e.g.,~\cite{paxos-made-epr},
we would henceforth take EPR
to refer to the extension of the original fragment from~\cite{bsr-epr}
to many-sorted logic.
The extension allows quantifier alternations and function symbols,
as long as no cycles are introduced,
see \Cref{sec:decidable} for a precise definition.
The proofs of decidability and small model property of EPR
carry over to the many-sorted extension.
}.
Many-sorted EPR already proved useful
in verifying systems from various domains,
e.g., distributed protocols~\cite{paxos-made-epr},
networking systems~\cite{vericon-networks},
and heap-manipulating programs~\cite{linked-lists-epr,heap-paths-epr}.
\FragmentShortName{} extends many-sorted EPR
in that it relaxes
some of the restrictions on quantifier alternations
and function symbols.
Namely, \FragmentShortName{} allows one sort,
equipped with a linear order,
to have cyclic functions
(i.e., functions mapping the sort to itself),
which is strictly beyond many-sorted EPR,
and can be useful in modeling verification problems.

There is a rich literature
on decidable fragments of first-order logic
(see e.g.~\cite{voigt-thesis,gurevich-book}
and references therein).
The \FragmentShortName{} fragment is unique in that landscape
as it extends EPR with cyclic functions,
and,
unlike monadic logics
(e.g.,~\cite{loeb1967-decidability,shelah1977decidability}),
allows a linear order;
to the best of our knowledge
it is not subsumed by any existing decidable fragment.

In summary, this paper makes the following contributions:
\begin{itemize}
	\item We introduce \emph{\structures{}},
	a finite representation of infinite structures,
	in which functions and relations
	are defined using LIA terms and formulas
	(\Cref{sec:templates});
	\item We provide an efficient way to find \aStructModel{}
	across a given (possibly infinite) search space,
	without an explicit enumeration,
	by reduction to LIA
	(\Cref{sec:finding});
	\item We identify a new decidable fragment of FOL,
	which extends many-sorted EPR,
	where the satisfiability problem can be solved
	by restricting the search space to a set of \structures{} of a bounded size, representing (possibly) \emph{infinite} models.
	(\Cref{sec:decidable});
	\item
We implement a tool called \ToolShortName{}
	(\ToolLongName{})
	that finds \structModels{} for FOL formulas
	using heuristically computed \templates{},
	and evaluate it on
examples
	from the domains
	of distributed protocols
	and of heap-manipulating programs (specifically, linked lists),
	which exhibit infinite counter-models.
	While existing tools fail to produce an answer,
	\ToolShortName{} finds \structModels{}
	for all 21 examples in our evaluation.
(\Cref{sec:evaluation}).
\end{itemize}

The rest of this paper is organized as follows.
\Cref{sec:overview} gives an informal overview
of the key ideas.
\Cref{sec:background} provides necessary background.
\Cref{sec:templates,sec:decidable,sec:finding,sec:evaluation}
present the main contributions listed above.
Finally, \Cref{sec:related} discusses related work
and \Cref{sec:conclusion} concludes.
\neta{if full}
\ifnofull
For technical proofs, see~\cite{infinite-needle-arxiv}.
\else
Technical proofs are deferred to \Cref{sec:proofs}.
\fi

\section{Overview}
\label{sec:overview}

In this section we use a simple example to demonstrate our approach
for representing infinite models and finding them automatically.
The example also demonstrates
the kind of verification problems expressible
in the decidable \FragmentShortName{} fragment.

\subsection{Motivation}
Quantification is prevalent when modeling systems
over unbounded domains in first-order logic.
As soon as quantifier alternations are introduced, infinite models,
which are usually not intended, may arise.
An example for the emergence of infinite models is
the Paxos~\cite{lamport-paxos} distributed protocol for consensus.
In~\cite{paxos-made-epr}, Paxos was verified in first-order logic,
by carefully rewriting and adapting the formalization
of both the protocol and its inductive invariant to ensure
that the verification conditions are in EPR,
avoiding problematic quantifier alternations.
However, this process requires creativity and can be quite tricky.
It is therefore desirable to verify Paxos using its natural first-order modeling,
avoiding rewrites.
Indeed, recently,~\cite{ic3po} considered
the natural FOL modeling of Paxos
as a benchmark for automatically inferring inductive invariants.
The invariants inferred match the natural hand-written ones~\cite{lamport-paxos},
but to obtain a complete mechanically-checked proof,
one needs to mechanically verify their inductiveness.
(Recall that in a typical use-case of invariant inference,
no reference inductive invariant is known.)
Unfortunately, the inductiveness of the inferred Paxos invariants
hinges on the property
that in every moment in time,
the sequence of rounds
(also known as ballots)
from the beginning of the protocol is finite,
i.e., induction on rounds is required.
This property is lost in the FOL modeling of the protocol.
As a result, the invariants are \emph{not} inductive in FOL,
and exhibit an infinite counter-model,
in which an infinite chain of rounds leads
to a violation of consistency.
This counter-model reveals the need for induction
over the sequence of rounds as part of the inductiveness proof.

The infinite counter-model is unintended, and perhaps unnatural,
but it is an unavoidable artifact of modeling in FOL.
Therefore, a method to describe and find such infinite models
is valuable for
deductive verification in FOL.
Presenting these models to the user
provides a better experience than
timing out,
as current tools do,
and allows the user to gain insight into the problem,
and refine
their modeling to eliminate these infinite models
(e.g., by adding instances of induction axioms).

\subsection{Running Example: The Echo Machine}

Since Paxos is too complex to be used as a running example,
we demonstrate our approach for representing and finding infinite counter-models on the verification problem
of an Echo Machine,
a simple system that captures concepts also found
in more complex examples, such as Paxos.

The Echo Machine is a reactive system
that starts by reading some value as input and echoing it,
and later, upon request, recalls a previously echoed value
and echoes it once more.
It is designed to satisfy the safety property that any
two values echoed by the machine
are identical.
We use the problem of verifying that
the Echo Machine satisfies this property as our running example.

\paragraph{Modeling in First-Order Logic}
We model the verification problem of the Echo Machine
in first order logic,
using an Ivy-like language~\cite{ivy-original,ivy-tool},
as presented in \Cref{fig:overview:ivy}.
The modeling introduces a $\Sort{value}$ sort and a $\Sort{round}$ sort, a constant $\Const{start}$ of sort $\Sort{round}$
to record the start round,
and a relation
$\Relation{echo} \colon \Sort{round} \times \Sort{value}$
to record
echoed values.

At first, modeling $\Sort{round}$ explicitly
as the natural numbers (or integers)
using the corresponding theory
might seem attractive.
However, current off-the-shelf \Solvers{} do not handle
combinations of uninterpreted functions with
interpreted numerical sorts well
in the presence of quantified formulas.
Indeed, in our example,
they get stuck and ultimately return ``unknown''.

We therefore abstract the $\Sort{round}$ sort
into an uninterpreted sort,
and introduce the $\llt$ relation,
axiomatized to be a strict linear order relation,
to represent the $<$ order over the natural numbers.
Additionally, the $\Const{start}$ constant is
axiomatized to be the least element in $\Sort{round}$,
according to $\llt$.

The behavior of the Echo Machine is modeled
as a first-order transition system,
where the state of the system is maintained by the
$\Relation{echo}$ relation,
and the transitions are
$\Action{echo\_start}$ and $\Action{echo\_prev}$.
The semantics of each transition
is given by a transition relation formula
over a ``double vocabulary'',
where the pre-state of a transition is denoted
by the plain $\Relation{echo}$ relation,
and the post-state is denoted by $\Relation{echo}'$.
For example, the transition relation for
$\Action{echo\_start}$ is:
\begin{equation}\label{eq:overview:transition-relation}
\exists \code{v}.\;
\parens{ \forall \code{V}.\; \neg\code{echo(start,V)} }
\land
\parens{
	\forall \code{R,V}.\; \code{echo}'\code{(R,V)}
	\leftrightarrow
	\parens{
          \code{echo(R,V)}
          \lor
	   \parens{ \code{R} = \code{start} \land \code{V} = \code{v} }
	}
}.
\end{equation}
That is, the transition is only enabled
when no value is echoed at $\Const{start}$,
as given by the $\code{\bfseries assume}$ statement
on line~\ref{line:ivy:assume-not-echo};
the transition updates the $\Relation{echo}$ relation
by adding a single pair $\parens{\code{start}, \code{v}}$,
as given by line~\ref{line:ivy:start-echo}.
The $\Action{echo\_prev}$ transition is enabled if no value
has been echoed at round $\code{r}$
(line~\ref{line:ivy:echo-prev-assume}),
and value $\code{v}$
has been echoed at some round $\code{R} \llt \code{r}$
(see line~\ref{line:ivy:prev-echoed},
$\code{prev\_echoed}$ definition);
it then echoes $\code{v}$ at round $\code{r}$
(line~\ref{line:ivy:echo-prev-update}).
The initial condition, that no value has been echoed, and the safety property, that all echoed values are the same,
are also written in FOL
(lines~\ref{line:ivy:init} and~\ref{line:ivy:safety},
respectively).

\begin{figure}[t]
	\begin{lstlisting}[language=ivy,multicols=2]
	sort round, value
	immutable individual start: round
	mutable relation echo: round @$\times$@ value
	immutable relation @$\llt$@: round @$\times$@ round
	axiom strict_linear_order(@$\llt$@)
	axiom @$\forall$@R. R @$\neq$@ start @$\Rightarrow$@ start @$\llt$@ R

	definition prev_echoed(r: round, v: value):
		@\label{line:ivy:prev-echoed}@@$\exists$@R. R @$\llt$@ r @$\land$@ echo(R, v)
		
	init @$\forall$@R,V. @$\neg$@echo(R, V) @\label{line:ivy:init}@@\SuppressNumber@
	@\ReactivateNumber@@\SuppressNumber@
	@\ReactivateNumber@
	transition echo_start(v: value):
		assume @$\forall$@V. @$\neg$@echo(start, V)   				@\label{line:ivy:assume-not-echo}@
		echo(start, v) @$\gets$@ true       	@\label{line:ivy:start-echo}@
	transition echo_prev(r: round, v: value):
		@\label{line:ivy:echo-prev-assume}@assume (@$\forall$@V. @$\neg$@echo(r, V))@\SuppressNumber@
			@$\land$@ prev_echoed(r, v)@\ReactivateNumber@
		@\label{line:ivy:echo-prev-update}@echo(r, v) @$\gets$@ true
	invariant [safety] @$\forall$@V@$_1$@,V@$_2$@.@\SuppressNumber@
		((@$\exists$@R.echo(R, V@$_1$@)) @$\land$@ (@$\exists$@R.echo(R, V@$_2$@)))
		@$\Rightarrow$@ V@$_1$@ = V@$_2$@@\label{line:ivy:safety}@ @\ReactivateNumber@
	invariant @$\forall$@R,V.@\label{line:ivy:inductive-invariant}@@\SuppressNumber@
		(echo(R, V) @$\land$@ R @$\neq$@ start) 
		@$\Rightarrow$@ prev_echoed(R, V)@\label{line:ivy:invariant}@ @\ReactivateNumber@
	\end{lstlisting}
\iflong \else \vspace{-0.4cm} \fi
\caption{The Echo Machine in pseudo-Ivy.}
\label{fig:overview:ivy}
\end{figure}

To establish that the system satisfies the safety property, an \emph{inductive invariant} that implies it is specified.
An inductive invariant is a property that holds in the initial states of the system and is preserved by every transition,
and hence holds in all the reachable states.
The verification problem is then to check that the conjunction of the provided invariants forms an inductive invariant.

\paragraph{Checking inductiveness as a satisfiability problem}
Verifying that a formula $\code{INV}$
is inductive amounts to checking
unsatisfiability of the
\emph{verification condition} (VC) formulas
$\code{INV} \wedge \code{TR}_i \wedge \neg \code{INV}'$
for every transition, where
$\code{TR}_i$ is the transition formula
and $\code{INV}'$
is obtained from $\code{INV}$ by substituting
all mutable symbols by their primed versions.
When the invariant is not inductive,
a model of the VC formula for
some transition is a \emph{counterexample to induction} (CTI)---a
pair of states where the pre-state
satisfies the invariant
but the post-state obtained after
executing the transition does not.\footnote{Some papers consider the pre-state
of the transition violating inductiveness of the invariant as a CTI.
In this paper,
we follow the convention that the CTI consists of
both the pre-state and the post-state.}

It is easy to see that the Echo Machine
satisfies its safety property,
but the property alone is not inductive,
as it is not preserved by the \code{echo\_start} transition:
starting from a state where one value has been echoed at round
$\neq$ \code{start},
the \code{echo\_start} transition
can echo a different value at round \code{start},
violating the property.
(SMT solvers are able to find such a CTI.)
To make the safety property inductive, the invariant in line~\ref{line:ivy:invariant} is added,
stating that every value that is echoed at round $\neq \Const{start}$
was previously echoed.
This is meant to ensure that
all echoed values are equal to the one echoed at
$\Const{start}$, which proves the safety property.\footnote{\label{note:paxos}The mechanism of the Echo Machine for ensuring
that a value is ``safe to echo,''
by introducing rounds and using
a previous round as a witness,
is reminiscent of the consensus mechanism in Paxos~\cite{lamport-paxos},
where a value is ``safe to propose''
if it was previously voted for by a member of a quorum
and none of the quorum members voted
for another value since.}

Despite its simplicity,
this example is beyond known decidable fragments of FOL
and existing \Solvers{} as well as resolution-based theorem provers
are unable to prove (or disprove) it.

\subsection{Infinite Counterexample to Induction}
In fact, the inability of solvers to handle this problem
is not 
incidental.
In contrast to the intuition above
(that holds when rounds correspond to natural numbers),
the strengthened invariant is still \emph{not}
inductive in the FOL semantics,
and a CTI for transition \code{echo\_start} still exists, as we explain next. However, the CTI  consists of infinitely many rounds,
which existing solvers have no way of representing,
let alone constructing a CTI with.

Before we explain the CTI,
we note that the pre-state must satisfy
the invariant of line~\ref{line:ivy:inductive-invariant}, that is,
for every round and value in the $\Relation{echo}$ relation
there should exist a previous round when the value was echoed.
This $\forall \exists$ quantifier alternation
leads to a Skolem function $\PreviousFun \colon
\Sort{round} \times \Sort{value} \to \Sort{round}$,
which we make explicit in the CTI.
Intuitively, the Skolem function maps each pair of
round \code{r} and value \code{v}
in the $\Relation{echo}$ relation
to the witnessing (lower) round
that made \code{v} safe to echo at round \code{r}.

The CTI, depicted in \Cref{fig:overview:rounds}, consists of an infinite decreasing sequence of rounds,
all of which are greater than the initial round \Const{start}.
The $\PreviousFun$ function is defined arbitrarily for $\Const{start}$ (and any value),
and as the predecessor w.r.t.\ $\llt$ for other rounds (and any value).
In the pre-state, all rounds except for \Const{start} have echoed the same value $v_1$;
we can see that this state indeed satisfies both of the invariants.
However, by executing \code{echo\_start} where a value \code{v} $= v_0 \neq v_1$ is chosen
for \Const{start},
we reach an unsafe state of the system.

\paragraph{Fixing the model based on the infinite CTI}
The infinite CTI stems from the fact that,
in the FOL semantics, each echoed value
being echoed previously does \emph{not} imply
that it was echoed at \Const{start}.
The CTI reveals that reaching this conclusion
requires induction on the rounds.
Indeed, when the user adds the following instance of the induction scheme:
\[
\forall\code{V}.\ \parens{ \exists\code{R}.\ \code{echo(R, V)} }
\Rightarrow
\exists \code{R}.\ \code{echo(R, V)} \land
\forall \code{R'}.\ \code{R'} \llt \code{R}
\Rightarrow \neg \code{echo(R', V)}
\]
stating that if some value is echoed,
then there is also a minimal round in which it is echoed,
the verification condition formulas become unsatisfiable,
and
\neta{changed:}
existing
solvers manage to prove it.

\paragraph{Our goal}
Unfortunately,
despite its simplicity,
to the best of our knowledge,
no existing solver is able to produce the infinite CTI.
As such, the user receives no feedback from the solver,
which either diverges attempting
to prove the verification conditions,
or gives up and returns ``unknown''.
Changing this unfortunate outcome is the goal of our work.

\begin{figure}[t]\centering \begin{subfigure}[t]{.48\textwidth}\centering \scalebox{0.85}{\begin{tikzpicture}[->,>=stealth',shorten >=1pt,auto,scale=0.8]

		\node[tnode]
			(start) at (1,1)
			{\Const{\figLfont{start}}};
		\node
			(n0) at (2,1)
			{\dots};
		\node[tnode]
			(n1) at (4,1)
			{};
		\node
			(n2) at (6,1)
			{\dots};
		\node[tnode]
		 	(n3) at (8,1)
		 	{};

		\node[vnode]
			(v0) at (1, 3)
			{$\Const{v}$};
		\node [above right = 0.5pt of v0]
			{$v_0$};

		\node[vnode]
			(v1) at (6,3)
			{};
		\node [above right = 0.5pt of v1]
			{$v_1$};

		\path
		(start) edge [loop left] node {\TinyPrevious} (start)
		(n1) edge node {\TinyPrevious} (n0)
		(n2) edge node {\TinyPrevious} (n1)
		(n3) edge node {\TinyPrevious} (n2)
		;

		\path[color=\PostStateColor]
		(start) edge [bend left] node [right] {\Relation{\figLfont{echo'}}} (v0)

		(n1) edge [bend left] node [left] {\Relation{\figLfont{echo'}}} ([yshift=6pt]v1.west)
		(n2) edge [bend left] node [left,yshift=-6pt] {\Relation{\figLfont{echo'}}} (v1)
		(n3) edge [bend right] node [left] {\Relation{\figLfont{echo'}}} ([yshift=-6pt]v1.east)
		;

		\path[color=\PreStateColor]
		(n1) edge [bend left] node [right] {\Relation{\figLfont{echo}}} ([yshift=-6pt]v1.west)
		(n2) edge [bend right] node [right,yshift=-6pt] {\Relation{\figLfont{echo}}} (v1)
		(n3) edge [bend right] node [right] {\Relation{\figLfont{echo}}} ([yshift=6pt]v1.east)
		;
	\end{tikzpicture}
	}
\caption{
An infinite CTI.
The interpretation of \Const{start} and  \code{v} are depicted as annotations of the corresponding elements.
\iflong
\else
Interpretations depicted in \PreStateColor{} are relevant for the pre-state, \PostStateColor{} for the post-state, and black for both.
\fi
The interpretation of $\PreviousFun$ is depicted as arrows from rounds to rounds,
with the intention that
for each round,
the image of the function is the same for both values.
The interpretation of $\llt$ orders the rounds from left to right.
}
\label{fig:overview:rounds}
\end{subfigure}
\hfill
\begin{subfigure}[t]{.48\textwidth}\centering
\scalebox{0.85}{
	\begin{tikzpicture}[->,>=stealth',shorten >=1pt,auto,scale=0.8]

		\node[tnode]
		(start) at (1,1)
		{\Const{\figLfont{start}}};
		\node [above right = 0pt of start,yshift=-3pt,xshift=-2pt]
			{$\alpha$};

		\node[tnode,accepting]
		(n1) at (5,1)
		{};
		\node [above right = 0pt of n1,yshift=-5pt]
			{$\beta$};
		\node [below = 0pt of n1]
		{\figLfont{$x \leq 0$}};

		\node[vnode]
		(v0) at (1, 3)
		{$\Const{v}$};
		\node [right = 0.5pt of v0]
		{$v_0$};

		\node[vnode]
		(v1) at (5,3)
		{};
		\node [right = 0.5pt of v1]
		{$v_1$};

		\path
		(start) edge [loop left] node {\figLfont{\TinyPrevious: $0$}} (start)
		(n1) edge [loop left] node {\figLfont{\TinyPrevious: $x_1 - 1$}} (n1)
		;

		\path[color=\OrderColor]
		(start) edge [bend right] node {\figLfont{$\llt$}} (n1)
		(n1) edge [loop right] node {\figLfont{$\llt$: $x_1 < x_2$}} (n1)
		;

		\path[color=\PostStateColor]
		(start) edge [bend left] node [left] {\Relation{\figLfont{echo'}}} (v0)
		(n1) edge [bend left] node [left] {\Relation{\figLfont{echo'}}} (v1)
		;

		\path[color=\PreStateColor]
		(n1) edge [bend right] node [right] {\Relation{\figLfont{echo}}} (v1)
		;
	\end{tikzpicture}
	}
\caption{\AStruct{} representing the infinite CTI from (a).
Double lines denote summary nodes,
		for which \iflong\else the subset formulas \fi are written underneath. Terms and formulas defining functions, respectively relations, are written on the edges,
		using the syntax $a\colon b$, where $a$
		is a function or relation and $b$ is a term or formula, respectively. The formula on the edge is omitted when it is $\top$;  and the entire edge is omitted when the formula is $\bot$.
	}
	\label{fig:overview:template}
\end{subfigure}
\caption{A CTI for \code{echo\_start}.
	Values are depicted as squares and
	rounds as circles.
	\oded{Maybe we can just use the same edges for echo and echo', and just label the edges as echo, echo', and also label the prev edges with prev,prev' (prev' would be below prev). Not sure if it's worth our time now, but we can consider it for the final version.}
}
\end{figure}

\subsection{\STructures{} for Compacting the Infinite Pattern}
\commentout{
 The key enabler of our approach for finding infinite counterexamples is the observation that,
though infinite, the CTI in \Cref{fig:overview:rounds} can be described compactly:
it has one unique \texttt{start} round,
in which no value is echoed in the pre-state and $v_0$ is echoed in the post-state;
and a repeating pattern of infinitely many rounds,
all echoing $v_1$ (in both the pre- and post-state),
and each pointing to the $\PreviousFun$ one,
as a witness for satisfying the $\FunDef{prev\_echoed}$ assumption.

We introduce \emph{templates} as a way to express this idea formally and precisely (\Cref{sec:templates}).
A template capturing the infinite model from \Cref{fig:overview:rounds} is depicted in \Cref{fig:overview:template}.
}

The key to our approach for finding infinite
\neta{changed:}
counter-models
is observing that
though infinite, the CTI in \Cref{fig:overview:rounds} can be described compactly:
it has one unique \texttt{start} round,
in which no value is echoed in the pre-state and $v_0$ is echoed in the post-state;
and a repeating pattern of infinitely many rounds,
each echoing $v_1$ (in both the pre- and post-state)
and pointing to the $\PreviousFun$ one
as a witness for satisfying the invariant at line~\ref{line:ivy:inductive-invariant}. Our \emph{\structures{}} (\Cref{sec:templates}) generalize this observation and formalize a way to express infinite models with similarly structured repetition.
\AStruct{} capturing the infinite model from \Cref{fig:overview:rounds} is depicted in \Cref{fig:overview:template}.

\paragraph{\Structure{} domain}
\Structures{} use ``regular nodes'' to represent single elements,
such as the \Const{start} round
($\alpha$ in \Cref{fig:overview:template}),
as well as the individual $v_0$ and $v_1$ values;
and ``summary nodes'' to represent (infinitely) repeating patterns in the model.
A summary node represents a (copy of a) subset of the integers,
described by some Linear Integer Arithmetic (LIA) formula.
In our example, all rounds besides \code{start} are grouped into
a summary node $\beta$,
which is defined to represent the integers $\leq 0$.

Each node in the \structure{} represents a set
of \emph{\concrete{} elements},
which together form the domain of the infinite model represented by the \structure{}.
\AConcrete{} element is a pair comprised
of a node and an integer ``index''.
For uniformity, we use a pair with index $0$ for regular nodes.
For example,
the \concrete{} elements for the \structure{} of \Cref{fig:overview:template}
are $\pair{\alpha,0}, \pair{\beta,0}, \pair{\beta,-1}, \pair{\beta,-2}, \cdots$.

\paragraph{Interpretation}
\AStruct{} defines the interpretation of functions over
its nodes
by specifying both the target node,
and a LIA term to select one of its \concrete{} elements.
This term may also refer to the index of the source nodes.
For the $\PreviousFun$ function in \Cref{fig:overview:template}:
\begin{align*}
	\PreviousFun(\alpha, v_0)
		= \PreviousFun(\alpha, v_1)
		= \pair{ \alpha, 0 }  &\qquad&
	\PreviousFun(\beta, v_0)
		= \PreviousFun( \beta, v_1)
		= \pair{ \beta, x_1 - 1 }
\end{align*}
where $x_1$ represents the index of the first argument.
This means that the $\PreviousFun$
of the (only) \concrete{} element formed by
the regular node $\alpha$ is the same node (for any value),
and the $\PreviousFun$ of each \concrete{} element
$\pair{\beta, x_1}$,
formed by the summary node $\beta$, is
the \concrete{} element of $\beta$ whose index is smaller by $1$,
i.e., $\pair{\beta, x_1 - 1}$.

Interpretations of relations are similarly defined
by specifying for each tuple of nodes,
which of their \concrete{} elements are in the relation.
This is done by assigning
a LIA formula over the indices of the nodes.
In our example, the \Relation{echo}, \Relation{echo'}, and $\llt$ relations
are defined as follows:
\begin{align*}
	\Relation{echo}(\beta, v_1) &= \top
		&   \Relation{echo'}(\alpha, v_0) &= \top
		&  \alpha \llt \beta &= \top
	\\
	&
		&   \Relation{echo'}(\beta, v_1) &= \top
		&  \beta \llt \beta &= x_1 < x_2
\end{align*}
where $x_1, x_2$ represent the indices of the first and second argument,
respectively,
and for all other tuples the relations are defined as $\bot$.
The definition $\alpha \llt \beta = \top$ means
\emph{all} \concrete{} elements of $\alpha$
are $\llt$-related to \emph{all} \concrete{} elements of $\beta$.
The definition $\beta \llt \beta = x_1 < x_2$ means
that among the \concrete{} elements associated with $\beta$,
elements are only $\llt$-related to the elements
whose integer indices are strictly larger than theirs,
forming an infinite $\llt$-decreasing sequence.

\paragraph{Model checking \structures{}}
How do we know that the infinite model represented by the \structure{},
also called the \emph{\concretization{} of the \structure{}},
indeed satisfies the formula of interest?
It turns out that the \emph{model checking} problem for \structures{}
reduces to checking LIA formulas
(\Cref{thm:templates:model-checking}).
Given the \structure{} in \Cref{fig:overview:template},
we transform the original VC formula
into a (quantified) LIA formula.
Universal quantifiers are transformed into
a combination of conjunction (over nodes)
and universal LIA quantifiers (over indices),
and similarly for existential quantifiers
(with disjunctions and existential LIA quantifiers).
Function and relation applications are transformed into the LIA terms and formulas that interpret them in the \structure{}.
Critically, the resulting LIA formula
has no uninterpreted functions or relations.
As we show in \Cref{sec:model-checking-enc},
the resulting LIA formula is true
(in LIA) if and only if the \concretization{} of the \structure{}
satisfies the original FOL formula.

For example, in the VC formula of the Echo Machine,
the anti-reflexivity axiom for ``$\llt$'' is expressed as
$
	\forall x\colon \Sort{round}.\ \neg
	\parens{ {\color{blue}{x \llt x}} }
$.
To check that the \concretization{} of the \structure{} satisfies
this sub-formula, it is translated into
$
	\forall x\colon \Sort{int}.\
		\parens{x = 0 \to \neg {\color{blue}{\bot}}}
			\land
		\parens{ x \leq 0 \to \neg \parens{ {\color{blue}{x < x}} }}
$,
which is true in LIA.
The first conjunct is generated from the regular node $\alpha$,
where the \concrete{} elements are characterized by $x=0$
and the definition of $\alpha \llt \alpha$ is $\bot$;
the second conjunct is generated from the summary node $\beta$,
where the \concrete{} elements are characterized by $x \leq 0$
and the definition of $\beta \llt \beta$ results in $x < x$.
In this way,
the translation ``bakes'' the nodes into the formula
and keeps only the LIA definitions
from the \structure{} to be checked in LIA.

\subsection{Finding \STructures{} via Symbolic Search}
The reducibility of the model-checking
problem for \structures{}
to LIA establishes its decdiability.
(Recall that LIA is decidable even with quantifiers.)
Our next goal is obtaining an efficient process,
by which to find \aStructModel{} (i.e., a satisfying \structure{})
for a given formula.
Enumeration is an example for a sound procedure
that searches for \aStructModel{},
but it is inefficient.
Instead, we introduce the concept of \aBeforeStructure{} \emph{\structAdj{} \template{}},
as a way to describe a family of \structures{},
by limiting the terms and formulas allowed in summary nodes
to some finite subset of the LIA language.
Such \aTemplate{} induces a search space of \structures{},
and in \Cref{sec:finding} we develop an efficient method
to symbolically consider all \structures{} within
that space
and search among them for a possibly infinite satisfying model
for a given formula.

Our search procedure encodes
the entire family of \structures{} induced by \aTemplate{}
using auxiliary Boolean and integer variables.
It transforms the original FOL formula into a
(quantified) LIA formula whose structure is similar to
the structure of the model checking formula
except for the use of the auxiliary variables.
The resulting formula is satisfiable in LIA whenever
at least one of the \structures{} in the family satisfies the original formula
(\Cref{thm:finding:finder}), in which case
the assignment to the auxiliary variables induces the sought \structure{}.

The symbolic encoding moves the complexity of finding the
right \structure{} in a given family into the hands
of existing \Solvers{},
which are able to solve LIA formulas quickly.
However, this symbolic search requires
specifying \aTemplate{}
in order to constrain the search space.
Rather than putting the burden of defining such \aTemplate{} on the user,
we propose a simple
(yet robust, see \Cref{sec:evaluation})
heuristic for computing a candidate \template{}
syntactically from the formula.
The \template{} we propose uses simple building blocks
for function terms and relation formulas:
all functions use the LIA terms $0$ and $x_i \pm k$;
unary relations use $\top, \bot,$ and $x_i = 0$,
binary relations additionally allow comparisons $x_i \bowtie x_j \pm k$,
and ternary relations additionally allow a
``between'' formula $x_i \leq x_j < x_k$
(see \Cref{fig:evaluation:heuristic} for the precise definition).

By using such \aTemplate{} and enumerating
over the number of nodes,
finding a satisfying \structure{} for the Echo Machine
takes less than one second.

\subsection{Decidable Fragment}
Rather than a fluke,
we show that in the case of the Echo Machine,
considering the proposed \templates{} is not just a heuristic.
We prove that for the verification problem of the Echo Machine
(and other problems obeying a syntactic restriction)
the set of \structures{} induced by
such \aTemplate{} is always sufficient,
provided that the formula is indeed satisfiable.
Note that for arbitrary satisfiable formulas
it is not
guaranteed that \aStructModel{} exists,
let alone one from a simple, predefined, family of \templates{}.
This is because not every infinite model
can be represented by \aStruct{},
and some satisfiable formulas are bound to have models
that are not the \concretization{} of any \structure{}.

\neta{todo: rephrase?}
That said, in \Cref{sec:decidable}
we identify
a new fragment of FOL,
the \emph{\FragmentLongName{} (\FragmentShortName)} fragment,
where all satisfiable formulas have a (possibly infinite) model
that is representable by \aStruct{},
thus making the satisfiability problem in this fragment decidable.
The VC formula of the Echo machine falls into \FragmentShortName{},
both before and after the instance of the induction scheme is added.

{\FragmentShortName} extends the many-sorted variant of the
Effectively PRopositional (EPR)
decidable fragment of FOL.
While many-sorted EPR forbids
all function definitions that form cycles between sorts,
{\FragmentShortName} extends it to allow one ``infinite'' sort
($\Sort{round}$ in our example)
that
is allowed to have cyclic functions
(such as $\PreviousFun$),
any number of
relations where the infinite sort appears at most once (such as the $\Relation{echo}$ relation),
and one binary relation,
axiomatically defined to be a strict linear order,
and conventionally denoted $\llt$.
Linear order is a common primitive
in many verification problems,
especially when abstracting the naturals or the integers in FOL,
or when verifying properties of linear data-structures
such as linked lists.
The ability to add cyclic functions to the vocabulary,
which is strictly beyond EPR,
makes the language more expressive
and well-suited for a wider variety
of verification problems.

We prove the decidability
of checking satisfiability of a formula in \FragmentShortName{}
by showing the fragment enjoys a ``small \structModel{} property,''
bounding the number of nodes in candidate \structModels{}
in terms of syntactic parameters of the formula,
and restricting
the subset of the LIA language needed
in function terms and relation formulas
(\Cref{thm:full-fragment-decidable}).
I.e., the small \structModel{} property
provides a cap for enumerating the number of nodes,
and induces a definition of candidate \templates{},
which is strictly covered by our heuristic:
for formulas in \FragmentShortName{}
only the $\llt$ relation may use comparisons, and other relations only require $\top$ and $\bot$.
This makes our symbolic search a decision procedure for \FragmentShortName{}:
when enumerating the heuristic \templates{} up to the bound,
if no \structModel{} is found,
the formula is necessarily unsatisfiable.
A notable property of our decidability proof is that
while the \structures{} are finite (and bounded in size),
the models they represent can be infinite.
Thus, \FragmentShortName{} does \emph{not} enjoy a finite model property,
as opposed to many decidable fragments, including EPR:
for EPR, the syntactic restrictions imply a bound
on the number of ground terms,
from which a \emph{finite} model can always be constructed.
In contrast, in the case of \FragmentShortName{},
the number of ground terms is infinite due to the cyclic functions,
and the models, accordingly, are potentially infinite.

\subsection{Applicability of \STructures{}}The ``small \structModel{} property'' establishes
that the simple \templates{} we consider
are complete for \FragmentShortName{}.
While the \FragmentShortName{} fragment
does not cover all of the examples we consider in this paper
(e.g., complex distributed protocols),
it does provide important theoretical justification for
designing the search space of \structures{}
and using our heuristic for computing \aTemplate{} for a formula.
These \templates{} turn out to be valuable
for a myriad of complex formulas
that lie outside \FragmentShortName{}.

We implemented our symbolic search,
combined with the heuristically computed \structAdj{} \templates{},
in our \ToolShortName{} tool
(\ToolLongName{})
and successfully applied our approach to
examples that admit infinite counter-models,
which can be eliminated by additional FOL axioms
(e.g., instances of an induction scheme).
The examples are taken
from the domains of distributed protocols
(including five variants of Paxos~\cite{lamport-paxos}
),
and of properties of linked lists~\cite{natural-proofs}.
All 21 examples in our evaluation,
are solved by \ToolShortName{},
which quickly finds infinite counter-models.
\Zzz~\cite{z3-smt-solver}, \cvc~\cite{cvc5-smt-solver}
and Vampire~\cite{vampire} all fail to solve these examples.

\commentout{
\paragraph{Beyond the decidable fragment}
Our technique for finding templates is a
decision procedure for the \FragmentShortName{} fragment.
Though it is
is not guaranteed to work for formulas outside of \FragmentShortName{}, our experiments show that it also succeeds to
automatically find templates for complex formulas outside the fragment.
In particular, our symbolic encoding supports searching
through broader families
of templates
specified by user-provided \emph{template skeletons}.
We implemented this procedure in our \ToolShortName{} tool
(\ToolLongName{})
and successfully applied our approach to examples that admit infinite counter-models, which can be eliminated by additional FOL axioms (e.g., instances of an induction scheme).
The examples are taken from a variety of domains:
from distributed protocols
(including two variants of Paxos~\cite{lamport-paxos}),
through axiomatic arithmetic
(non-commutative addition, e.g.,~\cite{vampire-integer-induction}),
to properties of linked lists
(inspired by~\cite{natural-proofs}).
All of these examples were captured either by $\Templatesepr$ or by one additional user-specified template skeleton.
While \ToolShortName{} quickly finds counter-models,
\Zzz~\cite{z3-smt-solver}, \cvc~\cite{cvc5-smt-solver} and Vampire~\cite{vampire} all fail to solve these examples.
\oded{do we really need user provided skeletons here, or is it all in $\TemplatesExtended$? Should we mention $\TemplatesExtended$ here? Right now it sounds like anything beyond $\Templatesepr$ is bespoke.}
}

\section{Background}
\label{sec:background}
\paragraph{First-Order Logic}
We use the usual definition of
many-sorted first-order logic (FOL)
with equality.
A vocabulary $\Sigma$ consists of
a set of sorts $\Sorts$,
and constant, function and relation symbols,
each with a sorted signature.
In the cases where $\Sigma$ has a single sort,
we will omit $\Sorts$ and the signatures of the symbols.
We denote by $\Variables$ the set of variables used in FOL (we assume that each variable has a designated sort,
and omit sorts when they are clear from the context).
Terms $t$ over $\Sigma$ are either constants $c$, variables $x$, 
or well-sorted function applications of the form
$f \parens{ t_1, \dots, t_k }$.
Formulas are defined recursively, where atomic formulas are either $t_1 \foleq t_2$
(we use $\foleq$ as the logical equality symbol)
or $R \parens{ t_1, \dots, t_k }$, where $R$ is a relation symbol,
and non-atomic formulas are built using Boolean connectives
$\neg, \land, \lor, \to$
and quantifiers $\forall, \exists$.
Given a formula $\varphi$ (term $t$),
$\FV(\varphi)$ ($\FV(t)$)
denotes the set of free variables
appearing in $\varphi$ ($t$). 
A term $t$ is \emph{ground} if $\FV(t) =\emptyset$.
A formula $\varphi$ is \emph{closed} if $\FV(\varphi) =\emptyset$.
Models and assignments are defined in the usual way.
A formula $\varphi$ is satisfiable 
if it has a satisfying model $M$,
denoted $M \models \varphi$.
We denote by $\top$ a quantifier-free closed tautology,
and by $\bot$ a contradiction.

\paragraph{Linear Integer Arithmetic}
The interpreted theory of Linear Integer Arithmetic (LIA)
is defined
over a vocabulary that includes
the sort of integers,
and standard constants, functions and relations
of linear arithmetic 
($0, 1, +, -, <$).
Many \Solvers{} support extensions of LIA
where all integers are considered constant symbols,
functions include multiplication and division by a constant,
and relations include short-hand comparisons ($>, \leq, \geq$),
and equality modulo a constant.
We adopt these extensions,
and denote by $\Terms_\subLIA$ the set of all LIA terms,
and by $\QF_\subLIA$ the set of all quantifier-free LIA formulas.
We distinguish 
between the LIA order $<$ and any axiomatized FOL order relation,
for which we typically use $\llt$.
We denote by $\models_{\LIA}$ satisfiability in
the standard model of integer arithmetic,
and recall that checking satisfiability of (quantified) LIA formulas
is decidable~\cite{lia-decidable}
and supported by \Solvers{}~\cite{z3-smt-solver,cvc5-smt-solver}.

\section{\STructures{} as Representations of Infinite Models}
\label{sec:templates}

In this section we introduce \structures{} as a way to represent
(certain) infinite structures.
We show how the model-checking problem for models represented by \structures{}
reduces to the problem of checking satisfiability of a LIA formula,
and is therefore decidable.

\subsection{\STructures{}}
Given a many-sorted vocabulary $\Sigma$ with a set of sorts $\Sorts$,
we define \aBeforeStructure{} \emph{\structure{}} 
as a triple $\struct = (\Domain, \Bounds, \Interp)$,
where $\Domain$ is the \emph{domain} of $\struct$,
$\Bounds$ is an assignment of \emph{bound formulas}
to the domain,
and $\Interp$ is an \emph{interpretation} of the symbols in $\Sigma$,
defined as follows.

The \emph{domain} of $\struct$ maps each sort $\sort \in \Sorts$ to
a finite, non-empty, set of nodes $\Domain(\sort)$,
also denoted $\Domain_\sort$,
such that the domains of distinct sorts are disjoint, i.e.,
$\Domain_\sort \cap \Domain_{\sort'} = \emptyset$
for any two sorts $\sort \neq \sort'$.
For each sort $\sort \in \Sorts$,
the domain $\Domain_\sort$ is partitioned into
``summary nodes'', denoted $\Domain^S_\sort$,
and ``regular nodes'', denoted $\Domain^R_\sort$,
such that $\Domain_\sort = \Domain^S_\sort \uplus \Domain^R_\sort$.
Intuitively, each summary node represents
a (partial) copy of the integers,
while each regular node represent an individual element.

The \emph{bound} assignment maps, for every sort $\sort$,
every node $n \in \Domain_\sort$
to a satisfiable \LIA{} formula $\Bounds(n) \in \QF_\LIA$
with at most one free variable,
$\FV \parens{ \Bounds(n) } \subseteq \braces{ x }$.
We further require that for regular nodes,
the bound formula is always $x \liaeq 0$\footnote{
	We assign a bound formula to regular nodes only
	for the uniformity of the presentation.
}.
We think of the bound formula as representing a subset
of the integers --- those that satisfy $\Bounds(n)$ ---
and use the formula and the subset of
$\Integers$ it represents interchangeably.
Intuitively, $\Bounds(n)$ restricts the copy of the integers
represented by $n$ to the elements in $\Bounds(n)$
(for regular nodes, this is always a single integer, $0$).

Given $z \in \Bounds(n)$,
we refer to $\pair{ n, z }$ as \aConcrete{} element of $n$,
and denote the union set of explicit elements 
for a set of nodes $N$ by
$\ElemsOf(N) = \braces{
	\pair{ n, z } \mid n \in N, z \in \Bounds(n)
}$.

The \emph{interpretation} function maps each symbol to its interpretation.
A constant symbol $c$ of sort $\sort$
is mapped to an element of some node of sort $\sort$:
$\Interp(c) \in \ElemsOf(\Domain_\sort)$, also denoted $c^\struct$.

The interpretation of a function symbol
$f \colon \sort_1 \times \dots \times \sort_k \to \sort$
is a function that maps nodes of the argument sorts
to an image node of the range sort,
accompanied by an image LIA term:
$\Interp(f) \colon
\Domain_{\sort_1} \times \dots \times \Domain_{\sort_k}
	\to
	\Domain_\sort \times \Terms_\LIA
$, also denoted $f^\struct$.
We further require that if
$f^\struct \parens{ n_1, \dots, n_k } = \pair{ n, s }$, then
$\FV(s) \subseteq \braces{ x_i \mid 1 \leq i \leq k }$,
and that
$\bigwedge_{i=1}^k \Bounds(n_i)[x_i/x] \models_\subLIA \Bounds(n)[s/x]$.
Intuitively, we assign a free variable $x_i$ to each
argument $n_i$,
representing the specific element of node $n_i$ to which
the function is applied.
The image term $s$ selects
an element of
the image node
for each combination of argument elements.
The requirement
$\bigwedge_{i=1}^k \Bounds(n_i)[x_i/x]
\models_\subLIA \Bounds(n)[s/x]$
ensures that whenever the argument elements
are elements of
the corresponding nodes
$n_1, \dots, n_k$,
the selected element is an element of $n$.
In particular,
if the image node $n$ is a regular node,
the term must evaluate to $0$.

The interpretation of a relation symbol
$R \colon \sort_1 \times \dots \times \sort_k$ is a
function that maps tuples of nodes of the relevant sorts to a LIA formula:
$\Interp(R) \colon
\Domain_{\sort_1} \times \dots \times \Domain_{\sort_k}
\to \QF_\LIA$,
also denoted $R^\struct$.
We further require that
$\FV\parens{ R^\struct(n_1, \dots, n_k) }
	\subseteq
\braces{ x_i \mid 1 \leq i \leq k }$.
Intuitively, the tuples of integers
that satisfy the LIA formula define the elements of nodes $n_i$
for which the relation holds.

\begin{example}
	\label{ex:templates:definition}
	The \structure{} of \Cref{fig:overview:template}
	is $\struct = \parens{ \Domain, \Bounds, \Interp }$,
	where
	$\Domain(\Sort{value}) = \braces{ v_0, v_1 }$,
	$\Domain(\Sort{round}) = \braces{ \alpha, \beta }$;
	$v_0, v_1, \alpha$ are regular nodes
	with $\Bounds(n) = x \liaeq 0$;
	$\beta$ is a summary node
	with
	$\Bounds(\beta) = x \leq 0$;
and
	\[
	\begin{footnotesize}
	\begin{array}{r@{\,}l;{3pt/1pt}r@{\,}ll;{3pt/1pt}r@{\,}ll}
		\multirow{2}{*}{$\Const{start}^\struct =$}
		&
		\multirow{2}{*}{$\pair{\alpha, 0}$}
		&
		\multirow{2}{*}{$\PreviousFun^\struct(n_1,n_2) = \Big\{$}
			& \pair{ \alpha, 0 }
			& n_1 = \alpha
		& \multirow{2}{*}{$\Relation{echo}^\struct(n_1,n_2) = \Big\{$}
			& \top
			& n_1 = \beta, n_2 = v_1
		\\ & &
			& \pair{ \beta, x_1 - 1 }
			& n_1 = \beta
		&
			& \bot
			& \text{otherwise}
		\\
		\multirow{3}{*}{$\Const{v}^\struct =$}
		&
		\multirow{3}{*}{$\pair{v_0, 0}$}
		&
		\multirow{3}{*}{$n_1 \llt^\struct n_2 = \Bigg\{$}
			& \top
			& n_1 = \alpha, n_2 = \beta
		& \multirow{3}{*}{$\Relation{echo'}^\struct(n_1,n_2) = \Bigg\{$}
			& \top
			& n_1 = \alpha, n_2 = v_0
		\\ & & 
			& x_1 < x_2
			& n_1 = n_2 = \beta
		&
			& \top
			& n_1 = \beta, n_2 = v_1
		\\ & & 
			& \bot
			& \text{otherwise}
		&
			& \bot
			& \text{otherwise}
	\end{array}
	\end{footnotesize}
	\]
\end{example}

An example of a \structure{}
with two summary nodes 
\neta{if full}
\ifnofull
can be found in~\cite{infinite-needle-arxiv}.
\else
appears 
in \Cref{fig:evaluation:non-commutative}
in \Cref{sec:eval-details}.
\fi

\paragraph{\Concretization{}}
\AStruct{} $\struct = (\Domain, \Bounds, \Interp)$ represents a (possibly) infinite
first-order structure, defined by its \concretization{} $M = \Concrete(\struct)$.

The domain of sort $\sort \in \Sorts$ in $M$, denoted
$\Domain^M_\sort$, is the set of 
\concrete{} elements $\ElemsOf(\Domain_\sort)$,
i.e., it
consists of all pairs $\pair{n, z}$ of nodes $n$ in $\Domain_\sort$
and integers $z$ that satisfy $\Bounds(n)$.
Thus, each summary node may contribute
infinitely many elements to the \concrete{} domain,
while each regular node contributes a single element.
Note that the requirements for bound formulas to be satisfiable
ensures that the explicit domains of all sorts are non-empty.

The interpretation of the symbols in $M$,
denoted $\alpha^M$ for symbol $\alpha$,
is defined as follows.
For a constant symbol $c$, $c^M = c^\struct$.
For a function symbol $f$ of arity $k$,
\iflong
\else
$f^M \parens{\pair{ n_1, z_1 }, \dots, \pair{ n_k, z_k }}
= \pair{ n, \bar v_\subLIA(s) }$,
where $\pair{ n, s } = f^\struct \parens{ n_1, \dots, n_k }$,
$v_\subLIA$ is an assignment such that
$v_\subLIA(x_i) = z_i$,
and $\bar v_\subLIA$ extends $v_\subLIA$
to all LIA terms in the standard way.
\fi
That is, the function maps the elements
$\pair{ n_1, z_1 }, \dots, \pair{ n_k, z_k }$
of the \concrete{} domain to an element
contributed by node $n$
that is determined by evaluating the term $s$
on the specific elements $z_i$ of nodes $n_i$.
For a relation symbol $R$,
$
R^M =
	\braces{
		\parens{ \pair{ n_1, z_1 }, \dots, \pair{ n_k, z_k } }
		\mid
		v_\subLIA \models_\subLIA R^\struct \parens{ n_1, \dots, n_k }
		\text{ where }
		v_\subLIA(x_i) = z_i \in \Bounds(n_i)
	}
$.
That is, the tuples of \concrete{} elements in the relation are determined
by evaluating the LIA formula $R^\struct \parens{ n_1, \dots, n_k }$
on the specific elements $z_i$ of nodes $n_i$.

\begin{example}
	\label{ex:templates:concretization}
	Let us \concretize{} the \structure{} 
	from \Cref{ex:templates:definition}.
	The \concrete{} domains are
	$\Domain^M(\Sort{value}) =
	\braces{ \pair{ v_0, 0}, \pair{ v_1, 0 } }$ and
	$\Domain^M(\Sort{round}) =
	\braces{ \pair{ \alpha, 0 } } \cup
	\braces{ \pair{ \beta, z } \mid z \leq 0 }$.
The \concretization{} of the constants is:
$\Const{start}^M = \pair{\alpha, 0}$ and
$\Const{v}^M = \pair{v_0, 0}$.
\iflong
\else
As for the function $\PreviousFun$ and the relations:
\begin{footnotesize}
\[
\begin{array}{l@{};{3pt/1pt}@{}r}
	\PreviousFun^M(\pair{ n, z }, v) =
	\left\{
	\begin{array}{ll}
		\pair{ \alpha, 0 } & n = \alpha \\
		\pair{ \beta, z - 1 } & n = \beta
	\end{array}
	\right.
	&
	\begin{array}{r@{\,}l}
		\llt^M &=
		\braces{
			\parens{ \pair{ \alpha, 0 }, \pair{ \beta, z } }
			\mid z \leq 0
		}
		\cup
		\braces{
			\parens{ \pair{ \beta, z_1 }, \pair{ \beta, z_2 } }
			\mid z_1 < z_2 \leq 0
		}
		\\
		\Relation{echo}^M &=
		\braces{
			\parens{ \pair{ \beta, z }, \pair{ v_1, 0 } }
			\mid z \leq 0
		}
		\\
		\Relation{echo'}^M &=
		\braces{
			\parens{ \pair{ \beta, z }, \pair{ v_1, 0 } }
			\mid z \leq 0
		}
		\cup
		\braces{ \parens{ \pair{ \alpha, 0 }, \pair{ v_0, 0} } }	
	\end{array}
\end{array}
\]
\end{footnotesize}
\fi

The resulting \concrete{} model is isomorphic
to the infinite model depicted in
\Cref{fig:overview:rounds}.
\end{example}

\begin{remark}
	\label{rem:why-we-need-bounds}
	In fact, for the Echo Machine,
	the degenerate bound formula
	$\Bounds(\beta) = \top$
	would produce \aStruct{}
	whose \concretization{} is a counter-model to the VC formula
	just as well
	(even though it is not isomorphic to \Cref{fig:overview:rounds}).
	One may wonder therefore if bound formulas are necessary,
	or do regular nodes
	(where $\Bounds(n) = x \liaeq 0$)
	and unbounded summary nodes
	(where $\Bounds(n) = \top$)
	suffice.
\iflong
\else
	An example
	that requires a summary node with a
	bound formula other than $\top$ (e.g., $x \geq 0$) 
	\neta{if full}
	\ifnofull
	is provided in~\cite{infinite-needle-arxiv}.
	\else
	can be found in \Cref{sec:examples}.
	\fi
\fi
\end{remark}

\subsection{Model Checking a \STructure{}}
\label{sec:model-checking-enc}

\AStruct{} $\struct$ represents \aConcrete{} model $M = \Concrete(\struct)$.
In this section we address the problem of checking if $\Concrete(\struct)$
satisfies a given FOL formula $\varphi$,
in which case, we also say that
$\struct$ satisfies $\varphi$
(with abuse of notation).
This is done by a transformation that ``bakes''
the \structure{} into the formula,
resulting in a \LIA{} formula that is true if and only if
$\Concrete(\struct) \models \varphi$.

We define the transformation recursively
(over the inductive structure of formulas).
As a result,
we must consider both closed formulas and formulas with free variables.
To that end, we first define assignments in \structures{}.

\paragraph{\StructAssignments{}}
If $\varphi$ has free variables,
then its value in $M$ depends also on an assignment
$v \colon \Variables \to \Domain^M$ (which respects the sorts).
We refer to such an assignment 
as \aBeforeConcrete{} \emph{\concrete{}} assignment.
Recall that $\Domain^M \subseteq \Domain^\struct \times \Integers$,
i.e., \concrete{} assignments map each variable to
a node paired with an integer.
For the purpose of defining the transformation to \LIA{},
we introduce \emph{\structAssignments} where \concrete{} integers
are replaced by integer variables.
In the sequel, let
$\Variables^\supLIA = \{ x^\supLIA \mid x \in \Variables\}$
be a set of integer variables
such that
each variable $x \in \Variables$ has a unique
\emph{symbolically corresponding} LIA variable
$x^\supLIA \in \Variables^\supLIA$.

\begin{definition}\label{def:templates:template-assignment}
Given \aStruct{} $\struct$, 
\aBeforeStructAssignment{} 
\emph{\structAssignment{}} is a mapping
$\structAss \colon \Variables \to \Domain^\struct \times \Variables^\supLIA$ 
such that for every variable
$x \in \Variables$, $\structAss(x) = \pair{n, x^\supLIA}$ for some $n \in \Domain^\struct$.
\end{definition}

That is, \aStructAssignment{} fixes the node $n$
to which each variable $x$ is mapped,
but leaves the element of $n$ symbolic,
captured by the integer variable $x^\supLIA$
that uniquely corresponds to $x$.

We denote by $\barStructAss$ the extension of $\structAss$
to arbitrary terms
over $\Sigma$.
The extended assignment $\barStructAss$
maps terms to nodes paired with
LIA terms over $\Variables^\supLIA$.
Formally,
$\barStructAss \colon \Terms \to \Domain^\struct \times \Terms_\LIA$
\iflong
\else
is defined by setting
$\barStructAss(x) = \structAss(x)$;
$\barStructAss(c) = c^\struct$;
and
$\barStructAss \parens{ f(t_1, \dots, t_k) }
= \pair{ n, s \brackets{ s_i / x_i } }$,
where $\barStructAss(t_i) = \pair{ n_i, s_i }$
and $f^\struct(n_1, \dots, n_k) = \pair{ n, s }$.
\fi
That is, if $\barStructAss$ interprets the terms $t_i$ as
$\pair{ n_i, s_i }$,
and the \structure{} defines
$f^\struct(n_1, \ldots, n_k) = \pair{n,s}$,
then $\barStructAss$ interprets $f(t_1, \ldots, t_k)$
as the node $n$ and the term obtained from $s$
by substituting $s_i$ for $x_i$.

\begin{example}
	\label{ex:templates:template-assignment}
	For the \structure{} defined in \Cref{ex:templates:definition},
	we can define an assignment $\structAss$ such that
	$\structAss(x_1) = \pair{ \beta, x^\supLIA_1 }$.
	Given the term
	$t = \PreviousFun \parens{ \PreviousFun \parens{ x_1 } }$,
	$\barStructAss(\struct) = \pair{ \beta, \parens{ x^\supLIA_1 - 1 } - 1 }$.
\end{example}

The correspondence between FOL variables and their LIA counterparts
allows us to ``decompose'' \aConcrete{} assignment $v$
into \aStructAssignment{}
where all integers are replaced by variables,
accompanied by a LIA assignment to $\Variables^\supLIA$,
which captures the \concrete{} integers from $v$.

\begin{definition}\label{def:templatization}
Given \aConcrete{} assignment $v \colon \Variables \to \Domain^M$,
its \emph{\templatization{}} is the \structAssignment{}
$\structAss \colon \Variables \to \Domain^\struct \times \Variables^\supLIA $ defined by
$\structAss(x) = \pair{ n, x^\supLIA }$ whenever $v(x) = \pair{ n, z }$.

The \emph{residual} assignment of $v$ is the LIA assignment
$v_\subLIA \colon \Variables^\supLIA \to \Integers$ defined by
$v_\subLIA(x^\supLIA) = z$ whenever $v(x) = \pair{ n, z }$.
\end{definition}

We are now ready to
define the recursive transformation
that reduces model checking of \structures{} in FOL formulas
to checking satisfiability of LIA formulas
by ``baking'' \aStruct{} and 
\aConcrete{} assignment into the FOL formula.

\iflong
\else
\paragraph{Model checking \structures{} by transforming FOL formulas to LIA}
Given a FOL formula $\varphi$, \aStruct{} $\struct$
and \aConcrete{} assignment $v$,
we use $\structAss$, the \templatization{} of $v$,
to translate $\varphi$ into a LIA formula.
When evaluated with the residual assignment, $v_\subLIA$,
this formula captures whether $\Concrete(\struct)$ and $v$
satisfy $\varphi$.
\fi
The transformation, $\TemplateTransform^\struct_{\structAss}(\varphi)$,
\iflong
\else
is defined in \Cref{fig:templates:transformation}.
\fi

\iflong
\else

\begin{figure}
\begin{minipage}{0pt}
	\begin{footnotesize}
		\vspace{-0.5cm}
		\begin{align*}
			\TemplateTransform^\struct_{v_\struct} \parens{ R \parens{ t_1, \dots, t_k } }
			& =	R^\struct \parens{ n_1, \dots, n_k } \brackets{ s_i / x_i }
\text{ where } \bar v_\struct (t_i) = \pair{ n_i, s_i }
			\\
			\TemplateTransform^\struct_{v_\struct} \parens{  t_1 \foleq t_2 }
			& =	s_1 \liaeq s_2 \text{ if } n_1 = n_2 \text{ and } \bot \text{ otherwise,}
\text{ where } \bar v_\struct (t_i) = \pair{ n_i, s_i }
			\\
			\TemplateTransform^\struct_{v_\struct} \parens{ \neg \varphi }
			& = \neg \parens{ \TemplateTransform^\struct_{v_\struct} (\varphi) }
			\\
			\TemplateTransform^\struct_{v_\struct} \parens{ \varphi \circ \psi }
			& = \TemplateTransform^\struct_{v_\struct}(\varphi)
			\circ \TemplateTransform^\struct_{v_\struct}(\psi)
\qquad \text{ for } \circ \in \braces{ \land, \lor, \to }
			\\
			\TemplateTransform^\struct_{v_\struct}
			\parens{ \forall x \colon \sort.\varphi }
			& =
			\forall x^\supLIA.\ \bigwedge_{ n \in \Domain_\sort} \parens{ \Bounds(n)[x^\supLIA/x] \to
				\TemplateTransform^\struct_{v_\struct \brackets{ \pair{ n, x^\supLIA } / x }}
				(\varphi) }
\\
\TemplateTransform^\struct_{v_\struct}
			\parens{ \exists x \colon \sort.\varphi }
			& =
			\exists x^\supLIA.\ \bigvee_{ n \in \Domain_\sort } \parens{ \Bounds(n)[x^\supLIA/x]  \land
				\TemplateTransform^\struct_{v_\struct \brackets{ \pair{ n, x^\supLIA } / x }}
				(\varphi)}
\end{align*}
	\end{footnotesize}
	\end{minipage}
\caption{
	Model-checking \structure{} $\struct$ and assignment $v$ by
	translation to LIA.
}
\label{fig:templates:transformation}
\end{figure}
 \fi

The transformation uses the
LIA terms and formulas that interpret
function and relation symbols in the \structure{}
as building blocks to transform the formula.
Specifically, if $\structAss$ interprets the terms $t_i$ as $\pair{n_i, s_i}$,
then
the atomic formula $R \parens{ t_1, \dots, t_k }$
is transformed to the LIA formula
$R^\struct \parens{ n_1, \dots, n_k }$, which the
\structure{} defines as the interpretation of $R$ on nodes
$n_1, \ldots, n_k$,
except that the variables $x_1, \ldots, x_k$
that may appear in $R^\struct \parens{ n_1, \dots, n_k }$
are substituted by the LIA terms $s_1, \ldots, s_k$.

Similarly, the atomic formula $t_1 \foleq t_2$
is transformed to the LIA formula
$s_1 \liaeq s_2$ if the nodes $n_1$ and $n_2$ are the same,
requiring that $t_1,t_2$ are mapped not only to the same node
but also to the same element of the node,
and transformed to $\bot$ otherwise.

The transformation of the Boolean connectives is straightforward.

A universal quantifier $\forall x$ is transformed into a
universal quantifier over an integer variable~$x^\supLIA$,
such that for every node $n$ in the \structure{} 
(captured by a conjunction over all nodes of the corresponding sort),
all the elements of $n$, captured by assignments to $x^\supLIA$
that satisfy the bound formula of $n$,
satisfy the body of the quantified formula.
Similarly for an existential quantifier.

Note that the quantifier structure of
$\TemplateTransform^\struct_{\structAss} (\varphi)$ is the same as that of $\varphi$,
except that quantification is over integer variables.
Further, if $\FV(\varphi) \subseteq \{x_1,\ldots,x_k\}$
then
$\FV(\TemplateTransform^\struct_{\structAss} (\varphi))
\subseteq \{x^\supLIA_1,\ldots,x^\supLIA_k\}$.
Moreover, as expected,
if two \structAssignments{} agree on all the free variables of $\varphi$,
then they result in the same \LIA{} formula.
In particular, for a closed formula,
the transformation does not depend on the \structAssignment{} used.
The following theorem, proven by structural induction on $\varphi$, summarizes the correctness of the transformation for
model checking \aStruct{}.

\begin{theorem}
	\label{thm:templates:model-checking}
	Let $\struct$ be \aStruct{}, $v$ \aConcrete{} assignment,
	$\structAss$ its \templatization{} and $v_\subLIA$ the residual assignment.
Then for every formula $\varphi$ over $\Sigma$ the following holds:
	\[
	\Concrete(\struct), v \models \varphi
	\iff
	v_\subLIA \models_\subLIA \TemplateTransform^\struct_{\structAss}(\varphi).
	\]
\end{theorem}

\begin{example}\label{ex:templates:transformation}
	Continuing with our running example, let us consider checking whether the \structure{} from \Cref{ex:templates:definition}
    satisfies the formula
	$\varphi =
	\forall x_2. \neg
	\parens{ {\color{blue}{
		\underline{\PreviousFun \parens{ \PreviousFun \parens{ x_1 }  }}
		\llt
		x_2}}
	}$
    under some \concrete{} assignment $v$
    where $v(x_1) = \pair{\beta, -7}$.
The transformation of $\varphi$ w.r.t.\ $\structAss$ is
	\[
	\TemplateTransform^\struct_{\structAss}(\varphi) =
    \forall x^\supLIA_2.\
	\parens{
		x_2^\supLIA \liaeq 0 \to \neg {\color{blue}{\bot}}
	} \land \parens{
		x_2^\supLIA \leq 0 \to \neg \parens{ {\color{blue}{\parens{ \underline{\parens{ x^\supLIA_1 - 1 } - 1 }} < x^\supLIA_2 }} }
	}.
	\]
(The blue color highlights the transformation of the atomic formula and the underline
depicts the transformation of the corresponding term.)
Recall that in this example $v_\subLIA(x_1^\supLIA) = -7$.
Therefore, $v_\subLIA \not \models \TemplateTransform^\struct_{\structAss}(\varphi)$,
which is expected  since $\Concrete(\struct), v \not \models \varphi$ (see \Cref{ex:templates:concretization}).
\end{example}

\subsection{Beyond Single-Dimension \STructures{}}
We illustrate one limitation of \structures{},
and sketch a possible way to overcome it.
The summary nodes we have seen thus far
represent a single dimension of infinity
(a single copy of $\Integers$),
and \structures{} can compactly describe
infinite structures with relatively simple repeating patterns
using those summary nodes.
However, 
they cannot represent
structures with more complex,
e.g., ``nested'',
repeating patterns.
One such example would be a structure
where domain elements are linearly ordered in a fashion
akin to
$\omega \times \omega$.

\begin{table}
\caption{Conjuncts of $\BeyondFormula$,
which has a model akin to
	$\omega \times \omega$.}
\vspace{-0.25cm}
	\begin{footnotesize}
		\begin{tabular}{@{}ll@{}}
			\toprule
			Conjunct & Meaning \\
			\midrule
			$ \forall x, y, z. \parens{ R(x, y) \land R(y, z) } \to R(x, z) $
			&
			$R$ is transitive
			\\
			$ \forall x. \neg R(x, x) $
			&
			$R$ is anti-reflexive
			\\
			$ \forall x, y. R(x, y) \lor R(y, x) \lor x \foleq y $
			&
			$R$ is linear
			\\
			$ \forall x. R(x, f(x)) $
			&
			$f$ is ``increasing''
			\\
			$ \forall x, y. R(x, y) \to \parens{ R(f(x), y) \lor f(x) \foleq y } $
			&
			$f$ is a ``successor''
			\\
			$ \forall x. P(x) \liff P(f(x)) $
			&
			$f$ preserves $P$
			\\
			$ \forall x. \exists y, z. R(x, y) \land R(x, z) \land P(y) \land \neg P(z) $
			&
			$P$ alternates infinitely
\\
			\bottomrule
		\end{tabular}
	\end{footnotesize}
	\label{tbl:templates:beyond}
\end{table}

Specifically, consider a vocabulary
$ \Sigma = \braces{ f(\cdot), P(\cdot), R(\cdot, \cdot) } $,
where $f$ is a unary function,
$P$ is a unary relation, and $R$ is a binary relation.
The formula $\BeyondFormula$,
defined by the conjunction of the formulas
in~\Cref{tbl:templates:beyond},
requires its models to include an infinite repetition
of infinitely many $P$'s,
followed by infinitely many $\neg P$'s.

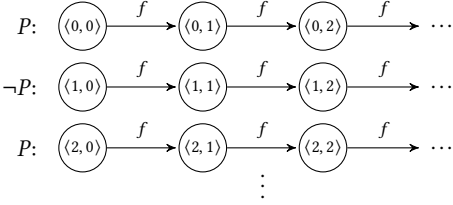
\begin{SCfigure}[3]
\begin{tikzpicture}[->,>=stealth',shorten >=1pt,auto,scale=0.8]

\tikzset{enode/.style={circle,draw,minimum size=18pt,inner sep=0mm,font=\tiny}}
\tikzset{fun/.style={font=\tiny}}
\tikzset{dots/.style={font=\small}}
\tikzset{rel/.style={font=\small}}

\node[rel,align = right]
at (2,3)
{ $\phantom{\neg} P$: };
\node[enode]
(n-1-1)
at (3,3)
{ $\pair{ 0, 0 }$ };
\node[enode]
(n-1+0)
at (5,3)
{ $\pair{ 0, 1 }$ };
\node[enode]
(n-1+1)
at (7,3)
{ $\pair{ 0, 2 }$ };
\node[dots]
(n-1+2)
at (9,3)
{ \dots };

\node[rel,align = right]
at (2,2)
{ $\neg P$: };
\node[enode]
(n+0-1)
at (3,2)
{ $\pair{ 1, 0 }$ };
\node[enode]
(n+0+0)
at (5,2)
{ $\pair{ 1, 1 }$ };
\node[enode]
(n+0+1)
at (7,2)
{ $\pair{ 1, 2 }$ };
\node[dots]
(n+0+2)
at (9,2)
{ \dots };

\node[rel, align = right]
at (2,1)
{ $\phantom{\neg} P$: };
\node[enode]
(n+1-1)
at (3,1)
{ $\pair{ 2, 0 }$ };
\node[enode]
(n+1+0)
at (5,1)
{ $\pair{ 2, 1 }$ };
\node[enode]
(n+1+1)
at (7,1)
{ $\pair{ 2, 2 }$ };
\node[dots]
(n+1+2)
at (9,1)
{ \dots };

\node[dots]
at (6,0.5)
{ \vdots };

\path
(n-1-1) edge node[fun] { $f$ } (n-1+0)
	(n-1+0) edge node[fun] { $f$ } (n-1+1)
	(n-1+1) edge node[fun] { $f$ } (n-1+2)

(n+0-1) edge node[fun] { $f$ } (n+0+0)
	(n+0+0) edge node[fun] { $f$ } (n+0+1)
	(n+0+1) edge node[fun] { $f$ } (n+0+2)

(n+1-1) edge node[fun] { $f$ } (n+1+0)
	(n+1+0) edge node[fun] { $f$ } (n+1+1)
	(n+1+1) edge node[fun] { $f$ } (n+1+2)
	;
\end{tikzpicture}\caption{Model $M$ for $\BeyondFormula$,
	which is linearly ordered in a fashion akin to
	$\omega \times \omega$.
	The relation $R$ is reflected by the ordering of the nodes,
	which are ordered top to bottom, left to right.
	$P$ alternates between the rows,
	i.e., either $P$ holds for an entire row
	or $\neg P$ holds for an entire row
	(as noted to the left).
	The successor function $f$ always stays within the same row,
	mapping each node to its immediate neighbor to the right.
	\vspace{-10pt}
}
\label{fig:templates:beyond}
\end{SCfigure} 
The conjunction $\BeyondFormula$ is satisfiable by the infinite model
$M = \parens{ \Domain^M, \Interp^M }$ ,
depicted in \Cref{fig:templates:beyond},
where $\Domain^M = \Nat \times \Nat$
and $\Interp^M$ is defined as
\[
\begin{footnotesize}
\begin{array}{c;{3pt/1pt}c}
	f^M \parens{ \pair{ a, b } } =  \pair{ a, b + 1 }
	&
	P^M =  \braces{ \pair{ a, b } \mid a \equiv 0 \mod 2 }
\\
	\multicolumn{2}{c}{R^M = \braces{\parens{ \pair{ a_1, b_1 }, \pair{ a_2, b_2 } }\mid a_1 < a_2 \lor \parens{ a_1 = a_2 \land b_1 < b_2 }}.
	}
\end{array}
\end{footnotesize}
\]
I.e.,
the domain consists of pairs of integers,
$f^M$ increments the second integer in each pair
(along the horizontal dimension in \Cref{fig:templates:beyond}),
$P^M$ holds whenever the first integer in a pair is even
(alternating along the vertical dimension in \Cref{fig:templates:beyond}),
and
$R^M$ is the lexicographical order over pairs of integers.

In order to represent
this structure
using \aStruct{},
infinitely many summary nodes would be needed.
However, \structures{} can be naturally generalized
to allow representation of such structures by
\begin{inparaenum}
\item
letting each summary node represent a subset of
$\Integers \times \Integers$ instead of $\Integers$,
and, accordingly,
\item
attaching to each summary node $n$ two LIA variables,
$x^1$ and $x^2$,
to be used in
bound formulas,
and in the LIA terms and formulas
that interpret the function and relation symbols.
The intention is that an assignment
$\brackets{ x^1 \mapsto z_1, x^2 \mapsto z_2 }$
represents the \concrete{} element $\pair{n, z_1, z_2}$ of $n$.
\end{inparaenum}

With this generalization,
the structure $M$ from above,
which satisfies $\BeyondFormula$,
can be represented by \aStruct{}
with a single summary node,
$T = \parens{ \Domain = \braces{ n }, \Bounds, \Interp }$,
where $\Bounds$ and $\Interp$ are defined as
\[
\begin{footnotesize}
\begin{array}{rcl;{3pt/1pt}rcl}
	\Bounds(n) & = & x^1 \geq 0 \land x^2 \geq 0
	&
	f^T(n) & = & \pair{ n, x^1_1, x^2_1 + 1 }
	\\
	P^T(n) & = & x^1_1 \equiv 0 \mod 2
	&
	R^T(n, n) & = & x^1_1 < x^1_2 \lor \parens{ x^1_1 = x^1_2 \land x^2_1 < x^2_2 } .
\end{array}
\end{footnotesize}
\]
where $x^j_i$ is the variable representing the
$j$'th copy of $\Integers$ associated
with the node used as the $i$'th argument
of a function / relation.
That is, the lower indices refer
to different arguments, as before,
while the upper indices refer to one of the copies
of $\Integers$ within a summary node.
$T$ represents an infinite model,
which is isomorphic to the model $M$ above.
(The domain
$\Domain^M = \Nat \times \Nat$
maps straightforwardly to the \concrete{} domain of $T$,
$\braces{
\pair{n, z_1, z_2} \in
\braces{ n } \times \Integers \times \Integers
\mid z_1 \geq 0 \land z_2 \geq 0
} = \braces{ n } \times \Nat \times \Nat $.)

We can further generalize
and let the multiplicity of the copies of $\Integers$ represented by each node
be a
parameter of either the \structure{} (as a whole)
or each summary node
(regular nodes can be thought of as 0-dimension summary nodes).
Both the \concretization{} and the translation 
at the heart of the model checking procedure for \structures{}
can be generalized to multi-dimensional \structures{} 
such that the model-checking theorem
(\Cref{thm:templates:model-checking})
still holds.
However,
note 
that with any generalization,
the fact
that \structures{}
are
encoded in a finite way
and that their model-checking is decidable
means that some formulas will not have \aStruct{}:
any recursively enumerable set of models,
for which model-checking is decidable,
cannot be complete since FOL satisfiability is not in RE.

For the remainder of the paper,
we exclusively use single-dimension \structures{},
since they effectively handle
the examples we consider.

  \section{Checking Satisfiability by Finding \STructures{}}
\label{sec:finding}
\commentout{
In this section,
we present our algorithm for finding templates
for formulas in
the decidable fragment \FragmentShortName{}.
\sharon{even if we don't swap the order of the sections, we need to revise this sentence}
More generally,
this algorithm describes an efficient way to traverse
a family of templates.
Efficiency is achieved by representing the entire family \emph{symbolically},
rather than enumerating and checking each template.
The idea is that given some formula $\varphi$,
we construct a LIA formula where a satisfying assignment
induces a satisfying template for $\varphi$.
For formulas in \FragmentShortName{} this process
gives us a decision procedure,
but it is sound for formulas outside \FragmentShortName{}
as well.}
We now present an algorithm
for finding a satisfying \structure{}
for a given FOL formula
within a given search space.
Broadly,
given a FOL formula, 
we leverage an SMT solver 
to \emph{symbolically} search 
for \aStructModel{} in a
(possibly infinite)
family of \structures{} 
given by \aBeforeTemplate{} \emph{\template{}} (defined below).
The idea is that given some FOL formula $\varphi$ and \aTemplate{},
we construct a LIA formula where a satisfying assignment
induces \aStructModel{} for $\varphi$.
This process is sound for any formula,
i.e., whenever \aStructModel{} is found,
its \concretization{}
satisfies the formula.
In \Cref{sec:decidable} we present a class of formulas,
for which this process is also complete:
i.e.,
for any formula
in that class,
there is a computable \template{},
such that
if the formula is satisfiable,
\aStructModel{} is guaranteed to be found
when searching using this \template{}.

\paragraph{\StructAdj{} \templates{}}
\commentout{To define the space of templates
over which the algorithm searches,
we use the notion of a template skeleton,
which fixes the nodes in the domain of each sort,
but leaves the bound formulas and
the interpretation of symbols under-specified.}

To define the search space of \structures{},
i.e., \aStruct{} family,
we use the notion of \aBeforeStructure{} \emph{\structAdj{} \template{}}.
\AStructAdj{} \template{}, similarly to \aStruct{}, 
fixes a finite set of nodes for each sort's domain,
but uses \emph{finite sets} of LIA terms and formulas
with additional free variables
as \emph{candidates} for the bound formulas and
the interpretation of functions and relations.
Formally:

\begin{definition}
	\ABeforeStructure{} \emph{\structAdj{} \template{}}
	for a vocabulary $\Sigma$
	is a triple
	$\finder = ( \Domain, \BoundsF, \InterpF )$
	where $\Domain$ is \aStruct{} domain,
	$\BoundsF$ is a \emph{finite} set of bound formulas
	for summary nodes,
	and $\InterpF$ defines the space of interpretations
	for function and relation symbols:
	for every function symbol $f \in \Sigma$,
	$\InterpF(f) \subseteq \Terms_\LIA$
	and for every relation symbol $R \in \Sigma$,
	$\InterpF(R) \subseteq \QF_\LIA$,
	such that $\InterpF(f)$ and $\InterpF(R)$ are finite.
\end{definition}
We distinguish two kinds of free variables
in bound formulas, function terms and relation formulas in $\finder$.
\emph{\StructAdj{} free variables} are the same as the free variables in \structures{}, 
used to identify \concrete{} elements in nodes.
Namely, they consist of $x$ in  bound formulas
and of $x_1, x_2, \dots$ in function terms and relation formulas, according to their arity (see \Cref{sec:templates}).
All other free variables are considered \emph{\template{} free variables},
conventionally denoted $\IntFlag, \IntFlag{}{1}, \IntFlag{}{2}, \dots$.
These variables are not part of \aStruct{} definition;
they are used
to capture infinitely many similar \LIA{} terms and formulas,
and are replaced by an interpreted LIA constant
(integer)
when the \template{} is instantiated.

\AStructAdj{} \template{} $\finder = ( \Domain, \BoundsF, \InterpF )$
represents the set of all \structures{}
$\struct$
over the domain $\Domain$,
where
bound formulas are taken from $\BoundsF$
for summary nodes
and are $x \liaeq 0$ for regular nodes,
constants are interpreted as regular nodes,
the terms and formulas in the interpretations of
function and relation symbols are 
taken from $\InterpF$,
with all \template{} free variables
substituted by integers.
Note that these 
integers
may be different 
in each use
of a term or formula.
For example, two nodes may use the same bound formula
$x \geq d \in \BoundsF$
with different substitutions, resulting in bounds
$x \geq 0, x \geq 1$.
The formal definition follows, where $\SkeletonFreeVariables(a)$
denotes the set of \template{} free variables appearing in a term or formula $a$.

\begin{definition}
\AStruct{} $\struct = \parens{ \Domain, \Bounds, \Interp }$
is \emph{represented by} \aStruct{} skeleton
$\finder = \parens{ \Domain, \BoundsF, \InterpF }$
if
\begin{inparaenum}[(i)]
\item for every regular node $n \in \Domain$,
$\Bounds(n) = x \liaeq 0$,
and for every summary node $n \in \Domain$,
$\Bounds(n) =
	\psi \brackets{ z_1 / \IntFlag{}{1}, \dots, z_m / \IntFlag{}{m} }$,
where $\psi \in \BoundsF$,
$z_1, \dots, z_m \in \Integers$,
and $\SkeletonFreeVariables(\varphi) = \braces{ \IntFlag{}{1}, \dots, \IntFlag{}{m} }$; similarly,
\item for every $n_1, \dots, n_k \in \Domain$ and
$f \in \Sigma$
(resp.\ $R \in \Sigma$),
$f^T \parens{ n_1, \dots, n_k } =
	\pair{ n, s \brackets{ z_1 / \IntFlag{}{1}, \dots, z_m / \IntFlag{}{m} } }$
(resp.\
$R^T \parens{ n_1, \dots, n_k } =
	\psi \brackets{ z_1 / \IntFlag{}{1}, \dots, z_m / \IntFlag{}{m} }$)
for some node $n \in \Domain$,
integers $z_1, \dots, z_m$ and
$s \in \InterpF (f)$
(resp.\ $\psi \in \InterpF (R)$)
s.t.\
$\SkeletonFreeVariables(t) = \braces{ \IntFlag{}{1}, \dots, \IntFlag{}{m} }$
(resp.\
$\SkeletonFreeVariables(\psi) = \braces{ \IntFlag{}{1}, \dots, \IntFlag{}{m} }$).
\end{inparaenum}
\end{definition}

In the sequel,
we refer to the \template{} and
the set of \structures{} it represents interchangeably,
and write $\struct \in \finder$
when $\struct$ is represented by $\finder$.

\begin{example}
To
	capture a family of \structures{} where
	all summary nodes represent a full copy of $\Integers$,
	we can define \aTemplate{} $\FinderOf{\struct_1}$
	where $\FinderOf{\Bounds_1} = \braces{ \top }$.
	The bound formula $\top$
	has neither \template{} free variables
	nor \structAdj{} free variables.
	A richer family of \structures{},
	where summary nodes may also represent
	the sets of positive or negative integers,
can be captured by \aTemplate{} $\FinderOf{\struct_2}$
	where $\FinderOf{\Bounds_2} = \braces{ \top, x \geq 1, x \leq -1 }$.
	The bound formulas $x \geq 1, x \leq -1$
	contain the \structAdj{} free variable $x$
	and no \template{} free variables.
	Finally, to capture a family of \structures{}
	where summary nodes may represent any half-space of the integers,
	we can use instead the bound formulas
	$x \geq \IntFlag, x \leq \IntFlag$,
	where $\IntFlag$ is \aTemplate{} free variable
	(and $x$ is still a \structAdj{} free variable).
	Specific \structures{} within this \template{} would replace
	$\IntFlag$ with specific \LIA{} (integer) constants;
	for example, $x \geq 3$ or $x \leq -2$.
\end{example}

\paragraph{\Structure{} finding through LIA encoding}
To check if \aStructAdj{} \template{} $\finder$
includes \aStructModel{} for a FOL formula,
we use an encoding similar to the one
used in \Cref{sec:model-checking-enc} for
model checking an individual \structure{}.
The key difference is that the bound formulas and
the interpretation are not (fully) specified in $\finder$,
and, instead, need to be found.
To account for this, the encoding introduces
Boolean auxiliary variables
that keep track of which interpretation and
bound formulas are used,
and LIA auxiliary variables
that keep track of the
LIA constants assigned to \template{} free variables,
such that
each \structure{} in $\finder$ is induced by
some assignment to the auxiliary variables. 

\paragraph{Auxiliary variables}
We use the following Boolean auxiliary variables:
	a variable
	$\GuardFlag{ n \colon \psi }$
	for any summary node $n$
	and any bound formula $\psi \in \BoundsF$,
	represents the decision
	to use $\psi$ as the bound formula for $n$;
a variable
	$\GuardFlag{ c \to n }$
	for any constant $c \in \Sigma$
	and any regular node $n$,
	represents the decision to
	define $c^T$ as $\pair{ n, 0 }$;
a variable
	$\GuardFlag{ f \parens{ n_1, \dots, n_k }  \to \pair{ n, s }}$
	for any function $f \in \Sigma$,
	any term $s \in \InterpF(f)$,
	and any nodes $n_1, \dots, n_k, n$,
	represents the decision to define
	$f^T \parens{ n_1, \dots, n_k }$ as $\pair{n,s}$,
	modulo the substitutions of the
	\template{} free variables in $s$; and
a variable
	$\GuardFlag{R \parens { n_1, \dots, n_k } \colon \psi}$
	for any relation $R \in \Sigma$,
	any formula $\psi \in \InterpF(R)$,
	and any nodes $n_1, \dots, n_k$,
	represents the decision to define
	$R^T \parens{ n_1, \dots, n_k }$ as $\psi$,
	modulo the substitutions of the
	\template{} free variables appearing in $\psi$.

To encode the possible substitutions of \template{} free variables,
allowing different substitutions in different contexts,
the following LIA auxiliary variables are used:
a variable
	$\IntFlag{n}$
	for any summary node $n \in \Domain$,
	and any \template{} free variable
	$\IntFlag$ appearing in any bound formula
	$\psi \in \BoundsF$;
a variable
	$\IntFlag{ f \parens{ n_1, \dots, n_k } }$
	for any function
	$f \in \Sigma$,
	nodes
	$n_1, \dots, n_k \in \Domain$,
	and any \template{} free variable
	$\IntFlag$ appearing in any term $s \in \InterpF(f)$;
and
a variable
	$\IntFlag{ R \parens{ n_1, \dots, n_k } }$
	for any relation
	$R \in \Sigma$,
	nodes
	$n_1, \dots, n_k \in \Domain$,
	and any \template{} free variable
	$\IntFlag$ appearing in any formula $\psi \in \InterpF(R)$.

\paragraph{Encoding}
Given a FOL formula $\varphi$ and
\aStructAdj{} \template{} $\finder$,
the ``\structure{} finding'' LIA formula
is $\varphikey \land \varphiaux$,
where $\varphikey$ encodes satisfaction of $\varphi$,
and $\varphiaux$ ensures
that the assignment to the Boolean auxiliary variables
induces a well-defined \structure{}.
We obtain $\varphikey$ by transforming
$\varphi$ in a similar way to
$\TemplateTransform^T_{v_T} (\varphi)$
(see \Cref{fig:templates:transformation}).
However, here
we use
the Boolean auxiliary variables as guards
and the LIA auxiliary variables as parameters
in order to consider different possibilities
for bounds, constants,
and function and relation applications.
This affects the transformation of
terms, atomic formulas and quantifiers.
For example,
for
$\BoundsF = \braces{  x \geq \IntFlag, \top }$,
the encoding of quantifiers will use
$
\parens{
	\GuardFlag{n \colon x \geq \IntFlag} \to x \geq \IntFlag{n}
}
\land
\parens{
	\GuardFlag{n \colon \top} \to \top
}
$
as the bound formula for a summary node $n$,
and simply $x \liaeq 0$ for regular nodes.
Atomic formulas
use
a conjunction of Boolean auxiliary variables,
accumulating the decisions
for the constants, functions and relations
that appear in them,
with the LIA auxiliary variables
substituting the \template{} free variables.
For example, transforming $R(f(c))$
involves case splits using conjunctions
of
auxiliary variables
for $c$, $f$ and $R$.
Lastly,
if $\varphi$
has any
free variables,
the transformation adds a disjunction
over all possible \structAssignments{} $\structAss$.

We use $\varphiaux$ to enforce
the following constraints:
\begin{inparaenum}[(i)]
	\item
	for every constant, function and relation symbol,
	exactly one Boolean auxiliary variable is true
	for each combination of argument nodes;
	\item
	for each node, exactly one of the Boolean bound variables is true;
\item
	for every constant,
	the Boolean auxiliary variables map it
	to a regular node;
	and
\item
	for every function and every combination of argument nodes,
	their image according to the
	Boolean auxiliary variables
	satisfies the selected bound formula
	of the image node,
	according to the Boolean bound variables.
\end{inparaenum}
There are no constraints for the \LIA{} auxiliary variables.
\neta{if full}
\ifnofull
(Details in~\cite{infinite-needle-arxiv}.)
\else
(Details in \Cref{sec:full-finder-transformation}.)
\fi

Ultimately, we can check if $\varphi$
is satisfied by some \structure{} in $\finder$
(and if so find the satisfying \structure{})
by checking the satisfiability of $\varphikey \land \varphiaux$,
which can be done using an SMT solver.

\begin{restatable}{theorem}{FindingTheorem}
\label{thm:finding:finder}
Given a formula $\varphi$ over $\Sigma$
and \aStructAdj{} \template{} $\finder$,
there exists \aStruct{} $\struct$ in $\finder$
and \aConcrete{} assignment $v$ for $\Concrete(\struct)$
such that $\Concrete(\struct), v \models \varphi$
iff there exists an assignment $v'$
such that
$v' \models_\LIA \varphikey \land \varphiaux$.
Furthermore, we can efficiently extract $\struct$ and $v$ from $v'$.
\end{restatable}

\section{Decidability via \STructures{}}
\label{sec:decidable}

In this section we introduce the 
\emph{\FragmentLongName{} (\FragmentShortName{})} fragment,
a new decidable fragment of FOL,
where satisfiable formulas always have \structModels{}.
\FragmentShortName{} extends the decidable
many-sorted EPR fragment.
We establish decidability of \FragmentShortName{}
by proving a novel ``small \structModel{} property'',
showing that every satisfiable \FragmentShortName{} formula
has a (possibly infinite) model representable
by \aStruct{} of a bounded size
that uses a small subset of LIA.
The proof of the ``small \structModel{} property''
identifies a computable \template{},
with which the symbolic search algorithm of \Cref{sec:finding}
is guaranteed to find \aStructModel{} for formulas in the fragment.
Combined with soundness of the symbolic search,
this yields a decision procedure for formulas in \FragmentShortName{}.

This section is organized as follows:
we start by giving necessary background on the EPR fragment,
followed by a formal definition of the \FragmentShortName{} fragment;
we then continue with the proof of decidability
which is done
by first considering single-sort formulas (\Cref{sec:single-sort-construction}),
and then generalizing to the many-sorted case (\Cref{sec:extended-fragment}).

\paragraph{EPR}
The Effectively PRopositional (EPR) fragment of many-sorted FOL
consists of formulas whose
\emph{quantifier-alternation graph} is acyclic.
The quantifier-alternation graph of a formula $\varphi$ is a directed graph $(V,E)$, where
the set of vertices $V$ is the set of sorts
and the set of edges $E$ includes an edge
from $\sort_1$ to $\sort_2$ whenever
\begin{inparaenum}[(i)]
	\item
	$\varphi$ includes a function symbol where
	$\sort_1$ is the sort of one of the arguments and
	$\sort_2$ is the range sort, or
\item
	$\varphi$ includes an $\exists y \colon \sort_2$ quantifier
	in the scope of a $\forall x \colon \sort_1$ quantifier
	(assuming $\varphi$ is in negation normal form; otherwise
    the definition is extended
	according to the standard translation to negation normal form).
\end{inparaenum}

Acyclicity of the quantifier-alternation graph ensures that after Skolemization,
there are finitely many ground terms.
That is, the Herbrand universe, as used in the proof of Herbrand's theorem, is finite.\footnote{Herbrand's theorem considers FOL without equality,
	however it extends to equality through its axiomatization,
	leading to the same bounds on domain sizes.
}
This leads to a small model property:
every satisfiable EPR formula has a model whose domain size is bounded by the number of ground terms
of its Skolemized version,
which also
makes satisfiability checking decidable.
Moreover, the number of ground terms can be effectively \nolinebreak computed.

\commentout{
Acyclicity of the quantifier-alternation graph ensures that
there are finitely many ground terms of each sort,
resulting in a small model property:
by Herbrand's theorem every satisfiable
universally quantified formula over such a vocabulary
has a model whose domain size is bounded
by the number of ground terms for each sort\footnote{Herbrand's theorem considers FOL without equality,
	however it extends to equality through its axiomatization,
	leading to the same bounds on domain sizes in FOL with equality.
}.
This, in turn,
makes the problem of checking satisfiability decidable.
Moreover, the number of ground terms can be effectively computed
based on the quantifier-alternation graph.
}

\paragraph{The \FragmentShortName{} fragment}
\FragmentShortName{} extends the EPR fragment
  by allowing a designated sort, $\sort^\infty$, thought of as an ``infinite'' sort,
  which may have self-loops in the quantifier-alternation graph and whose domain
  does not have a finite model property.
  That is, some satisfiable \FragmentShortName{} formulas are only satisfied by models where the domain of $\sort^\infty$ is infinite.
  However, \FragmentShortName{} restricts the use of $\sort^\infty$ 
  such that models can always be represented using \structures{} from a restricted family, 
  essentially by restricting the use of $\sort^\infty$ to be monadic 
  except for one (optional) binary relation $\llt$ 
  that is axiomatized to be a strict linear order over $\sort^\infty$.
\commentout{
\sharon{add something like we had before: \FragmentShortName{} relaxes the acyclicity requirement of EPR:}
Formulas\sharon{removed "vocabulary of formulas"} in \FragmentShortName{}
may include
a designated sort, denoted $\sort^\infty$,
thought of as an ``infinite'' sort,
which may be equipped with a binary relation $\llt$,
axiomatized to be a strict linear order.
\FragmentShortName{} allows the quantifier-alternation graph
to have  self-loops at $\sort^\infty$,
but restricts the use of $\sort^\infty$
in the formula in other ways.
}
Formally, an \FragmentShortName{} formula is
$\Allt \land \psi$,\footnote{If $\psi$ does not include
$\sort^\infty$ and/or $\llt$, then $\Allt$ may be omitted.
Still, in that case, any model of $\psi$ can be extended to a model
of $\Allt$. Hence, adding $\Allt$ does not restrict the set of models.
For simplicity of the presentation,
we therefore assume that $\llt$ and its axiomatization
are always included in \FragmentShortName{} formulas.
} where:
\begin{compactenum}
	\item
	\label{enum:decidable:axiom-restriction}
	$\Allt$
	encodes the axioms of
	a strict linear order for $\llt$
	(anti-reflexivity $x \nllt x$,
	transitivity $x \llt y \land y \llt z \to x \llt z$,
	and linearity $x \llt y \lor y \llt x \lor x \foleq y$).

	\item
	\label{enum:decidable:variable-restriction}
	$\psi$ uses only one logical variable, $x$,
	of sort $\sort^\infty$.
	(Variables of other sorts are not restricted.)

	\item
	\label{enum:decidable:graph-restriction}
	In the quantifier alternation graph of $\psi$
	the only cycles
	are self-loops at $\sort^\infty$,
	and the only outgoing edges
	from $\sort^\infty$ are to $\sort^\infty$.

	\item
	\label{enum:decidable:unary-predicates-restriction}
	Functions and relations
	other than $\llt$
	used in $\psi$
	have at most one argument of sort
	$\sort^\infty$.

	\item
	\label{enum:decidable:nesting-restriction}
The only non-ground term of sort $\sort^\infty$
	that appears in $\psi$
	as an argument to functions is $x$.
	(That is, functions whose range sort is $\sort^\infty$ cannot be nested in non-ground terms.)
\end{compactenum}
For example,
the Echo Machine's
invariants
(\Cref{fig:overview:ivy})
and transition formulas
(e.g., \Cref{eq:overview:transition-relation})
are in \FragmentShortName{},
while the following 
are not:
$\forall x, y \colon \sort^\infty.\;
x \llt y \to f(x) \llt f(y)$
(violates~\ref{enum:decidable:variable-restriction});
$\forall x \colon \sort^\infty.\;\exists y \colon \sort.\; R(x, y)$
(violates~\ref{enum:decidable:graph-restriction});
$\forall x \colon \sort^\infty.\; Q(x, x)$
(violates~\ref{enum:decidable:unary-predicates-restriction});
and
$\forall x \colon \sort^\infty.\; x \llt f(f(x))$
(violates~\ref{enum:decidable:nesting-restriction}).

\paragraph{The $\Templatesepr$ \template{} family}
To establish decidability, we prove that
satisfiable \FragmentShortName{} formulas have \structModels{},
and that
all \structures{}
needed to satisfy formulas in \FragmentShortName{}
can be represented by \templates{} from the following family,
denoted $\Templatesepr$, where:
\begin{inparaenum}[(i)]
	\item bound formulas for all summary nodes are
	$\top$,
\item all function terms are
	$0$ or $x_1 + k$
	for some $k \in \Integers$, and
\item all relation formulas are degenerate ($\top$ or $\bot$),
	except for $\llt$,
	where they can also be $x_1 < x_2$ or $x_1 \leq x_2$.
\end{inparaenum}
Furthermore,
given an \FragmentShortName{} formula,
we can compute a
sufficient
\emph{finite subset} of $\Templatesepr$,
which we can enumerate.
Formally:

\begin{theorem}\label{thm:full-fragment-decidable}
There exists a computable function
$\Function{} \colon \textnormal{\FragmentShortName{}} \to \powerset{\Templatesepr} $,
such that for every formula $\varphi \in \textnormal{\FragmentShortName{}}$,
$\Function(\varphi)$ is finite,
and $\varphi$ is satisfiable iff
there exists \aTemplate{} $\finder \in \Function(\varphi)$,
and \aStruct{} $\struct \in \finder$ therein,
such that $\Concrete(\struct) \models \varphi$.
\end{theorem}

\begin{corollary}\label{cor:decidability}
Satisfiability is decidable for formulas in \FragmentShortName{}.
\end{corollary}

We prove \Cref{thm:full-fragment-decidable} in two steps:
firstly, we discuss the case where
there are no sorts other than $\sort^\infty$
(\Cref{sec:single-sort-construction}),
and secondly, we generalize our construction to many sorts
(\Cref{sec:extended-fragment}).

\subsection{Step 1: Single Sort Construction}
\label{sec:single-sort-construction}
In this first step of the proof,
we consider
the simpler case
of \FragmentShortName{} formulas
where the only sort appearing in a formula
is the infinite sort.
Let $\varphi = \Allt \wedge \psi$
be an \FragmentShortName{} formula
where the only sort is $\sort^\infty$.
(Note that in the case of a single sort,
all function and relation symbols in $\varphi$,
except for $\llt$, must be unary.)
The single-variable requirement ensures that $\psi$ can be Skolemized
into an equi-satisfiable formula $\psi'$ by
introducing a Skolem constant for every existential quantifier
(in negation normal form),
even if it is in the scope of a universal quantifier.

Let $G$ denote the set of ground terms,
$\ell$ the number of function symbols,
and $m$ the number of unary relations that appear in $\psi'$.
We show that if $\varphi$ is satisfiable,
there exists \aTemplate{} $\finder \in \Templatesepr$
and \aStruct{} $\struct \in \finder$ therein,
such that
$\Concrete(\struct) \models \varphi$.
Further, $\finder$ (and $\struct$) has at most
$\abs{G}$ regular nodes
and
$\BellNumber(\ell + 1) \cdot 2^{m(\ell + 1)} \cdot (\abs{G} + 1)^{\ell + 1}$
summary nodes,
where $\BellNumber(n)$ is the $n$'th ordered Bell number
counting the number of weak orderings on a set of n elements.
In addition, all function terms are $0$ or $x_1 + k$
for some $-\ell \leq k \leq \ell$.
The bounds on the number of nodes
and on $k$
restrict the set of relevant \templates{} in $\Templatesepr$
to a finite subset,
which defines $\Function(\varphi)$
in \Cref{thm:full-fragment-decidable}.

Skolemization ensures that the models of $\varphi = \Allt \wedge \psi$ are the same as the models of $\Allt \wedge \psi'$ when the latter are projected onto the original vocabulary.
It therefore suffices to reason about models of $\varphi' = \Allt \wedge \psi'$.
Let $M = \parens{ \Domain^M, \Interp^M }$ be a model for $\varphi'$.
We construct \aStruct{} $\struct = \parens{ \Domain^\struct, \Bounds^\struct, \Interp^\struct }$
such that $\Concrete(\struct) \models \varphi'$
and therefore $\Concrete(\struct) \models \varphi$.
(Note that $\Concrete(\struct)$ need not be isomorphic to $M$.)

\paragraph{Domain}
To define the domain of $\struct$,
we first partition the elements of $\Domain^M$
to equivalence classes according to whether or not
they satisfy the following atoms in $M$:
\begin{enumerate}[label={($\Domain$-\arabic*)}]
	\item
	\label{enum:domain:element}
	\emph{Element atoms}:
	these consist of \emph{order} element atoms,
	of the form $x \bowtie g$
    where $\bowtie \, \in \braces{ \llt,\foleq,\lgt }$\footnote{
		We write $x \lgt g$ for $g \llt x$.
	}
	and $g \in G$;
	and \emph{relation} element atoms,
	of the form $R(x)$, where $R$ is a unary relation symbol.

	\item
	\label{enum:domain:image}
	\emph{Image atoms}:
	these consist of \emph{order} image atoms,
	of the form $f(x) \bowtie g$;
    and \emph{relation} image atoms,
    of the form $R(f(x))$, where $f$ is a unary function symbol.

    \item
    \label{enum:domain:mixed}
	\emph{Mixed atoms}:
	$t \bowtie t'$,
    where
    $t,t' \in \braces{ x } \cup \braces{ f(x) \mid f \text{ is a unary function symbol} }$.
    \oded{maybe rethink the name ``mixed atoms'' later}
\end{enumerate}
Each class is labeled with the set of atoms
satisfied by its elements.
We introduce a node in the domain of the \structure{}, 
$\Domain^\struct$,
for each class in the partition.
Classes labeled by an atom of the form
$x \foleq g$
for some ground term $g$
are regular nodes
(i.e., with bound formula $x \liaeq 0$),
and the rest are summary nodes with bound formula $\top$.
Note that for every ground term $g \in G$,
exactly one node is labeled $x \foleq g$.
This means that each ground term corresponds
to a single regular node (a
regular node may correspond to multiple ground terms that are equal in $M$).
In particular, the number of regular nodes is bounded by $\abs{G}$.
\oded{can shorten a line by editing the above paragraph}

The number of summary nodes is bounded by the number of feasible combinations of atoms:
combinations of mixed atoms are bounded by $\BellNumber(\ell + 1)$,
combinations of relation (element and image) atoms are bounded by $2^{m(\ell + 1)}$,
and combinations of order (element and image) atoms are bounded by $(\abs{G} + 1)^{\ell + 1}$
(see discussion of \emph{segments} delimited by ground terms below).

For each element $e \in \Domain^M$
we use $\tau(e)$ for the node in $\Domain^\struct$ that corresponds to
its class,
and for each node $n \in \Domain^\struct$
we use $E(n) = \braces{ e \in \Domain^M \mid \tau(e) = n }$ for the set of $\Domain^M$ elements in its class.
(For regular nodes, $E(n)$ is a singleton.)
We assume an arbitrary strict linear order on the nodes, denoted $\ll$,
which we use as a tiebreaker
when defining the interpretation of $\llt$ and the function symbols.

Note that the partitioning takes into account all the atoms
that may appear as non-ground atomic formulas
in $\psi'$.
Thus, grouping elements to classes according to atoms
and defining a node for each class
allows us to define \aStruct{}
where the \concrete{} elements of $n$ in $\Concrete(\struct)$ satisfy
exactly the same atomic formulas as $E(n)$ in $M$.
This, in turn, ensures that $\Concrete(\struct)$ satisfies $\varphi'$.

\para{Interpretation of constants}
For a constant $c$, we define $c^\struct$ to be $c^\struct = \pair{ n, 0 }$, where $n$ is the regular node labeled $x \foleq c$.
Note that $n = \tau \parens{c^M}$.

\para{Interpretation of unary relations}
Unary relations are determined to be either $\top$ or $\bot$ for each node based on the relation element atoms that label it.
Namely, $R^\struct(n)$ is $\top$ if $R(x)$ labels $n$ and $\bot$ otherwise, for every unary relation $R$ and node $n$.

\para{Interpretation of $\llt$}
Defining $\llt^\struct$ is not as straightforward,
as it is not fully determined by the labeling.
Since each ground term in $G$
has a corresponding regular node,
the labeling by order element atoms essentially dictates an order on all the ground terms (and accordingly the regular nodes), and places each summary node in a \emph{segment} delimited by ground terms.
For example, if there are 3 ground terms $g_1, g_2, g_3$ and the labeling dictates $g_1 \llt g_2 \foleq g_3$, then one regular node corresponds to $g_1$, another to $g_2 \foleq g_3$,
and each summary node is either smaller than $g_1$, between $g_1$ and $g_2 \foleq g_3$, or greater than $g_2 \foleq g_3$.
Therefore, the only remaining question is how to order summary nodes that are in the same segment.
We use this observation to distinguish between the following cases in the definition of $n \llt^\struct n'$:

	\begin{enumerate}[label={($\Order$-\arabic*)}]
		\item
		\label[definition]{enum:decidable:order-regular}
If either of $n,n'$ is a regular node, then it corresponds to a ground term $g$
and $n \llt^\struct n'$ is determined to be either $\top$ or $\bot$ by the order induced by the labeling.
For example, if $n$ is labeled $x \foleq g$ and $n'$ is labeled $x \lgt g$ then $n \llt^\struct n'$ is $\top$.

		\item
		\label[definition]{enum:decidable:order-ground}
If both $n$ and $n'$ are summary nodes that belong to different segments, $n \llt^\struct n'$ is determined by the labeling via transitivity.
That is, if there exists some ground term $g$ such that $n$ is labeled $x \llt g$ and $n'$ is labeled $x \lgt g$ (or vice versa),
then $n \llt^\struct n'$ is $\top$ (respectively, $\bot$).

		\item
		\label[definition]{enum:decidable:order-otherwise}
For summary nodes $n$ and $n'$ that are in the same segment,
we use the standard $<$ order over the integers to
determine the order between $\pair{n,z}$
and $\pair{n',z'}$,
and use $\ll$ as a tiebreaker when  $z = z'$.
Formally, we define $n \llt^\struct n'$ to be $x_1 \leq x_2$ if $n \ll n'$ and $x_1 < x_2$ otherwise.
\end{enumerate}

\ref{enum:decidable:order-otherwise} means that within each segment that includes summary nodes, the ordering
of the (infinitely many) elements in $\Concrete(\struct)$  always alternates between elements from  different summary nodes.
For example, if $n_1$ and $n_2$ are two summary nodes in the same segment with $n_1 \ll n_2$, then
the ordering in $\Concrete(\struct)$ is
$\cdots \llt^{\Concrete(\struct)} \pair{n_1,-1} \llt^{\Concrete(\struct)} \pair{n_2,-1} \llt^{\Concrete(\struct)} \pair{n_1,0} \llt^{\Concrete(\struct)} \pair{n_2,0} \llt^{\Concrete(\struct)} \pair{n_1,1} \llt^{\Concrete(\struct)} \pair{n_2,1} \llt^{\Concrete(\struct)} \cdots$.
For an illustration of $\llt^{\Concrete(\struct)}$
\neta{if full}
\ifnofull
see~\cite{infinite-needle-arxiv}.
\else
see \Cref{sec:decidable-order-illustration}.
\fi
We note that this ordering
~is independent of $M$.
That is, the ordering would be alternating as above even if $\forall e \in E(n_1), e' \in E(n_2). \, e \llt^M e'$.

\commentout{
Consider two nodes, $n$ and $n'$. If either of them is a regular node, then it corresponds to a ground term,
and the interpretation of $n \llt n'$ is determined to be either $\top$ or $\bot$ by the labeling.
Moreover, the interpretation of $n \llt n'$ may be determined by the labeling via transitivity even if both $n$ and $n'$ are summary nodes. That is, if there exists some ground term $g$ such that the labeling dictates $n \llt g \land g \llt n'$ or $n' \llt g \land g \llt n$.
The labeling essentially dictates an order on all the ground terms, and places each summary node in a \emph{segment} delimited by ground terms.
For example, if there are 3 ground terms $g_1, g_2, g_3$, the labeling may dictate $g_1 \llt g_2 \foleq g_3$, and then each summary node is either smaller than $g_1$, between $g_1$ and $g_2 \foleq g_3$, or greater than $g_2 \foleq g_3$.
Therefore, the only remaining question is how to order summary nodes that are in the same segment.
For $n$ and $n'$ that are in the same segment,
we use the standard $<$ order over the integers to
determine the order between $\pair{n,z}$
and $\pair{n',z'}$,
and use $\ll$ as a tiebreaker when  $z = z'$.
Formally, we define $n \llt^\struct n'$ to be $x_1 \leq x_2$ if $n \ll n'$ and $x_1 < x_2$ otherwise.
This means that within each segment delimited by the ground terms,
$\Concrete(\struct)$ has either none or infinitely many elements, and the ordering always alternates between elements that come from different summary nodes.
For example, if $n_1$ and $n_2$ are two summary nodes in the same segment with $n_1 \ll n_2$, then
the ordering in $\Concrete(\struct)$ is
$\cdots \llt \pair{n_1,-1} \llt \pair{n_2,-1} \llt \pair{n_1,0} \llt \pair{n_2,0} \llt \pair{n_1,1} \llt \pair{n_2,1} \llt \cdots$.
We note that this ordering, which the rest of our construction depends on, is independent of $M$.
That is, the ordering would be alternating as above even if $\forall e \in E(n_1), e' \in E(n_2). \, e \llt^M e'$.
}

\para{Interpretation of unary functions}
The most intricate step in our construction is showing that it is possible to define an interpretation of the functions that is consistent with the labeling.
For a unary function symbol $f$,
we define the node in the image of $f^\struct$
according to the image atoms of $f$
and the corresponding term
according to the mixed atoms.

Specifically,
for a node $n$ labeled with a set $L$ of image atoms for $f$,
let $n'$ be a node labeled with a set $L'$ of element atoms,
such that
\begin{inparaenum}[(i)]
	\item
	\label{enum:interp:matching-atoms}
	$\braces{ a\brackets{ f(x) / x } \mid a \in L' } = L$
	(i.e., the element atoms of $n'$ agree
	with the $f$-image atoms of $n$), and
\item
	if $x \foleq f(x)$ labels $n$, then $n' = n$;
	otherwise, $n'$ is the least according to $\ll$
	among all nodes that satisfy~(\ref{enum:interp:matching-atoms}).\footnote{Such a node must exist
        (consider $\tau \parens{ f^M(e) }$
        for some $e \in E (n)$).
        }\footnote{We use the
    minimal according to $\ll$ only as a way to get a deterministic choice,
    since the choices made for different functions should be  consistent (e.g., if $n$ is labeled $f(x) \foleq g(x)$).
    This is not related to the role of $\ll$ as a tiebreaker for $\llt^\struct$.}
\end{inparaenum}
Furthermore, to define the term in $f^\struct(n)$ while ensuring consistency with the
  mixed atoms that label $n$ (e.g., $f(x) \llt g(x)$, $f(x) \foleq g(x)$),
  we observe that the mixed atoms partition $S = \braces{ x } \cup \braces{ h(x) \mid h \text{ is a unary function symbol} }$
  into linearly ordered classes of ``equal'' terms.
  This order is isomorphic to $\{0,\dots,N\}$ for some natural number $N < \abs{S}$,
  and we let $\mu$ be the mapping that assigns to each $t \in S$ the number in $\{0,\ldots,N\}$ isomorphic to its class.
For example, if $S = \braces{x, g(x), h(x)}$ and $n$ is labeled with $g(x) \llt x$ and $h(x) \foleq x$ then
  $\mu(g(x)) = 0$, and $\mu(x) = \mu(h(x)) = 1$.
  Finally, we let
  $k = \mu(f(x)) - \mu(x)$ be the (possibly negative) ``distance'' between the classes of $f(x)$ and $x$ according to the aforementioned order, and define $f^\struct(n)$ using $k$ as follows:
\begin{small}
	\begin{subequations}
	\begin{empheq}[left={f^\struct(n) = \left\{}, right={\right.}]{align}
		\tag{$\Function$-1}\label{eq:decidable:f-regular}
		&\pair{ n', 0 } && \text{ if $n'$ is a regular node, }
\\
		\tag{$\Function$-2}\label{eq:decidable:f-summary}
		&\pair{ n', x_1 + k } && \text{ otherwise. }
	\end{empheq}
	\end{subequations}
	\end{small}

The use of $k$ ensures function interpretations 
are consistent with $n$'s mixed atoms.
Specifically, if $x \foleq f(x)$ labels $n$ then $n'=n$ and $k=0$,
and if $g(x) \llt h(x)$ labels $n$ then 
the $k$ used for $g$ is smaller than the $k$ used for $h$.
If the image nodes of both $g(x)$ and $h(x)$ are summary nodes
within the same segment, 
the choice of $k$
ensures the the correct ordering of the images,
otherwise, the order is guaranteed by the
ordering of the segments.

The following lemma summarizes the correctness of the construction 
\neta{if full}
\ifnofull
(see proof in~\cite{infinite-needle-arxiv}).
\else
(see proof in \Cref{sec:proofs:decidable}).
\fi

\begin{lemma}
\label{lem:decidable:single-sort-construction}
For any \FragmentShortName{} formula $\varphi$
where the only sort is $\sort^\infty$,
and any model thereof,
if
$\struct$
is the \structure{} obtained
by the above construction,
then $\Concrete(\struct) \models \varphi$.
\end{lemma}

\sidecaptionvpos{figure}{t}
\begin{SCfigure}
	\centering
	\scalebox{0.725}{
	\begin{tikzpicture}[->,>=stealth',shorten >=1pt,auto,scale=0.7]

		\node[tnode] (cM) at (1,1)
		{\smaller $c^M$};

		\node[tnode] (a1) at (3,1)
		{\smaller $R^M$};

		\node[tnode] (a2) at (5,1)
		{};

		\node[tnode] (a3) at (7,1)
		{};

		\node[tnode] (a4) at (9,1)
		{\smaller $R^M$};

		\node (a-rest) at (11,1)
		{\dots};

		\node[tnode] (b-1) at (-1,1)
		{};

		\node[tnode] (b-2) at (-3,1)
		{\smaller $R^M$};

		\node[tnode] (b-3) at (-5,1)
		{\smaller $R^M$};

		\node (b-rest) at (-7,1)
		{\dots};

		\path
		(cM) edge node {\tiny $f$} (a1)
		(a1) edge node {\tiny $f$} (a2)
		(a2) edge [bend right = 45] node [below] {\tiny $f$} (a4)
		(a3) edge [bend left = 45] node [above] {\tiny $f$} (a-rest.north)
		(a4) edge node [above] {\tiny $f$} (a-rest)

		(b-1) edge node [above] {\tiny $f$} (b-2)
		(b-2) edge [bend left = 60] node [below] {\tiny $f$} (b-rest.south)
		(b-3) edge node [above] {\tiny $f$} (b-rest)
		;

		\path[-,dashed]
			(-7.5,-1) edge node [above] {$M$} node [below] {$\struct$} (11.5,-1.25);

		\node[tnode] (cT) at (1,-2.5)
		{\smaller $c^\struct$};

		\node[tnode] (f-cT) at (3,-2.5)
		{\smaller $R^\struct$};

		\node[tnode, accepting] (a-neg-R) at (5,-2.5)
		{};

		\node[tnode, accepting] (a-R) at (9,-2.5)
		{\smaller $R^\struct$};

		\node[tnode, accepting] (b-neg-R) at (-1,-2.5)
		{};

		\node[tnode, accepting] (b-R) at (-3,-2.5)
		{\smaller $R^\struct$};

		\path
		(cT) edge node {\tiny $f:0$} (f-cT)
		(f-cT) edge node {\tiny $f:0$} (a-neg-R)
		(a-neg-R) edge [bend left] node [below] {\tiny $f: x_1 + 1$} (a-R)
		(a-R) edge [bend left] node [above] {\tiny $f: x_1 + 1$} (a-neg-R)

		(b-neg-R) edge [bend right] node [above] {\tiny $f: x_1 - 1$} (b-R)
		(b-R) edge [bend right] node [below] {\tiny $f: x_1 - 1$} (b-neg-R)
		;
	\end{tikzpicture}
	}
	\caption{An infinite model $M$
		and its constructed \structure{} $\struct$.
		The interpretation of $\llt$ orders
		elements from left to right;
		adjacent summary nodes are ordered alternatingly.
		Bound formulas for summary nodes are $\top$.\neta{add nodes, add mapping from model to template}
	\vspace{-15pt}
	}
	\label{fig:decidable:model}
\end{SCfigure}
\begin{example}
	\label{ex:decidable:construction}
	Consider the model $M$
	for
	$\varphi = \Allt \land
	\neg R(c)
	\land R(f(c))
	\land
	\forall x.
	\parens{ R(x) \liff \neg R(f(x)) }
	\land \parens { c \lleq x \liff x \llt f(x) }
	\land x \not\foleq f(x)$
	depicted in \Cref{fig:decidable:model}~(top).
The construction above yields the \structure{} $\struct$
	shown in \Cref{fig:decidable:model}~(bottom).
The formula has two ground terms, $c$ and $f(c)$,
	so $\struct$ has two regular nodes.
As for summary nodes,
	observe that in $M$ elements fall into two segments
	($M$ has no elements ``between'' $c$ and $f(c)$);
moreover, the elements in each segment
	fall into one of two equivalence classes,
	yielding a total of four summary nodes in $\struct$.
\neta{still relevant?}
    \sharon{so far great, but what about demonstrating def of $f$ and $R$?}

	Note that $\Concrete(\struct)$ and $M$ are not isomorphic.
	In particular,
	in $M$ there are two elements,
	adjacent according to $\llt^M$,
	that both satisfy $R(x)$
	(e.g., the two top-left elements drawn in \Cref{fig:decidable:model}),
	whereas $\struct$ completely regularizes the pattern
	of alternating $R(x)$ and $\neg R(x)$ elements.
\end{example}

\commentout{
\begin{remark}
	The proof of
    \iflong
\else
    \Cref{lem:decidable:construction-correctness} (see \Cref{lem:decidable:total-strict-order})
    \fi
	reveals that $\Logic$
	(and \FragmentShortName{})
	can be extended to different axiomatic orderings,
	with minimal changes to the construction above.
	In particular, we can allow a semi-linear order,
	where instead of linearity we have the weaker property
	that
	$\forall x,y,z.\
		\parens{ x \llt y \land x \llt z } \to
		\parens{ y \foleq z \lor y \llt z \lor z \llt y }
	$.
	This opens the door to describe
	a wider range of verification problems,
	with other notions of ``order''.
	\neta{see appendix for details?}
        \oded{do we want to keep this remark? Is it even true? If the order is not linear, are we sure everything still works? The segments might not be defined now, no? I'm not sure this remark helps us unless we actually present other order axiomatizations for which the construction holds. I don't think it's correct to just generally say it generalizes to other axiomatizations, since it depends on the details of each axiomatization, no?}
\end{remark}
}

\neta{revised:}\oded{I liked the previous version more, but it's not critical. I guess for me it seems easier to first understand that other atoms can be encoded/don't break the construction, and then think about why nested mixed atoms would break the construction.}\sharon{I don't remember what this comment is about. can we discuss?}

\begin{remark}
	The restriction \FragmentShortName{}
	imposes on nesting of $\sort^\infty$ functions
	is crucial for the construction,
	but it is only
	essential for mixed atoms,
	where it disallows
	atoms like $f(f(x)) \llt x$.
	Such atoms
	would invalidate the construction
	that defines $f^\struct(n)$
	solely according to the labeling of $n$.
Note that an atom like $R(f(f(x))))$ can be
	simulated in \FragmentShortName{}
	as $R'(f(x))$ where $R'$ is a fresh unary relation symbol
	axiomatized by $\forall x. R'(x) \liff
	R(f(x))$. Atoms like $f(f(x)) \llt g$ can be similarly encoded.
\end{remark}

\subsection{Step 2: Generalizing to Many Sorts}\label{sec:extended-fragment}
In the remainder of this section,
we generalize the construction of the previous section
to many sorts,
thus proving decidability of the \FragmentShortName{} fragment.
The generalization is done by refining
the partitions induced by the atoms defined in
\labelcref{enum:domain:element,,enum:domain:image,,enum:domain:mixed}.

Given $\varphi = \Allt \wedge \psi$, we first transform $\psi$
to $\psi'$ as in \Cref{sec:single-sort-construction}.
In particular, all existential quantifiers over sort $\sort^\infty$ are Skolemized
by functions from sorts other than $\sort^\infty$ to $\sort^\infty$.
We observe that for each sort
$\sort \neq \sort^\infty$
only finitely many ground terms can be constructed,
since the quantifier-alternation graph of $\psi'$ is acyclic except for self-loops at $\sort^\infty$
and $\sort^\infty$ has no outgoing edges to other sorts.
Hence, given a model $M \models \varphi' = \Allt \wedge \psi'$,
we can apply Herbrand's construction
(adapted to handle equality)
to obtain a model $M' \models \varphi'$
where the domain of each sort
$\sort \neq \sort^\infty$
is bounded by the number of ground terms
(as in many-sorted EPR).
Each element in these domains
will become a regular node in the \structure{} $\struct$,
setting bounds on the domain sizes of all sorts
$\sort \neq \sort^\infty$ in $\struct$.
The interpretation of all constant,
relation and function symbols that do not include $\sort^\infty$
in $\struct$ will be exactly as in $M'$ \sharon{candidate to remove for space (recently added):}
(using only $\top$ and $\bot$ for relation formulas and $0$ for  function terms).

Next, we address the elements of sort $\sort^\infty$ in $M'$.
These are grouped into finitely many summary and regular nodes in $\struct$
by a variant of the construction from \Cref{sec:single-sort-construction},
where we use the ground terms of the other sorts to refine the partition
of the elements of $\sort^\infty$.
Specifically:\sharon{removed itemize for shortening (and some additional butchering)}

\begin{enumerate}[labelindent=\parindent,itemindent=2.75\parindent,leftmargin=0pt,itemsep=0pt,label=$(\arabic*)$,topsep=0pt]
\item
We treat all (finitely many) instantiations
of multi-ary functions with (one) argument and range sort $\sort^\infty$
on the ground terms of sorts $\sort \neq \sort^\infty$
as unary functions from $\sort^\infty$ to $\sort^\infty$,
and extend the sets of image- and mixed atoms to refer to them.
For example, given 
$f \colon \sort^\infty \times \sort \to \sort^\infty$,
we add the atoms $x \llt f(x,g)$,
$x \foleq f(x,g)$,
and $x \lgt f(x,g)$
as mixed atoms
for each ground term $g$ of sort $\sort$.
\neta{also put image atoms in example?}

\item
We treat all (finitely many) instantiations of functions
with range $\sort^\infty$ and no argument of sort $\sort^\infty$
on the ground terms of the other sorts
as additional ground terms of sort $\sort^\infty$,
and extend the set of order (element and image) atoms to refer to them.
For example, given $h \colon \sort' \to \sort^\infty$ and $f$ as before,
we add 
the atoms $x \bowtie h(g')$
and $f(x, g) \bowtie h(g')$
for all ground terms $g',g$ of
sorts $\sort', \sort$, resp.

\item 
We treat all (finitely many) instantiations of
relation symbols other than $\llt$ that include $\sort^\infty$
on the ground terms of the other sorts
as additional unary relation symbols of sort $\sort^\infty$, and add
the corresponding relation (element and image) atoms.
For example, given $R \colon \sort^\infty \times \sort'$,
we add 
the atoms $R(x, g')$ and $R(f(x,g), g')$
for all ground terms $g',g$ of sorts
$\sort',\sort$, resp.
\end{enumerate}

The interpretation of the constant,
relation and function symbols that include $\sort^\infty$
in $\struct$ is defined similarly to \Cref{sec:single-sort-construction}.
Hence, we obtain \aStruct{} $\struct$,
such that $\Concrete(\struct) \models \varphi$;
the domain of each sort
$\sort \neq \sort^\infty$
consists only of regular nodes,
whose number is bounded by the number of
ground terms of the corresponding sort;
and the domain of $\sort^\infty$ consists of summary and regular nodes,
whose number is bounded by the equivalence classes of
the refined partition.
The function terms
and relation formulas
used in $\struct$ are restricted
as in \Cref{sec:single-sort-construction}.
Thus,
we have a computable finite subset of $\Templatesepr$
which includes \aTemplate{} $\finder$
such that $\struct \in \finder$.
Relating back to $\Cref{thm:full-fragment-decidable}$,
for any formula $\varphi$,
we define
$\Function(\varphi)$
to be the finite subset of $\Templatesepr$
that meets the restrictions above.

\section{Implementation and Evaluation}
\label{sec:evaluation}

We implemented the ideas presented in this paper
in a tool
named \ToolShortName{}
(\ToolLongName).\footnote{We plan to open-source
	the implementation
	once the paper is published.
}
\ToolShortName{}
allows the user both to specify \structures{} manually
and model-check them against FOL formulas,
and to automatically find \aStructModel{}
for a given formula.
The \structModel{} finding procedure
uses automatically-computed ``heuristic'' \templates{},
or a user-provided \template{}.
\ToolShortName{}
is implemented in Python,
takes input in SMTLIB syntax~\cite{smtlib-standard},
and uses \cvc{}~\cite{cvc5-smt-solver}
as the backend for checking LIA formulas
(per \Cref{thm:templates:model-checking,thm:finding:finder}).
\commentout{
\ToolShortName{} currently requires the user to
provide the number of nodes of each sort,
including the number of regular and summary nodes for the infinite sort.
This requirement can be lifted using an enumeration strategy and possibly parallelization.
}

We evaluated \ToolShortName{}'s \structModel{}
finding capabilities with the heuristic \templates{}
on a host of examples
with very positive results ---
\ToolShortName{} found \structModels{}
for all of the examples,
and many of the \structModels{}
are
captured by the \template{} family $\Templatesepr$ of \FragmentShortName{}.
We compare against the \Solvers{} \cvc{}
and \Zzz{}~\cite{z3-smt-solver},
and the resolution-based theorem prover Vampire~\cite{vampire},
all of which diverge when run on these examples.

\begin{figure}[h]
\vspace{-8pt}
	\begin{minipage}{0.73\textwidth}
	\begin{minipage}{0.35\textwidth}
	\begin{footnotesize}
		\[
		\begin{array}{r@{\hspace{0.1cm}}c@{\hspace{0.1cm}}l}
			\BoundsH & = & \braces{ \top, x \geq \IntFlag, x \leq \IntFlag }
			\\
			\InterpH(f) & = & \braces{ 0, x_1, x_1 + 1, x_1 - 1 }
\end{array}
		\]
	\end{footnotesize}
	\end{minipage}
	\begin{minipage}{0.45\textwidth}
	\begin{footnotesize}
		\[
		\begin{array}{r@{\hspace{0.1cm}}c@{\hspace{0.1cm}}l}
			\InterpH(R) & = &
			\left\{
			\begin{array}{c}
				\top, \bot, 
				x_1 < x_2, x_1 \leq x_2, x_1 \geq x_2, x_1 > x_2, \\
				x_1 \liaeq 0, x_1 \liaeq x_2, x_1 \liaeq x_2 + 1, x_1 \liaeq x_2 - 1, \\
				x_1 \leq x_2 < x_3, x_3 \leq x_2 < x_1, x_1 \liaeq x_2 \liaeq x_3
			\end{array}
			\right\}
		\end{array}
		\]
	\end{footnotesize}
	\end{minipage}
	\end{minipage}
	\vspace{-8pt}
	\caption{The heuristic \template{} family $\Heuristic$,
		parameterized by domain $\Domain$.
}
	\label{fig:evaluation:heuristic}
	\vspace{-8pt}
\end{figure}

\paragraph{Heuristic \templates{}}
\ToolShortName{} uses the heuristic \template{} family $\Heuristic$
defined in \Cref{fig:evaluation:heuristic},
and a domain enumeration strategy
for the number of regular and summary nodes.
The design of $\Heuristic$ was
motivated by the
$\Templatesepr$ \template{} family of \FragmentShortName{},
extended\footnote{\TemplatesWord{} for \FragmentShortName{}
	may include function terms $x \pm k$
	where $1 < k \leq \ell$
	(see \Cref{sec:single-sort-construction}).
	However,
in our benchmarks
	$\ell = 1$.
}
to support functions and relations
of greater arity than
what the syntactic restrictions of \FragmentShortName{} allow.
The $\Heuristic$ family uses bound formulas
with \template{} free variables
($x \geq \IntFlag, x \leq \IntFlag$)
to represent half-space summary nodes,
but avoids them in function terms and relation formulas
since they lead to harder LIA queries for the solver, and did not provide a benefit
for the examples
we considered.
The enumeration strategy prioritizes \templates{}
with homogeneous domain sizes
and smaller total domain size.

\paragraph{Optimizations}
\ToolShortName{} implements the translations to LIA
developed in \Cref{sec:templates,sec:finding},
with several optimizations.
First, when searching for a satisfying \structure{} for a given formula,
the interpretation of one constant from each sort
is arbitrarily fixed to the first regular node of that sort,
under an arbitrary node ordering.
This symmetry breaking reduces the search space
but does not hurt completeness
since any \structure{} is isomorphic
 to one of the \structures{} considered.
Second, since $\llt$ is assumed to be a strict linear order,
we have that for $x \neq y$, $x \llt y$ iff $y \nllt x$.
Thus, for every two nodes
we define auxiliary variables for $\llt$
only for one of the two possible permutations
(and derive the other one).
Third,
for the interpretation of functions
and relations, we consider a term or formula
with \aStructAdj{} free variable
$x_i$ only for arguments $n_1, \dots, n_k$
where $n_i$ is a summary node.
For example,
for a unary function $f$ the terms $x_1, x_1 + 1, x_1 - 1$
will only be considered
when the argument is a summary node;
for regular nodes the term $0$ will always be used.
This is in line with what \FragmentShortName{} guarantees
in \structModels{},
and indeed works for examples both in
and outside of \FragmentShortName{}.

\paragraph{Evaluation}
We evaluated \ToolShortName{}
in comparison to
\Zzz{}, \cvc{} and Vampire.
Current solvers
do not detect infinite models,
and
prior work on
SMT-based verification
circumvented them
in various ways.
For example,~\cite{bounded-horizon}
avoids infinite counter-models
by manually adding axioms,
and~\cite{ic3po}
checks
only bounded instances
and compares inferred invariants
to hand-written ones.
As there is no standard benchmark
suite
for infinite models,
we curated a set of 21 examples
of
distributed protocols
and programs manipulating
linked lists,
based on prior work
where FOL is used as an abstraction
and infinite models may arise.
We run the tools on these examples
using
a
Macbook Pro with an Apple M1 Pro CPU
and 32 GB RAM,
with
\Zzz{} version 4.12.2,
\cvc{} version 1.0.8,
and Vampire version 4.5.1.
The results are summarized in \Cref{tbl:evaluation:results}.
For each example,
the table indicates
if the example is in \FragmentShortName{},
\commentout{if the template found can be represented by a skeleton from $\Templatesepr$,}
the number of regular and summary nodes in the infinite sort,
the numbers of nodes in other sorts,
and the run time for finding the \structModel{}.
Running time was averaged across 10 runs.

\ToolShortName{} is able to
effectively find \structModels{}
for
all 21 examples.
In contrast, \Zzz{}, \cvc{} and Vampire
are unable to conclude that these examples are satisfiable,
and consistently time out after 10 minutes
or return ``unknown''
on all of these examples.
(We do not expect the results to be different
for any higher timeout threshold.)
To demonstrate the general usefulness of
\structures{} for representing and finding infinite models
we also include
an anecdotal example where we find
a non-standard model of non-commutative addition.
We briefly describe each example in the sequel;
the formal modeling
and depictions of some of the \structures{}
\neta{if full}
\ifnofull
are provided in~\cite{infinite-needle-arxiv}.
\else
are in \Cref{sec:examples}.
\fi

\begin{table}
\caption{Evaluation results for \ToolShortName{}. We report the mean and standard deviation 
		of run time
		across 10 runs
		with different random seeds.
``-'' denotes timeout after 10 minutes.
\neta{$\dagger$ mix with lia}
	}
	\vspace{-0.25cm}
	\begin{scriptsize}\begin{minipage}[t]{0.5\textwidth}\vspace{0pt}\hspace{-15pt}\begin{tabular}{@{}lcHcrr@{}HHH}
			\toprule
			Example 
& \parbox{0.7cm}{In \FragmentShortName{}}
& \parbox{1.2cm}{Template \\ belongs to $\Templatesepr$}
& \parbox{1cm}{ $\sort^\infty$ Size\\(reg./sum.) }
& \parbox{0.6cm}{Others' Sizes}
			& \multicolumn{1}{c}{Time (s)}
			& \Zzz{}
			& \cvc{}
			& Vampire
			\\
			\arrayrulecolor{black!30}
\midrule
			Echo Machine & Yes & Yes & 1/1 & 2 
			& $0.18 \pm 0.01$ 
			& - & - & -
			\\
			Voting protocol & No & Yes & 2/1 & 3, 2, 2 
			& $45.28 \pm 6.95$
			& - & - & -
			\\
			Simple Paxos & No & Yes & 2/1 & 2, 2, 2 
			& $12.72 \pm 0.14$
			& - & - & -
			\\
			Implicit Paxos & No & Yes & 2/1 & 3, 2, 2
			& $133.09 \pm 41.43$
			& - & - & -
			\\
			Paxos & No & No & 2/1 & 3, 3, 2 
			& $79.69 \pm 22.22$
			& - & - & -
			\\
			Flexible Paxos & No & No & 2/1 & 3, 3, 1, 1
			& $27.51 \pm 0.20$
			& - & - & -
			\\
			Ring Leader & No & Yes & 0/1 & n/a 
			& $0.08 \pm 0.00$
			& - & - & -
			\\
			Line Leader & No & Yes & 2/1 & n/a 
			& $1.59 \pm 0.08$
			& - & - & -
\\ 
			
			\arrayrulecolor{black}
			\bottomrule
		\end{tabular}
	\end{minipage}\begin{minipage}[t]{0.35\textwidth}\vspace{0pt}
		\begin{tabular}{@{}l@{\,\,}cHcHr@{}HHH@{}}
		\toprule
		Example 
& \parbox{0.7cm}{In \FragmentShortName{}}
& \parbox{1.2cm}{Template \\ belongs to $\Templatesepr$}
& \parbox{1cm}{ $\sort^\infty$ Size\\(reg./sum.) }
& \parbox{0.65cm}{Others' Sizes}
		& \multicolumn{1}{c}{Time (s)}
		& \Zzz{}
		& \cvc{}
		& Vampire
		\\
		\arrayrulecolor{black!30}
\midrule
List length
			$\dagger$ & No & No & 1/1 & n/a
			&  $0.09 \pm 0.00$
			& - & - & -
			\\ 
List segment (const)
			& No & No & 1/1 & n/a
			& $0.17 \pm 0.01$
			& - & - & -
			\\ 
List segment (var)
			& No & No & 1/1 & n/a
			& $0.25 \pm 0.01$
			& - & - & -
			\\ 
List segment (order)
			& No & No & 2/1 & n/a
			& $0.97 \pm 0.04$
			& - & - & -
			\\ 
List segment (rev)
			& No & No & 1/1 & n/a
			& $0.54 \pm 0.07$
			& - & - & -
			\\ 
Doubly-linked list
			& No & - & 1/1 & n/a
			& $0.16 \pm 0.01$
			& - & - & -
			\\ 
Doubly-linked length
			& No & No & 1/1 & n/a
			& $0.16 \pm 0.01$
			& - & - & -
			\\ 
Doubly-linked segment
			& No & - & 1/1 & n/a
			& $0.35 \pm 0.02$
			& - & - & -
			\\ 
Reverse list
			& No & No & 2/1 & n/a
			& $0.86 \pm 0.02$
			& - & - & -
			\\ 
Sorted list length
			$\dagger$ 
			& No & No & 1/1 & n/a
			& $0.09 \pm 0.00$
			& - & - & -
			\\ 
Sorted list
			& No & No & 1/1 & n/a
			&  $0.10 \pm 0.00$
			& - & - & -
			\\ 
Sorted list segment
			& No & No & 1/1 & n/a
			& $0.20 \pm 0.01$
			& - & - & -
			\\
Sorted list max
			& Yes & Yes & 1/1 & n/a
			& $0.06 \pm 0.00$
			& - & - & -
			\\
			\arrayrulecolor{black}
			\bottomrule
		\end{tabular}
	\end{minipage}
	\end{scriptsize}
	\label{tbl:evaluation:results}
\end{table} 
\para{Echo Machine}
The running example from \Cref{sec:overview},
inspired by the Paxos protocol.

\para{Paxos protocol and variants}
These are distributed consensus protocols for agreeing on a value
among sets of ``acceptors'' and ``values''
of unknown size.
Our FOL modeling follows~\cite{ic3po}.
We present 5 variants of Paxos:
abstract Paxos
(``Voting'' protocol~\cite{lamport-voting}),
simple and implicit Paxos~\cite{ic3po},
regular Paxos~\cite{lamport-paxos},
and flexible Paxos,
which considers a weaker quorum intersection
assumption~\cite{flexible-paxos}.
The infinite counter-models resemble that of the Echo Machine,
where rounds play a similar role.

\para{Line Leader}
A simple distributed protocol to elect a leader
for a set of nodes ordered in a line.

\para{Ring Leader}
This example models a leader election protocol for a set of nodes
ordered in a ring topology~\cite{ring-leader},
and tries to prove a simplified form of termination
following~\cite{ivy-original,bounded-horizon}.
The infinite counterexample essentially represents an infinite ring.
While the partial models presented in~\cite{bounded-horizon}
suggest to the user that an infinite counterexample exists,
\ToolShortName{} is able to find one explicitly.

\para{Linked Lists}
\neta{changed}
The right-hand side of \Cref{tbl:evaluation:results}
contains examples of verification of linked lists properties
(e.g., that reversing a list does not change
the set of elements within it).
The first 12 examples
consist of all of the
linked lists examples
from~\cite{natural-proofs}.
There,
linked lists are defined inductively,
and the least-fixed-point semantics
is approximated in FOL
by manually adding induction axioms.
We remove the added axioms,
and show that indeed infinite counter-models arise.
For two of the examples
which mix LIA and uninterpreted FOL together
\neta{cleaner way to use dagger}
(example 1 and 10, marked by $\dagger$),
we model the LIA sort as an uninterpreted sort with a single,
unbounded summary node.
This hints at the future possibility of combining \structures{} with theories.
\neta{TWIT: theories with infinite templates}\sharon{cool!}
The last example (``Sorted list max'')
was inspired by~\cite{natural-proofs},
but proves a different property of sorted linked lists:
that the last element is larger than the first.

\para{Axiomatic Arithmetic}
For the standard axiomatization of
Presburger arithmetic with successor and addition
($s(x)\neq0$, $s(x){=}s(y) \to x {=} y$, $x+0=x$, $x+s(y) = s(x+y)$),
it is well known that proving commutativity of addition
(by induction)
requires auxiliary lemmas. Interestingly,
a counterexample demonstrating this
can be represented by \aStruct{}.
For this example,
the user must provide two additional
function terms $x_1 + x_2, x_1 + x_2 + 1$.
Using these, \ToolShortName{}
finds \aStruct{} that has \emph{two} summary nodes
in about 5 seconds
\neta{if full}
\ifnofull
(see~\cite{infinite-needle-arxiv}).
\else
(see \Cref{fig:evaluation:non-commutative} in \Cref{sec:examples}).
\fi

\section{Related Work}
\label{sec:related}
\para{\indent Reasoning over specific infinite structures}
Several seminal works~\cite{buchi-mso-s1s,buchi-deciable-mso,rabin-mso-s2s}
consider monadic second-order logic (MSO)
over the specific (infinite) structures
of lines (S1S) and trees (S2S).
These works use finite state automata to represent
infinite models.
Linear Temporal Logic
(LTL)~\cite{vardi-ltl-automata,vardi-verification-ltl,ltl-sat-checking}
also considers verification in the context of
the specific infinite structure of the naturals.
\cite{klin-smt-infinite}
develops a verification framework over
the naturals and the rationals.
In contrast,
we aim to discover infinite counter-models
for VCs
that are meant to be valid over all structures.

\para{Representing infinite and unbounded structures}
\cite{automatic-structures,automata-infinite-structures,gradel-finite-infinite}
introduce automatic structures,
which represent infinite structures
using automata and regular languages.
However, they do not
automatically find a satisfying structure for a given formula,
nor identify a decidable fragment where representable satisfying structures
are guaranteed to exist.
\neta{added:}
The finite representation of infinite terms of co-datatypes 
introduced in~\cite{co-datatypes} resemble summary nodes.
However, summary nodes represent infinitely many different elements
while an infinite term represents a single element.
\Structures{} are inspired by
TVLA~\cite{tvla-popl,tvla-sas},
but whereas in TVLA summary nodes
represent an unbounded \emph{finite}
number of elements,
we use summary nodes to represent
\emph{infinitely} many elements.
The use of LIA terms and formulas
in \structures{} resembles symbolic automata~\cite{automata-modulo-theories},
where transitions carry predicates
over rich theories.
Our construction of \aStruct{} in \Cref{sec:decidable}
bears some resemblances
to predicate abstraction~\cite{predicate-abstraction}.
However, here it is only part of the decidability proof.

\para{Constructing infinite models}
Existing \Solvers{}~\cite{z3-smt-solver,cvc5-smt-solver,yices-smt-solver}
and
resolution-based theorem provers~\cite{vampire,spass}
cannot construct infinite models.
Resolution can sometimes terminate when
the clause set is saturated
even if the counter-model is infinite.
\neta{added}
\cite{finite-local-theories}
uses constrained clauses
in order to represent possibly infinite sets of saturated clauses.
\cite{bachmair-original,bachmair-infinite-models}
present theoretical constructions
of counter-models from saturated sets of clauses,
but these constructions
are not implemented in existing tools.
Recently,~\cite{saturated-triggers}
showed that under certain restrictions,
trigger-based quantifier instantiation
can be complete
and used to prove satisfiability
without an explicit construction of the model.
The model evolution calculus~\cite{model-evolution-calculus,model-evolution-implementation}
constructs Herbrand models
based on a generalization of DPLL,
where the interpretation of relations
is implicitly defined by a finite set of
possibly non-ground literals.
In contrast,
our approach represents the domain
and the interpretation explicitly,
using LIA formulas and terms.
Our approach is therefore complementary,
and can give the user insight
when other tools fail.
In the context of higher-order logic
and inductive datatypes,~\cite{blanchette-nonstandard-models}
describes a way to obtain
finite \emph{fragments} of nonstandard, infinite models,
via a specific first-order abstraction.
Our approach finds entire infinite models
for arbitrary first-order formulas,
not
tied to
a specific abstraction of inductive definitions,
and yields a decidable fragment.
\neta{added:}
Recently,~\cite{infinite-model-synthesis}
experimented with using syntax-guided synthesis~\cite{sygus}
in order to construct infinite models.
Our use of \templates{} resembles syntax-guided synthesis,
but the partition of the domain via summary nodes allows us
to define more complex structures in a simpler way
(using simpler LIA terms).

\para{Decidable fragments of FOL in Verification}
Existing verification approaches based on FOL
either restrict verification conditions to have finite models~\cite{paxos-made-epr,modularity-for-decidability,complete-instantiation-smt},
or use finite partial models
that do not fully satisfy the formula
but satisfy some quantifier
instantiations~\cite{cex-guided-model-synthesis,incomplete-verification-engines,bounded-horizon}.
Complementary to these approaches,
we aim to explicitly find certain infinite models
via a finite representation.
Coherent uninterpreted programs~\cite{coherent-programs,verification-modulo-axioms}
are another FOL-based approach
where programs are restricted such that
safety verification is decidable.
In contrast, we target a richer formalism
(in fact, Turing complete),
focusing
on VC checking
rather than
safety verification.
There is a rich literature
on classical decidable fragments
of FOL~\cite{voigt-thesis,gurevich-book,epr-decidable,ackermann-decidable,goedel-decidable,gurevich1976-decidable,shelah1977decidability,loeb1967-decidability,mortimer1975-decidable}.
Most of these fragments have a finite model property,
and many cannot axiomatize a linear order;
to the best of our knowledge,
\FragmentShortName{} is not subsumed
by any known decidable fragment.
Also related are decidability results
for finite satisfiability
(e.g.,~\cite{shelah1977decidability,danielski-unfo});
these fragments are similarly incomparable to \FragmentShortName{}.
Moreover, \structures{} are useful beyond \FragmentShortName{} and extend the applicability of \Solvers{},
which use the standard semantics of FOL.

\commentout{
Some decidable fragments~\cite{bsr-epr,complete-instantiation-smt} circumvent the problem
of infinite models by
restricting formulas
such that
finite counter-models
are guaranteed to
exist~\cite{modularity-for-decidability,paxos-made-epr}.
Another existing approach has been to
use finite partial models
that do not fully satisfy the formula
but satisfy some quantifier
instantiations~\cite{cex-guided-model-synthesis,incomplete-verification-engines,bounded-horizon}.
Complementary to these approaches,
we aim to directly find certain infinite models.
The decidable fragment we introduce
extends the existing EPR fragment
by allowing an ``infinite'' sort
with cyclic self-functions
and an axiomatized order relation.
Various quantified axioms of order were also used in~\cite{verification-modulo-axioms}
as a decidable extension of the class of coherent programs~\cite{coherent-programs}.
Coherent programs are defined with uninterpreted functions,
but, whereas~\cite{verification-modulo-axioms} only allows quantifiers in \emph{specific} order axioms,
and cannot support arbitrary EPR axioms,
we also allow quantifiers in the transitions and the properties,
and our decidable fragment strictly extends EPR.
Many classic decidable fragments~\cite{epr-decidable,ackermann-decidable,goedel-decidable,gurevich1976-decidable,shelah1977decidability}
cannot axiomatize a linear order.
In~\cite{loeb1967-decidability} there are no binary relations at all,
and~\cite{mortimer1975-decidable} lacks functions.
Also related are decidability results
for satisfiability over finite models
(e.g.,~\cite{shelah1977decidability,danielski-unfo}).
These fragments are incomparable to \FragmentShortName{}.
Moreover, templates are useful beyond the decidable fragment
and extend the applicability of \Solvers{}, which
allow the use of a rich first-order language assuming the standard semantics.
}

\section{Conclusion}
\label{sec:conclusion}
This paper introduced \structures{}
as an efficient way to represent and discover infinite models of FOL formulas.
We presented a symbolic search algorithm to efficiently
traverse vast (possibly infinite) families of \structures{},
encoded as \structAdj{} \templates{},
and designed a robust heuristic to construct such \templates{}.
Naturally, by restricting \structures{} via \aTemplate{}
to use a finite subset of
LIA,
we are severely limiting the kind of \structures{} that can be found.
Using empirically-proven \templates{}
strikes a balance between efficiency and expressivity.
Our evaluation of \ToolShortName{}
demonstrates the effectiveness of the technique
for finding infinite counter-models.
Beyond the empirical evaluation,
we also made a theoretical contribution:
we identified a new decidable fragment of FOL,
which extends EPR,
where satisfiable formulas always have
a (possibly infinite) model representable
by \aStructModel{} of a bounded size
that is captured by the aforementioned heuristic \templates{}.
Our technique can also be extended and
fine-tuned to different domains
using specialized \structAdj{} \templates{}.
We consider this a first step,
and posit that automatic methods
for finding infinite counter-models can have
wide uses in verification,
in particular for guiding the user
(or as part of a CEGAR-style algorithm)
towards amending the modeling,
e.g., by adding instantiations of induction
or other axiom schemes.
Though in some cases \Solvers{} and theorem provers
can use induction to find a
refutation~\cite{induction-for-smt,automating-induction-smt,vampire-integer-induction,vampire-recursive-induction,vampire-reflection-induction},
these mechanisms are incomplete
and require heuristics~\cite{kaufmann-acl2,kaufmann-acl2-case-studies}.
Infinite counter-models obtained by integrating with \ToolShortName{}
could provide hints
for instantiating the induction scheme. 

\ifnocomments
\ifoverpage
\pagebreak
\ifnum 26<\thepage\relax
\ClassError{}{OVER PAGE LIMIT}
\fi
\fi
\fi

 \section*{Data-Availability Statement}
An extended version of this paper, 
with proofs and additional examples 
can be found at~\cite{infinite-needle-arxiv}.
An artifact for reproducing the results is available 
at~\cite{infinite-needle-artifact},
and the latest version of the tool can be obtained
at~\cite{infinite-needle-artifact-latest}.
 \begin{acks}
	We thank the anonymous reviewers and the artifact evaluation committee 
	for comments which improved the paper.
	We thank Raz Lotan and Yotam Feldman for insightful discussions and comments.
	The research leading to these results has received funding from the
	European Research Council under the European Union's Horizon 2020 research and
	innovation programme (grant agreement No [759102-SVIS]).
	This research was partially supported by the Israeli Science Foundation (ISF) grant No.\ 2117/23.
\end{acks}
\bibliography{includes/references.bib}
\appendix
\ifnofull\else
	\clearpage

\section{Proofs}
\label{sec:proofs}
\subsection*{Proofs for \nameref{sec:templates} (\Cref{sec:templates})}
\iflong
\else
\begin{lemma}\label{lem:templates:term-evaluation}
	Let $\struct$ be \aStruct{}, $v$ \aConcrete{} assignment,
	$\structAss$ its \templatization{}, and $v_\subLIA$ the residual assignment.
	Then for every term $t$ over $\Sigma$,
	if
	$\bar v(t) = \pair{ n, z } \in \Domain^C$,
	then
	$\barStructAss(t) = \pair{ n, s } \in \Domain^\struct \times \Terms_\LIA$
	for some LIA term $s$ such that $\bar v_\subLIA(s) = z$.
\end{lemma}
\fi
\begin{proof}[Proof for \Cref{lem:templates:term-evaluation}]
	Structural induction over $t$.
\end{proof}

The following technical lemmas will be useful in the sequel:
\begin{lemma}
	$
	\parens{ v \brackets{ \parens{ n, z } / x } }_\struct
	=
    \structAss \brackets{ \parens{ n, x^\LIA } / x }.
	$
\end{lemma}

\begin{lemma}
	$
	\parens{ v \brackets{ \parens{ n, z } / x } }_\subLIA
	=
	v_\subLIA \brackets{ z / x^\LIA }.
	$
\end{lemma}
\begin{proof}[Proofs]
	Trivial.
\end{proof}

\begin{proof}[Proof for \Cref{thm:templates:model-checking}]
	Structural induction over $\varphi$.
\end{proof}

\subsection*{Proof for \nameref{sec:finding} (\Cref{sec:finding})}
\begin{proof}[Proof for \Cref{thm:finding:finder}]
	Structural induction over $\varphi$.
\end{proof}

\subsection*{Proofs for \nameref{sec:decidable} (\Cref{sec:decidable})}
\label{sec:proofs:decidable}
\paragraph{Correctness of the construction}
\iflong
\else
To prove \Cref{lem:decidable:single-sort-construction},
we prove the following lemmas.
First, we show that $\struct$ is well-defined.
Since all bound formulas are degenerate,
all associated terms of function will always be in bounds.
What is left to prove is that $f^\struct$ is well-defined:

\begin{lemma}
	\label{lem:decidable:f-well-defined}
	For every node $n$,
	there exists some $n'$ such that
	$
	E(n') \cap \braces{ f^M(e) \mid e \in E(n) } \neq \emptyset
	$.
	Consequently, $f^\struct$ is well-defined.
\end{lemma}

Next, we show that elements of $M$
and \concrete{} elements of $\struct$
satisfy the same atomic formulas:
\begin{lemma}
	\label{lem:decidable:M-CT-equivalence}
	For every node $n \in \Domain^\struct$, element $e \in E(n)$,
	assignments $v$, $v'$ for $M$ and $\Concrete(\struct)$ respectively,
	and atom $a$,
	we have that
	$M, v \brackets{ e / x } \models a$
	iff
	$\Concrete(\struct), v' \brackets{ \pair{ n, z } / x } \models a$
	for every $z \in \Integers$.
\end{lemma}
Since formulas in \FragmentShortName{} have only universal quantifiers,
\Cref{lem:decidable:M-CT-equivalence} implies that $M$ and $\Concrete(\struct)$ 
are \FragmentShortName{}-equivalent.
Finally, we show that $\llt^{\Concrete(\struct)}$ is a strict total order:
\begin{lemma}
	\label{lem:decidable:total-strict-order}
	If $M \models \Allt$
	then $\Concrete(\struct) \models \Allt$.
\end{lemma}
\fi

\begin{proof}[Proof for \Cref{lem:decidable:f-well-defined}]
	Trivial, since this holds for any $\tau\parens{ f^M(e) }$
	where $e \in E(n)$.
\end{proof}

For proving \Cref{lem:decidable:M-CT-equivalence},
we first prove the following:
\begin{lemma}
	\label{lem:decidable:f-marking-connection}
	For any node $n$, for any element atom $a$,
	if $n$ is marked by $a[ f(x) / x ]$, then $f^\struct(n)$ is marked by $a$.
\end{lemma}
\begin{proof}
	Let $f^\struct(n) = \pair{ n', t }$.
	Let $a$ be some element atom
	such that $n$ is marked by $a [ f(x) / x ]$.
	Let $e \in E(n)$ be an element such that $f^M(e) \in E(n')$
	By definition, $M, v[e / x] \models a[ f(x) / x ]$,
	or equivalently, $M, v [ f^M(e) / x ] \models a$.
	Since element atoms agree for all class elements of a node,
	$n'$ is marked by $a$.
\end{proof}

\begin{lemma}
	\label{lem:decidable:x-marking-satisfy}
	For any node $n$,
	for any relation element atom $a$,
	if $n$ is marked by $a$,
	then for any assignment $v$, for any $i \in \Integers$,
	$\Concrete(\struct), v[ \pair{ n, i } / x] \models a$.
\end{lemma}
\begin{proof}
	Trivial.
\end{proof}

\begin{corollary}
	\label{cor:decidable:f-marking-satisfy}
	For any relation element atom $a$,
	if $n$ is marked by $a [ f(x) / x ]$,
	then for any assignment $v$, for any $i \in \Integers$,
	$\Concrete(\struct), v [ \pair{ n, i } / x] \models a [ f(x) / x ]$.
\end{corollary}

\begin{lemma}
	\label{lem:decidable:correct-marking}
	Note that for any element $e$ in $M$,
	by definition of the partitions,
	$\tau(e)$ is marked by an atom $a$
	iff
	$e$ satisfies that atom.
\end{lemma}
\begin{proof}
	By definition of the partitions.
\end{proof}

\begin{lemma}
	\label{lem:decidable:closed-literals}
	For any constant $c$,
	for any element atom $a$,
	$M \models a[c / x]$
	iff $\tau \parens{ c^M }$ is marked by $l$
	iff $\Concrete(\struct) \models a \brackets{ c / x }$.
\end{lemma}
\begin{proof}
	Let $n = \tau \parens{ c^M }$, then $E(n) = \braces{ c^M }$.
	$M \models a[c / x] \iff$ for any assignment $v$,
	$M, v[ c^M / x ] \models a$.
	If $M, v[ c^M / x ] \models a$, then $n$ is marked by $a$.
\end{proof}

\begin{lemma}
	\label{lem:decidable:self-function-marking}
	Let $S = \braces{ x } \cup \braces{ h(x) \mid h \text{ is a unary function symbol} }$
	be the set of mixed-atom terms.
	For any node $n$ and terms $t, t' \in S$
	if $n$ is marked by $a = t \bowtie t'$,
	for some ${\bowtie} \in \braces{  \prec, \liaeq, \succ}$,
	then for any assignment $v$, for any $i \in \Integers$,
	$\Concrete(\struct), v [ \parens{ n, i } / x ] \models a$.
\end{lemma}
\begin{proof}
	Let us denote the interpretation of the terms $t, t'$
	for $x \mapsto \pair{n, i}$:
	$\pair{m, j} = \overline{\brackets{ x \mapsto \pair{n, i} }} (t)$,
	$\pair{m', j'} = \overline{\brackets{ x \mapsto \pair{n, i} }} (t')$.
	
	\paragraph{In case either $m$ or $m'$ is a regular node}
	We consider the case for $m$ 
	(and the case for $m'$ is symmetrical).
	Since $m$ is a regular node, 
	then $j = 0$ (from \Cref{eq:decidable:f-regular})
	and there exists a ground term $g$ such that
	$m$ is labeled by $x \foleq g$ and
	$g^{\Concrete(\struct)} = \pair{m, 0}$.

	Necessarily, $n$ is marked by
	$g \bowtie t'$, and
	from \Cref{lem:decidable:f-marking-connection}
	we get that $m'$ is marked with $g \bowtie x$.
	From \Cref{enum:decidable:order-regular},
	$\pair{m, 0} \bowtie^{\Concrete(\struct)} \pair{m', j'}$,
	and therefore $\Concrete(\struct), v [ \parens{ n, i } / x ] \models a$.
	
	\paragraph{In case both $m$ and $m'$ are summary nodes}
	Let us denote by ${\bowtie_\subLIA} \in \braces{ <, =, > }$ 
	the comparison symbol that matches $\bowtie$.
	Since $m, m'$ are summary nodes, 
	according to \Cref{eq:decidable:f-summary},
	$j = i + k, j' = i + k'$ 
	then, since $\mu$ is defined according to the atoms
	of $n$ ($a$ among them),
	$k \bowtie_\subLIA k'$
	(and therefore $j \bowtie_\subLIA j'$).
	
	If there exists no ground term $g$ such that
	$m$ is marked by $x \prec g$ and $m'$ by $x \succ g$
	or vice versa (i.e., $m$ and $m'$ are in the same segment),
	then according to \Cref{enum:decidable:order-otherwise},
	$\pair{m, j} \bowtie^{\Concrete(\struct)} \pair{m', j'}$.
	
	Otherwise, there exists a ground term $g$ as above,
	and let us assume w.l.o.g. that 
	$m$ is marked by $x \prec g$ and $m'$ by $x \succ g$,
	thus (from \Cref{lem:decidable:correct-marking})
	$\pair{m, j} 
		\prec^{\Concrete(\struct)} 
		g^{\Concrete(\struct)} 
		\prec^{\Concrete(\struct)} 
		\pair{m', j'}
	$.
	$n$ is necessarily marked with the atoms
	$t \prec g$ and $t' \prec g$,
	which must agree with $a$
	(since $\prec$ is strict total order
	in the original model $M$),
	$\Concrete(\struct), v [ \parens{ n, i } / x ] \models a$.
\end{proof}

\begin{proof}[Proof for \Cref{lem:decidable:M-CT-equivalence}]
Follows from \Cref{lem:decidable:f-marking-connection,lem:decidable:x-marking-satisfy,cor:decidable:f-marking-satisfy,lem:decidable:self-function-marking,lem:decidable:correct-marking,lem:decidable:closed-literals}.
\end{proof}

\begin{proof}[Proof for \Cref{lem:decidable:total-strict-order}]
Follows from \Cref{enum:decidable:order-regular,enum:decidable:order-ground,enum:decidable:order-otherwise}
and strict total order properties of $\llt^M$.
\end{proof}

 	\iflong\else

\section{Examples}
\label{sec:examples}

\subsection*{Example for \nameref{sec:templates} (\Cref{sec:templates}): Half-space Summary Nodes}
\begin{example}
	Unbounded summary nodes,
	i.e., summary nodes whose bound formula is $\top$
	that represent a full copy of $\Integers$,
	are not always sufficient in order to represent 
	satisfying models of formula.
	As an example where a strict subset of $\Integers$ is needed,
	consider the vocabulary $\Sigma = \braces{ c, f(\cdot), \llt }$,
	and the formula
	$
	\varphi = \Allt \land \forall x .\
	c \lleq x \wedge x \llt f(x) \wedge \forall y.\ x \llt y \to f(x) \lleq y
	$
	where $t_1 \lleq t_2$ is a shorthand for $t_1 \llt t_2 \vee t_1 \approx t_2$.
	The only model of $\varphi$ is infinite, hence there must be at least one summary node $n$.
	Assuming that $c$ is interpreted as a regular node, and the interpretation of $f(c)$ is some element $\parens{n,z}$ of $n$,
	then $z$ has to be the minimal element in the subset associated with $n$,
	otherwise, we could always find an element between $c$ and $f(c)$.
	Therefore, in this example, $n$ must be bounded by $x \geq z$.
	(Adding a regular node for $f(c)$ does not solve the problem, only shifts it.)
\end{example}

\subsection*{Example for the Order Construction of \nameref{sec:decidable} (\Cref{sec:decidable})}
\label{sec:decidable-order-illustration}
\begin{figure}
	\scalebox{0.725}{
		\begin{tikzpicture}[->,>=stealth',shorten >=1pt,auto,scale=0.67]
			\tikzset{ground/.style={minimum size=30pt}}
			\tikzset{cnode/.style={circle,draw,minimum size=26pt,inner sep=0pt}}
			
			\node (predots) at (-6.5,1)
			{ \dots };
			
			\node[cnode] (s1-1) at (-4.5,1)
			{ $\pair{S_1,0}$ };
			
			\node[cnode] (s1-2) at (-2.5, 1)
			{ $\pair{S_1,1}$ };
			
			\node (g1-predots) at (-0.5,1)
			{ \dots };
			
			\node[cnode,ground] (g1) at (1,1)
			{ $\bm{\pair{G_1,0}}$ };
			
			\node (g1-postdots) at (2.5,1)
			{ \dots };
			
			\node[cnode] (s2-1) at (3.5,2)
			{ $\pair{S_2,0}$ };
			
			\node[cnode] (s3-1) at (4.5,0)
			{ $\pair{S_3,0}$ };
			
			\node[cnode] (s2-2) at (5.5,2)
			{ $\pair{S_2,1}$ };
			
			\node[cnode] (s3-2) at (6.5,0)
			{ $\pair{S_3,1}$ };
			
			\node (g2-predots) at (7.5,1)
			{ \dots };
			
			\node[cnode,ground] (g2) at (9,1)
			{ $\bm{\pair{G_2,0}}$ };
			
			\node (g2-postdots) at (10.5,1)
			{ \dots };
			
			\node[cnode] (s4-1) at (11.5,2.5)
			{ $\pair{S_4,0}$ };
			
			\node[cnode] (s5-1) at (12.5,1)
			{ $\pair{S_5,0}$ };
			
			\node[cnode] (s6-1) at (13.5,-0.5)
			{ $\pair{S_6,0}$ };
			
			\node[cnode] (s4-2) at (14,2.5)
			{ $\pair{S_4,1}$ };
			
			\node[cnode] (s5-2) at (15,1)
			{ $\pair{S_5,1}$ };
			
			\node[cnode] (s6-2) at (16,-0.5)
			{ $\pair{S_6,1}$ };
			
			\node (g3-predots) at (17,1)
			{ \dots };
			
			\node[cnode,ground] (g3) at (18.5,1)
			{ $\bm{\pair{G_3,0}}$ };
			
			\node[cnode,ground] (g4) at (20.5,1)
			{ $\bm{\pair{G_4,0}}$ };
			
			\path
			(predots) edge (s1-1)
			(s1-1) edge (s1-2)
			(s1-2) edge (g1-predots)
			
			(g1-postdots.north) edge (s2-1)
			(s2-1) edge (s3-1)
			(s3-1) edge (s2-2)
			(s2-2) edge (s3-2)
			(s3-2) edge (g2-predots.south)
			
			(g2-postdots.north) edge (s4-1)
			(s4-1) edge (s5-1)
			(s5-1) edge (s6-1)
			(s6-1) edge (s4-2)
			(s4-2) edge (s5-2)
			(s5-2) edge (s6-2)
			(s6-2) edge (g3-predots.south)
			
			(g3) edge (g4)
			;
			
			\path[-,very thick]
			(1,3.5) edge (g1)
			(g1) edge (1,-1.5)
			
			(9,3.5) edge (g2)
			(g2) edge (9,-1.5)
			
			(18.5,3.5) edge (g3)
			(g3) edge (18.5,-1.5)
			
			(20.5,3.5) edge (g4)
			(g4) edge (20.5,-1.5)
			;
			
		\end{tikzpicture}
	}
	
	\caption{\Concretization{} of \aStruct{} constructed as in \Cref{sec:single-sort-construction}.
	Each circle depicts an element.
	The labels $S_i, G_i$ denote different sets of atoms
	which correspond to different nodes in the \structure{}.
	I.e., each $S_i$ and $G_j$ is a unique set of
	element atoms \ref{enum:domain:element},
	image atoms \ref{enum:domain:image}, and
	mixed atoms \ref{enum:domain:mixed}.
	The sets $G_1, G_2, \dots$ include at least one atom
	of the form $x \foleq g$,
	where $g$ is some ground term,
	thus refer to regular nodes.
	The sets $S_1, S_2, \dots$ do not include any atom
	of the form $x \foleq g$,
	thus refer to summary nodes.
	The arrows between the elements depict 
	$\llt^{\Concrete\parens{\struct}}$
	by denoting immediate successors.
	Note that the order between the \concrete{} elements of,
	for example, $S_4, S_5$ and $S_6$ 
	follows a fixed pattern.
	}
	\label{fig:decidable:construction}
\end{figure}  
An illustration of the interpretation of $\llt$,
as defined in \labelcref{enum:decidable:order-regular,enum:decidable:order-ground,enum:decidable:order-otherwise},
is shown in \Cref{fig:decidable:construction}.
The elements of $S_2$ are all greater than the elements $S_1$,
which means,
according to \ref{enum:decidable:order-ground},
that there exists some ground term $g$
such that $x \llt g \in S_1$ and $x \lgt g \in S_2$
(and $x \foleq g \in G_1$, which delimits the two segments).
The order of the elements of $S_4, S_5$ and $S_6$
indicates,
according to \ref{enum:decidable:order-otherwise},
that $S_4 \ll S_5 \ll S_6$, arbitrarily.

 	\fi

\section{Evaluation Details}
\label{sec:eval-details}
\paragraph{Echo Machine}Defining
\begin{footnotesize}
	\begin{align*}
		\code{prev\_echoed}[\code{T}, \code{V}] &= 
		\PreviousFun(\code{T}) \llt \code{T}
		\land
		\Relation{echo}(\PreviousFun(\code{T}), \code{V})
		\\
		\Action{echo\_start}[\code{V}] &=
		\parens{
			\forall \code{V}'.
			\neg \Relation{echo}(\Const{start}, \code{V}')
		}
		\\
		&\land
		\parens{
			\forall \code{T}', \code{V}'. 
			\Relation{echo}'(\code{T}', \code{V}')
			\leftrightarrow
			\parens{
				\parens{ \code{T}' \approx \Const{start} \land \code{V}' \approx \code{V}}
				\lor
				\parens{
					\code{T}' \not\approx \Const{start}
					\land
					\code{V}' \not\approx \code{V}
					\land
					\Relation{echo}(\code{T}', \code{V}')
				}
			}
		}
		\\
		\Action{echo\_prev}[\code{T}, \code{V}] &=
		\parens{
			\forall \code{V}'.
			\neg \Relation{echo}(\code{T}, \code{V}')
		}
		\land
		\code{prev\_echoed}[\code{T}, \code{V}] \\
		&\land
		\parens{
			\forall \code{T}', \code{V}'. 
			\Relation{echo}'(\code{T}', \code{V}')
			\leftrightarrow
			\parens{
				\parens{ \code{T}' \approx \code{T} \land \code{V}' \approx \code{V}}
				\lor
				\parens{
					\code{T}' \not\approx \code{T}
					\land
					\code{V}' \not\approx \code{V}
					\land
					\Relation{echo}(\code{T}', \code{V}')
				}
			}
		}
		\\
		\code{transition} &=
		\exists \code{T}, \code{V}.
		\Action{echo\_start}[\code{V}]
		\lor
		\Action{echo\_prev}[\code{T}, \code{V}]
		\\
		\code{safety} &=
		\forall \code{T}, \code{T}', \code{V}, \code{V}'.
		\parens{
			\Relation{echo}(\code{T}, \code{V})
			\land
			\Relation{echo}(\code{T}', \code{V}') 
		} \to
		\code{V} \approx \code{V}'
		\\
		\code{aux\_invariant} &=
		\forall \code{T}, \code{V}
		\Relation{echo}(\code{T}, \code{V})
		\to
		\code{T} \approx \Const{start} 
		\lor \code{prev\_echoed}[\code{T}, \code{V}],
	\end{align*}
\end{footnotesize}
the formula we check is
\[
\Allt 
\land \code{safety} 
\land \code{aux\_invariant} 
\land \code{transition} 
\land \neg \code{safety}'
.
\]
The \structModel{} found by \ToolShortName{}
is depicted in \Cref{fig:overview:template}.

\paragraph{Ring Leader}Defining
\begin{footnotesize}
	\begin{align*}
		\code{safety} &=
		\parens{ \exists \code{N}. \Relation{leader}(\code{N}) }
		\lor
		\parens{ \exists \code{N}.
			\neg \Relation{sent}(N)
		}
		\lor
		\parens{
			\exists \code{N}, \code{N}'.
			\Relation{pending}(\code{N}, \code{N}')
		}
		\\
		\code{aux\_invariant} &=
		\forall \code{N}.
		\parens{ 
			\Relation{sent}(\code{N}) 
			\land 
			\neg \Relation{leader}(\code{N}) 
		} \to
		\parens{
			\Relation{pending}(\code{N}, \FunDef{target}(\code{N}))
			\lor
			\code{N} \llt \FunDef{target}(\code{N})
		},
	\end{align*}
\end{footnotesize}
where $\FunDef{target}$ is a Skolem function,
the formula we check is
\[
\Allt 
\land \code{aux\_invariant} 
\land \neg \code{safety}
.
\]
The \structModel{} found by \ToolShortName{}
is depicted in \Cref{fig:examples:ring-leader}

\begin{figure}
	\centering
	\begin{tikzpicture}[->,>=stealth',shorten >=1pt,auto]
		\node[tnode, accepting]
		(n) at (1,1)
		{};
		\node [below = 0pt of n]
		{\tiny $x \geq 0$};
		
		\node [above right = 0pt of n,xshift=-5pt]
		{\tiny $\alpha$};
		
		\path
		
		(n) edge [loop left] 
		node {\tiny $\FunDef{\tiny target}: x_1 + 1$} (n)
		
		(n) edge [loop right] node {\tiny $\llt: x_1 < x_2$} (n)
		;
	\end{tikzpicture}
	\caption{The \structModel{} found by \ToolShortName{} for the Ring Leader example.
		The formulas for the relations are 
		$\Relation{sent}(\alpha) = \top$,
		$\Relation{leader}(\alpha) = \bot$,
		and
		$\Relation{pending}(\alpha, \alpha) = \bot$.}
	\label{fig:examples:ring-leader}
\end{figure}

\paragraph{Sorted List}Defining
\begin{footnotesize}
	\begin{align*}
		\code{linked\_list} &=
		\forall \code{N}.
		\parens{ \code{N} \approx \Const{nil} }
		\liff
		\parens{ \Next(\code{N}) \approx \code{N} }
		\\
		\code{sorted\_def} &=
		\forall \code{N}.
		\Sorted(\code{N}) \liff
		\parens{
			\code{N} \approx \Const{nil}
			\lor
			\parens{
				\Sorted \parens{ \Next (\code{N}) }
				\land
				\Next(\code{N}) \llt \code{N}
			}
		}
		\\
		\code{property} &=
		\forall \code{N}.
		\Sorted(\code{N}) \to \Const{nil} \lleq \code{N}
		,
	\end{align*}
\end{footnotesize}
the formula we check is
\[
\Allt 
\land \code{linked\_list} 
\land \code{sorted\_def} 
\land \neg \code{property}
.
\]
The \structModel{} found by \ToolShortName{}
is depicted in \Cref{fig:examples:sorted}.

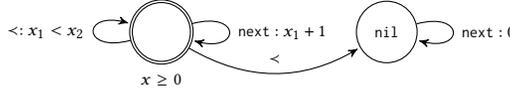
\begin{figure}
	\centering
	\begin{tikzpicture}[->,>=stealth',shorten >=1pt,auto]		
		\node[tnode, accepting]
		(n) at (1,1)
		{};
		\node [below = 0pt of n]
		{\tiny $x \geq 0$};
		
		\node[tnode]
		(nil) at (4,1)
		{ \tiny $\Const{\tiny nil}$ };

		\path
		
		(n) edge [loop right] 
		node {\tiny $\FunDef{\tiny next}: x_1 + 1$} (n)
		
		(nil) edge [loop right] 
		node {\tiny $\FunDef{\tiny next}: 0$} (nil)
		
		(n) edge [loop left] node {\tiny $\llt: x_1 < x_2$} (n)
		(n) edge [bend right] node {\tiny $\llt$} (nil)
		;
	\end{tikzpicture}
	\caption{The \structModel{} found by \ToolShortName{} for the Sorted List example.
		For all nodes, the formula for $\Sorted$ is $\top$.}
	\label{fig:examples:sorted}
\end{figure}

\paragraph{List Segment}Defining
\begin{footnotesize}
	\begin{align*}
		\code{nil\_def} &=
		\parens{ \code{N} \approx \Const{nil} }
		\liff
		\parens{ \Next(\code{N}) \approx \code{N} }
		\\
		\code{list\_def} &=
		\forall \code{N}.
		\List(\code{N}) \liff
		\parens{ 
			\code{N} \approx \Const{nil} 
			\lor 
			\List(\Next(\code{N}))
		}
		\\
		\code{lseg\_def} &=
		\forall \code{N}, \code{N}'.
		\ListSegment(\code{N}, \code{N}')
		\liff
		\parens{
			\code{N} \approx \code{N}'
			\lor
			\ListSegment \parens{ \Next(\code{N}), \code{N}' }
		}
		\\
		\code{property} &=
		\forall \code{N}, \code{N}'.
		\parens{
			\ListSegment(\code{N}, \code{N}')
			\land
			\List(\code{N}')
		} \to
		\List(\code{N})
		,
	\end{align*}
\end{footnotesize}
the formula we check is
\[
\code{nil\_def}
\land \code{list\_def}
\land \code{lseg\_def}
\land \neg \code{property}
.
\]
The \structModel{} found by \ToolShortName{}
is depicted in \Cref{fig:examples:lseg}.

\begin{figure}
	\centering
	\begin{tikzpicture}[->,>=stealth',shorten >=1pt,auto]		
		\node[tnode, accepting]
		(n) at (1,1)
		{};
		\node [below = 0pt of n]
		{\tiny $x \leq 0$};
		\node [above right = 0pt of n,xshift=-5pt]
		{ \tiny $\alpha$ };
		
		\node[tnode]
		(nil) at (6,1)
		{ \tiny $\Const{\tiny nil}$ };

		\path
		
		(n) edge [loop right] 
		node {\tiny $\FunDef{\tiny next}: x_1 - 1$} (n)
		
		(nil) edge [loop right] 
		node {\tiny $\FunDef{\tiny next}: 0$} (nil)
		
		(n) edge [loop left]
		node {\tiny $\Relation{\tiny lseg}$}
		(n)
		
		(nil) edge [loop left]
		node {\tiny $\Relation{\tiny lseg}$}
		(nil)
		
		(n) edge [bend right]
		node {\tiny $\Relation{\tiny lseg}$}
		(nil)
		
		(nil) edge [bend right]
		node {\tiny $\Relation{\tiny lseg}$}
		(n)
		;
	\end{tikzpicture}
	\caption{The \structModel{} found by \ToolShortName{} for the List Segment example.
		The $\List$ formulas are $\List(\Const{nil}) = \top$
		and $\List(\alpha) = \bot$.}
	\label{fig:examples:lseg}
\end{figure}

\paragraph{Axiomatic Presburger Arithmetic}Defining
\begin{footnotesize}
	\begin{align*}
		A_1 &=
		\forall x. s(x) \not\approx \Const{zero}
		\\
		A_2 &=
		\forall x, y. s(x) \approx s(y) \to x \approx y
		\\
		A_3 &=
		\forall x. x + \Const{zero} \approx x
		\\
		A_4 &=
		\forall x,y. x + s(y) \approx s(x + y)
		\\
		\code{inductive\_commutativity} &=
		\forall x,y. 
		\parens{ x + y \approx y + x } 
		\to 
		\parens{ x + s(y) \approx s(y) + x }
		,
	\end{align*}
\end{footnotesize}
the formula we check is
\[
A_1 \land A_2 \land A_3 \land A_4 
\land \neg \code{inductive\_commutativity}
.
\]
The \structModel{} found by \ToolShortName{}
has 3 regular nodes and 1 summary node,
and the different regular nodes all behave as 
neutral elements 
(i.e., for all them it holds that
$\forall x. x + e \approx x$
).
We refine our modeling with the additional axiom
\[
A_5 = \forall x, y. x + y \approx x \to y \approx \Const{zero}.
\]
This eliminates the three different ``zeros''
and instead \ToolShortName{} finds \aStructModel{}
where there are two summary nodes,
representing a nonstandard model of arithmetic
where there two parallel ``number lines''.
See \Cref{fig:evaluation:non-commutative}.
This infinite model can be eliminated,
by adding the lemma
\[
A_6 = \forall x, y. s(x) + y \approx s(x + y),
\]
which makes the formula unsatisfiable,
proving the commutativity of addition.

\begin{figure}
	\centering
	\begin{subfigure}{0.45\linewidth}
		\centering
		\begin{minipage}[c]{1\textwidth}\mbox{}\\[-\baselineskip]
			\scalebox{0.9}{
				\begin{tikzpicture}[->,>=stealth',shorten >=1pt,auto]
					\node[tnode]
					(zero) at (1,1)
					{\tiny $\Const{\figLfont{zero}}$};
					
					\node[tnode,accepting]
					(alpha) at (3,1)
					{};
					\node [below = 0pt of alpha]
					{\figLfont{$x \geq 0$}};
					\node [above right = 0pt of alpha]
					{$\alpha$};
					
					\node[tnode,accepting]
					(beta) at (5.5,1)
					{};
					\node [below = 0pt of beta]
					{\figLfont{$x \geq 0$}};
					\node [above right = 0pt of beta]
					{$\beta$};
					
					\path
					(zero) edge node {\figLfont{$s:0$}} (alpha)
					
					(alpha) edge [loop above] node [above] 
					{\figLfont{$s: x_1 + 1$}} (alpha)
					
					(beta) edge [loop above] node [above] 
					{\figLfont{$s: x_1 + 1$}} (beta)
					;
					
\end{tikzpicture}
			}
		\end{minipage}
	\end{subfigure}
	\begin{subfigure}{0.45\linewidth}
		\centering
		\begin{minipage}[c]{1\textwidth}\mbox{}\\[-\baselineskip]
			\begin{footnotesize}
				\begin{tabular}{@{}r|ccc@{}}
					\hline
					$+$ & $\Const{zero}$ & $\alpha$ & $\beta$
					\\
					\hline
					$\Const{zero}$
					& $\pair{\Const{zero},0}$
					& $\pair{\alpha,x_2}$
					& $\pair{\beta,x_2}$
					\\
					$\alpha$
					& $\pair{ \alpha,x_1 }$
					& $\pair{ \alpha, x_1 + x_2 + 1 }$
					& \cellcolor{yellow}$\pair{ \beta, x_1 + x_2 }$
					\\
					$\beta$
					& $\pair{ \beta,x_1 }$
					& \cellcolor{yellow}$\pair{ \beta, x_1 + x_2 + 1 }$
					& $\pair{ \alpha, x_1 + x_2 + 1 }$
					\\
					\hline
				\end{tabular}
			\end{footnotesize}
		\end{minipage}
	\end{subfigure}
	\vspace{-5pt}
	\caption{
		Non-commutative addition.
		Note the difference between the highlighted cells.\neta{vspace here}}
	\label{fig:evaluation:non-commutative}
	\vspace{-5pt}
\end{figure}
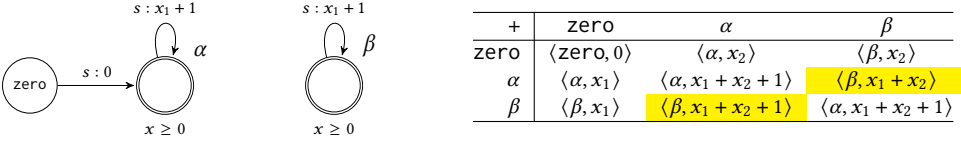  	

\section{Technical Construction for Finding \STructures{}}
\label{sec:full-finder-transformation}

In this section, we fill out the technical details of
the finding \structures{} algorithm,
presented in \Cref{sec:finding}.

Recall that the model checking encoding of
$\varphi$
(over vocabulary $\Sigma$)
w.r.t.\ \aStruct{} $\struct$ and \aConcrete{} assignment
$v \colon \Variables \to \Domain^{\Concrete(\struct)}$
is defined w.r.t.\ to \aStructAssignment{}
$\structAss \colon \Variables \to \Domain \times \Variables^\supLIA$
obtained by the \templatization{} of $v$
(see \Cref{def:templatization}).
In the model finding setting,
\aConcrete{} assignment is not given,
but needs to be found together with the \structure in $\finder$.
Each \structAssignment{} $\structAss$
is therefore used as a symbolic representation
of all the \concrete{} assignments
whose \templatization{} is $\structAss$.
Note that the domain of $\finder$ is fully specified,
thus \aStructAssignment{} is well-defined w.r.t.\ $\finder$.

In the model checking encoding,
\aStructAssignment{} $\structAss$ is extended to any term over
$\Sigma$ based on a \emph{specific} \structure{} $\struct$.
Similarly,
the model finding encoding extends \aStructAssignment{}
$\structAss$ according to \aStructAdj{} \template{} $\finder$
to any term $t$ over $\Sigma$ via
$v_{\struct, \finder}$,
which reflects the possible interpretations of $t$
in \emph{any} \structure{} in $\finder$.
Formally,
$v_{T, \finder} \colon \Terms
\to \powerset{\Prop \times \Domain \times \Terms_\LIA}$,
is defined below,
where $\Prop$ denotes the set of
all propositional formulas over
the auxiliary Boolean variables, and
the triple $\parens{ p, n, s } \in v_{T, \finder}(t)$
is understood as
``when $p$ is true,
it induces the interpretation $\pair{ n, s }$ for $t$''.

\begin{align*}
	v_{T, \finder} (x)
	&= \braces{ \parens{ \top, n, s } }
	\quad \text { where }
	\structAss(x) = \pair{n, x^\supLIA}
	\text{ and }
	s = x^\supLIA \text{ if } n \in \Domain^S, s = 0 \text{ otherwise }
	\\
	v_{T, \finder} (c)
	&= \braces{
		\parens{ \GuardFlag{ c \to n }, n, 0 }
		\mid n \in \Domain^R  } \\
v_{T, \finder} \parens{ f(t_1, \dots, t_k) }
	&= \braces{
		\begin{array}{l}
			\parens{
				\bigwedge^k_{i = 1} p_i
				\land
				\GuardFlag{ f \parens{ n_1, \dots, n_k } \to \pair{n, s} },
				n,
				s \brackets{ s_i / x_i }
				\brackets{ 
					\IntFlag{f \parens{ n_1, \dots, n_k } }{j} / \IntFlag{}{j} 
				} 
			} \mid
			\\
			\qquad
			\parens{ p_i, n_i, s_i } \in v_{T, \finder} (t_i),
			n \in \Domain, s \in \InterpF(f)
\end{array}
	}
\end{align*}

Therefore, for each term $t$ over $\Sigma$,
$v_{T, \finder}$ defines the potential set of
interpretations $\hat \structAss(t)$ of $t$
in \structures{} represented by $\finder$,
where each interpretation is ``guarded''
by the assignment to the auxiliary Boolean variables
under which it is obtained.

\iflong \else
\begin{figure}[t]
\begin{framed}
			\vspace{-0.25cm}
			\begin{align*}
				\FinderTransform^\finder_{\structAss} \parens{ R \parens{ t_1, \dots, t_k } }
				& = \bigwedge_{
					\tiny
					\begin{array}{c}
						\parens{ p_i, n_i, s_i } \in v_{T, \finder} (t_i),
						\\
						\varphi \in \InterpF(R)
				\end{array}
			}
			\parens{
				\bigwedge_i p_i
				\land
				\GuardFlag{R \parens { n_1, \dots, n_k } \colon \varphi}
			}
			\to \varphi \brackets{ s_i / x_i }
			\brackets{ \IntFlag{ R \parens { n_1, \dots, n_k } }{j} / \IntFlag{}{j} }
			\\
			\FinderTransform^\finder_{\structAss} \parens{  t_1 \approx t_2 }
			& =	\bigwedge_{
				\tiny
				\begin{array}{c}
					\parens{ p, n, s } \in v_{T, \finder} (t_1),
					\\
					\parens{ p', n', s' } \in v_{T, \finder} (t_2)
				\end{array}
			} \parens{ p \land p' } \to \Eq \parens{ n, s, n', s' }
\\
			&\omit\hfill $\displaystyle
			\text{ where } \Eq \parens{ n, s, n', s' } = \left\{
			\begin{array}{ll}
				\bot & \text{ if } n \neq n' \\
				\top & \text{ if } n = n' \in \Domain^R\\ s \approx s' & \text{ otherwise }
			\end{array}
			\right.$
			\\
			\FinderTransform^\finder_{\structAss} \parens{ \neg \varphi }
			& = \neg \parens{ \FinderTransform^\finder_{v_{T, \finder}} (\varphi) }
			\\
			\FinderTransform^\finder_{\structAss} \parens{ \varphi \circ \psi }
			& = \FinderTransform^\finder_{\structAss} (\varphi)
			\circ \FinderTransform^\finder_{\structAss}(\psi)
			\\
			\FinderTransform^\finder_{\structAss} \parens{ \forall x.\varphi }
			& =
			\forall x^\supLIA.\ \bigwedge_{ n \in \Domain}  \parens{ 
				\BoundsF(n)
					\brackets{ x^\supLIA/x }
					\brackets{ \IntFlag{n}{j} / \IntFlag{}{j} }
				\to
			\FinderTransform^\finder_{\structAss \brackets{ \pair{ n, x^\supLIA } / x }}
			(\varphi)}
\\
		\FinderTransform^\finder_{\structAss} \parens{ \exists x.\varphi }
		& =
		\exists x^\supLIA.\ \bigvee_{ n \in \Domain_\sort} \parens{
			\BoundsF(n)
				\brackets{ x^\supLIA/x } 
				\brackets{ \IntFlag{n}{j} / \IntFlag{}{j} }
			\land
			\FinderTransform^\finder_{\structAss \brackets{ \pair{ n, x^\supLIA } / x }}
			(\varphi)}
\end{align*}
\vspace{-0.25cm}
\end{framed}
\caption{Parameteric transformation by \structAdj{} \template{} $\finder$.
	\\
	\neta{$\BoundsF(n)$ abuse of notation}
}
\label{fig:finding:transformation}
\end{figure}
\fi

To capture satisfaction of $\varphi$ in
\aStruct{} in $\finder$ defined by
some assignment to the auxiliary variables
w.r.t.\ some \concrete{} assignment represented
by the \structAssignment{}t $\structAss$ we use the transformation
$\FinderTransform^\finder_{\structAss}(\varphi)$,
\iflong
\else
defined in \Cref{fig:finding:transformation}.
\fi
The difference compared to the transformation $\FinderTransform^T_{\structAss}(\varphi)$  used for model checking
is in the transformation of atomic formulas,
which for every term $t$ over $\Sigma$ considers the \emph{set} of interpretations in $v_{T, \finder} (t)$,
as opposed to $\hat \structAss(t)$, and in the use of the parametric bound formulas.

\begin{example}
Let us consider an example where
the vocabulary has a single sort, a single unary function $\Next$,
and a single binary relation symbol $\llt$,
and a consider \aTemplate{}
$\finder = \parens{ \Domain, \BoundsF, \InterpF }$,
whose domain includes two summary nodes 
$\Domain = \braces{ n_1, n_2 }$,
there are only degenerate bound formulas
$\BoundsF = \braces{ \top }$,
and the interpretations are
$\InterpF(\Next) = \braces{ 0, x_1 + \IntFlag }$,
and
$\Interp(\llt) = \braces{ \top, \bot, x_1 \leq x_2, x_1 < x_2 }$.
Let $\structAss$ be \aStructAssignment{} where
$\structAss(x) = \pair{ n_1, x^\supLIA }$.
Then the parametrically transformed formula for
$x \llt \Next (x)$
is the conjunction of the following:
\[
\left\{
\begin{array}{l}
\vdots
\\
\parens{
\GuardFlag{\Next(n_1) \to \pair{ n_2,0 }}
\land
\GuardFlag{ \llt(n_1, n_2) \colon \top}
} \to \top
\\
\parens{
\GuardFlag{\Next(n_1) \to \pair{ n_2, 0 }}
\land
\GuardFlag{ \llt(n_1, n_2) \colon \bot}
} \to \bot
\\
\parens{
\GuardFlag{ \Next(n_1) \to \pair{ n_2, 0 } }
\land
\GuardFlag{ \llt(n_1, n_2) \colon <}
} \to x^\supLIA < 0
\\
\parens{
\GuardFlag{\Next(n_1) \to \pair{ n_2 , 0 }}
\land
\GuardFlag{ \llt(n_1, n_2) \colon \leq}
} \to x^\supLIA \leq 0
\\
\vdots
\\
\parens{
\GuardFlag{\Next(n_1)\to \pair{ n_2, x_1 + \IntFlag } }
\land
\GuardFlag{ \llt(n_1, n_2) \colon <}
} \to x^\supLIA < x^\supLIA + \IntFlag{ \Next(n_1) }
\\
\vdots
\end{array}
\right.
\]
\end{example}

Finally, to state that $\varphi$ is satisfied by \aStruct{} in $\finder$ induced by the auxiliary variables
under some \concrete{} assignment represented 
by some \structAssignment{}, we define
\[
\varphikey = \bigvee_{\structAss \in V_\finder^\varphi}
\GuardFlag{ \structAss } \land \FinderTransform^\finder_{\structAss},
\]
where $V_\finder^\varphi$
denotes the set of different \structAssignments{}
to the free variables of $\varphi$ in $\finder$
(where a fixed arbitrary assignment is chosen to the remaining variables),
and $\GuardFlag{ \structAss }$ is an auxiliary Boolean variable,
representing the choice of $\structAss$.
This set is finite as each \structAssignment{} corresponds to a node assignment.
If $\varphi$ is closed,
$V_\finder^\varphi$ consists of a single (arbitrary) \structAssignment{}.

\FindingTheorem*

\begin{proof}
Similarly to the proofs for
\Cref{lem:templates:term-evaluation,thm:templates:model-checking}
\end{proof}
 \fi
\end{document}